\documentclass[galaxies,review,accept,moreauthors,pdftex,12pt,a4paper]{mdpi}

\usepackage{amsmath,amssymb}
\usepackage{soul}
\lastpage{465}
\doinum{10.3390/galaxies2030410}
\pubvolume{2}
\pubyear{2014}
\history{Received: 30 May 2014; in revised form: 7 July 2014 / Accepted: 8 July 2014 / \linebreak Published: 28 July 2014}

\Title{Generalized Curvature-Matter Couplings in Modified Gravity}

\Author{Tiberiu Harko $^{1}$ and Francisco S.N.~Lobo $^{2,}$*}

\address{%
$^{1}$ Department of Mathematics, University College London, Gower Street, London WC1E 6BT, UK; E-Mail: t.harko@ucl.ac.uk \\
$^{2}$ Centro de Astronomia e Astrof\'{\i}sica da Universidade de Lisboa, Campo Grande,  Edif\'{\i}cio~C8~1749-016~Lisboa, Portugal}

\corres{ E-Mail: flobo@cii.fc.ul.pt; \linebreak Tel.: +351-217-500-986; Fax: +351-217-500-977.} 

\abstract{
In this work, we review a plethora of modified theories of gravity with generalized curvature-matter couplings. The explicit nonminimal couplings, for instance, between an arbitrary function of the scalar curvature $R$ and the Lagrangian density of matter, induces a non-vanishing covariant derivative of the energy-momentum tensor, implying non-geodesic motion and, consequently, leads to the appearance of an extra force. Applied to the cosmological context, these curvature-matter couplings lead to interesting phenomenology, where one can obtain a unified description of the cosmological epochs. We also consider the possibility that the behavior of the galactic flat rotation curves can be explained in the framework of the curvature-matter coupling models, where the extra terms in the gravitational field equations modify the equations of motion of test particles and induce a supplementary gravitational interaction. In addition to this, these models are extremely useful for describing dark energy-dark matter interactions and for explaining the late-time cosmic acceleration.}


\def\be{\begin{equation}}
\def\ee{\end{equation}}
\def\bea{\begin{eqnarray}}
\def\eea{\end{eqnarray}}

\newcommand{\f}[2]{\frac{#1}{#2}}

\def\bl{\begin{align}}
\def\el{\end{align}}

\begin{document}\vspace{-12pt}


\section{Introduction}

Recent observations of supernovae, together with the
 Wilkinson Microwave Anisotropy Probe (WMAP)
 and  Sloan Digital Sky Survey (SDSS) 
 data, lead to the remarkable conclusion that our
universe is not just expanding, but has begun to accelerate
\cite{1,2,3}. The resolution of this fundamental question is extremely
important for theoretical cosmology, looking beyond the standard
theory of gravity. The standard model of cosmology has favored the
dark energy models as fundamental candidates responsible for the
cosmic expansion. However, it is clear that these questions
involve not only gravity, but also particle physics. String theory
provides a synthesis of these two branches of physics and is
widely believed to be moving towards a viable quantum gravity
theory. One of the key predictions of string theory is the
existence of extra spatial dimensions. In the braneworld
scenario, motivated by recent developments in string theory, the
observed three-dimensional universe is embedded in a
higher-dimensional spacetime \cite{4}. The new degrees of freedom belong to the gravitational sector and can be responsible for the late-time cosmic acceleration.

Indeed, detailed theoretical and phenomenological
analysis of the relation between the effective field yielding
dark energy, non-canonical Lagrangians and non-linear gauge
kinetic functions have been extensively explored in the literature. Note that generalizations of the action
functional can be approached in several ways. For instance,
prescriptions consist in replacing the linear scalar curvature
term in the Einstein--Hilbert action by a function of the scalar
curvature, $f(R)$, or by more general scalar invariants of the
theory. This class of theories is often termed higher-order gravity
theories~\cite{5,7,8}. In this context, infra-red modifications of
General Relativity have been extensively explored, including four-dimensional
modifications to the Einstein--Hilbert action, and the
consistency of various candidate models has been extensively analyzed. All modified gravity
theories induce observational signatures at the post-Newtonian
level, which are translated by the parameterized post-Newtonian
(PPN) metric coefficients arising from these extensions of  General Relativity (GR). Thus,
generalizations of the Einstein--Hilbert Lagrangian, including
quadratic Lagrangians that involve second order curvature invariants
have also been extensively explored \cite{5}. While the latter modified
theories of gravity offer an alternative explanation to the
standard cosmological model \cite{6}, it offers a paradigm for nature fundamentally distinct
from dark energy models of cosmic acceleration, even those that
perfectly mimic the same expansion history. It is a fundamental question to
understand how one may observationally and theoretically differentiate these modified theories of
gravity from dark energy models.
Thus, one should test these models against large-scale structure and lensing, astrophysical and laboratory measurements, as well as laboratory and space-based equivalence principle experiments. These tests from the Solar System, the large-scale structure and lensing essentially restrict the range of allowed modified gravity models and, thus, offer a window into understanding the perplexing nature of the cosmic acceleration, and of gravity itself.

In this review, we consider an interesting possibility,
which includes non-minimal couplings between the scalar curvature
and the matter Lagrangian density, introduced in \cite{Bertolami:2007gv}, and its extensions and generalizations. The specific cases of a curvature coupling to dark energy \cite{Nojiri:2004bi,Nojiri:2004bi2,Nojiri:2004bi3}, the Maxwell field~\cite{Bamba:2008ja} and an explicit curvature-Yang--Mills coupling \cite{Bamba:2008xa} was also explored in the context of inflation and of the late-time cosmic acceleration.
Indeed, in the context of $f(R)$ modified theories of gravity, it was shown that an explicit coupling between an arbitrary function of the scalar curvature $R$ and the Lagrangian density of matter induces a non-vanishing covariant derivative of the energy-momentum, implying non-geodesic motion and, consequently, leads to the appearance of an extra force \cite{Bertolami:2007gv}. The latter extra force is orthogonal to the four-velocity, and the corresponding acceleration law was obtained in the weak field limit. Connections with  Modified Newtonian Dynamics (MOND) 
 and with the Pioneer anomaly were further discussed.
These curvature-matter coupling theories include the more
evolved generalizations of $f(R,L_m)$ \cite{Harko:2010mv}
and $f(R,T)$ gravities \cite{Harko:2011kv}. Amongst
other features, these models allow for an explicit breaking of the
equivalence principle (EP), which is highly constrained by Solar System experimental tests
\cite{Faraoni,BPT06}, by imposing a matter-dependent deviation
from geodesic motion.

Note that the weak equivalence principle is considered one of the pillars of General Relativity, and in fact, even a sizable part of the modified-gravity community considers this principle as truly fundamental~\cite{Will:2014xja}. This fact has to be stressed, because it demonstrates somehow a limitation for this class of theories.
However, it has been recently reported, from
data of the Abell Cluster A586, that the interaction of dark matter
and dark energy does imply the violation of the equivalence
principle~\cite{BPL07}. Notice that the violation of the
equivalence principle is also found as a low-energy feature of
some compactified version of higher-dimensional theories.
Indeed, as emphasized by Thibault Damour, it is important to note that the EP is not one of the ``universal'' principles of physics \cite{Damour:2001fn}. It is a heuristic hypothesis, which was introduced by Einstein in 1907, and used by him to construct his theory of General Relativity. In modern language, the (Einsteinian)   Equivalence Principle (EP)  consists in assuming that the only long-range field with gravitational-strength couplings to matter is a massless spin-2 field. Modern unification theories, and notably String Theory, suggest the existence of new fields (in particular, scalar fields: ``dilaton'' and ``moduli'') with gravitational-strength couplings. In most cases, the couplings of these new fields violate the EP. If the field is long-ranged, these EP violations lead to many observable consequences, such as the variation of fundamental ``constants'', the non-universality of free fall and the relative drift of atomic clocks, amongst others. The best experimental probe of a possible violation of the EP is to compare the free-fall acceleration of different materials. Further tests of this principle remain important and relevant for new physics and do indeed strongly restrict the parameters of the considered theory~\cite{Damour:1996xt, Damour:2010rp}. However, it is important to note that, in this context, the EP does not in principle rule out the specific~theory.

More specifically, $f(R,L_m)$ gravity further generalizes $f(R)$ gravity by assuming that the gravitational Lagrangian is given by an arbitrary function of the Ricci scalar $R$ and of the matter Lagrangian $L_m$ \cite{Harko:2010mv}. This may also be considered a maximal extension of the Einstein--Hilbert action. Here, we use the term ``maximal extension'' in a strict mathematical sense, and we define it as ``an element in an ordered set that is followed by no other''. For~instance, if we consider the set $\{{\rm Grav\; Theor}\}$ of gravity theories as given by $\{{\rm Grav \;Theor\}}=\left\{{\rm standard\; general\; relativity}, f(R)\; {\rm gravity},{\rm linear \;curvature-matter\; coupling},...,f\left(R,L_m\right)\right\}$, the $f\left(R,L_m\right)$ gravity theory represent the maximal extension of the set of all gravitational theories constructed in a Riemann space and with the action depending on the Ricci scalar and the matter Lagrangian only. In \cite{Harko:2010mv}, the gravitational field equations of the $f\left(R,L_m\right)$ gravity theory in the metric formalism were obtained, as well as the equations of motion for test particles, which follow from the covariant divergence of the energy-momentum tensor. The equations of motion for test particles can also be derived from a variational principle in the particular case in which the Lagrangian density of the matter is an arbitrary function of the energy-density of the matter only. In \cite{Harko:2010mv}, the Newtonian limit of the equation of motion was also considered, and a procedure for obtaining the energy-momentum tensor of the matter was presented. In particular, the gravitational field equations and the equations of motion for a particular model in which the action of the gravitational field has an exponential dependence on the standard general relativistic Hilbert--Einstein Lagrange density were also derived.
The $f(R,L_m)$ gravitational theory was further generalized by considering the novel inclusion of a scalar field and a kinetic term constructed from the gradients of the scalar field, respectively \cite{Harko:2012hm}. Specific models with a nonminimal coupling between the scalar field and the matter Lagrangian were further explored. We emphasize that these models are extremely useful for describing an interaction between dark energy and dark matter and for explaining the late-time cosmic acceleration.

In the context of $f(R,T)$ modified theories of gravity, the gravitational Lagrangian is given by an arbitrary function of the Ricci scalar $R$ and of the trace of the energy-momentum tensor $T$ \cite{Harko:2011kv}. The gravitational field equations in the metric formalism were obtained, and it was shown that these field equations explicitly depend on the nature of the matter source. The field equations of several particular models, corresponding to some explicit forms of the function $f(R,T)$, were also presented. Furthermore, in \cite{Harko:2011kv}, the Newtonian limit of the equation of motion was also analyzed, and constraints on the magnitude of the extra-acceleration were obtained by analyzing the perihelion precession of the planet Mercury in the framework of the present model. An interesting specific case, namely the $f(R,T^\phi)$ model, was also analyzed in detail, where $T^\phi$ is the trace of the stress-energy tensor of a self-interacting scalar field. The cosmological implications of the model were briefly considered. We refer the reader to~\cite{Harko:2011kv} for more details.

The non-minimally curvature-matter-coupled $f(R,T)$ gravitational theory was further generalized by considering the inclusion of a contraction of the Ricci tensor with the matter energy-momentum tensor~\cite{Haghani:2013oma,Odintsov:2013iba,Haghani:2014ina}. We emphasize that examples of such couplings can be found in the Einstein--Born--Infeld theories \cite{deser} when one expands the square root in the Lagrangian. An interesting feature of this theory is that in considering a traceless energy-momentum tensor, \textit{i.e.}, $T=0$, the field equations of $f(R,T)$ gravity reduces to those of $f(R)$ gravity theories, while the presence of the $R_{\mu\nu}T^{\mu\nu}$ coupling term still entails a non-minimal coupling to matter. In \cite{Haghani:2013oma}, the Newtonian limit of the $f(R,T,R_{\mu\nu}T^{\mu\nu})$ gravitational theory was considered, and an explicit expression for the extra-acceleration, which depends on the matter density, was obtained in the small velocity limit for dust particles. The so-called Dolgov--Kawasaki instability \cite{Dolgov:2003px} was also analysed in detail, and the stability conditions of the model with respect to local perturbations was obtained. A particular class of gravitational field equations can be obtained by imposing the conservation of the energy-momentum tensor \cite{Haghani:2013oma}. In this context, the corresponding field equations for the conservative case was also derived by using a Lagrange multiplier method, from a gravitational action that explicitly contains an independent parameter multiplying the divergence of the energy-momentum tensor \cite{Haghani:2013oma}. The cosmological implications of the model were extensively investigated for both the conservative and non-conservative cases, and several classes of analytical solutions were obtained. In \cite{Odintsov:2013iba}, the  Friedmann-Lema\^{i}tre-Robertson-Walker (FRLW) 
 cosmological dynamics for several versions of the $f(R,T,R_{\mu\nu}T^{\mu\nu})$ gravity theory was also considered. The reconstruction of the above action was explicitly analyzed, including the numerical reconstruction for the occurrence of the $\Lambda$CDM (Lambda-Cold Dark Matter) universe. De Sitter universe solutions in the presence of non-constant fluids were also presented, and the problem of matter instability was further~discussed.

All of the above gravitational modifications are based on the curvature description of gravity. However, an interesting and rich class of modified gravity arises by modifying the action of the equivalent torsional formulation of General Relativity. The latter approach has been denoted the ``Teleparallel Equivalent of General Relativity'' (TEGR) \cite{Unzicker:2005in,TEGR1,TEGR2,Hayashi:1979qx,JGPereira,Maluf:2013gaa} and consists of replacing the torsion-less Levi--Civita connection by the curvature-less Weitzenb{\"{o}}ck one and using the vierbein instead of the metric as the fundamental field. In this formulation, the Lagrangian of the theory is constructed by contractions of the torsion tensor. Thus, in analogy to $f(R)$ gravity, if one wishes to modify gravity in this formulation, the simplest approach would be in generalizing the torsion scalar $\mathcal{T}$ to an arbitrary function $f(\mathcal{T})$ \cite{Ferraro:2006jd1,Ferraro:2006jd2,Linder:2010py}.

Now, in the context of TEGR models, one may also construct an extension of $f(\mathcal{T})$ gravity with the inclusion of a non-minimal torsion-matter coupling in the action \cite{Harko:2014sja}. The resulting theory is a novel gravitational modification, since it is different from both $f(\mathcal{T})$ gravity, as well as from the non-minimal curvature-matter-coupled theory. The cosmological application of this new theory proves to be very interesting. In particular, an effective dark energy sector was obtained, where the equation-of-state parameter can be quintessence or phantom-like, or cross the phantom-divide, while for a large range of the model parameters, the Universe results in a de Sitter, dark-energy-dominated, accelerating phase. Additionally, early-time inflationary solutions were also obtained, and thus, one can provide a unified description of the cosmological history. We refer the reader to \cite{Harko:2014sja} for more details.

In addition to the latter, non-minimal torsion-matter coupling, an alternative extension of $f(\mathcal{T})$ gravity was explored by considering a general coupling of the torsion scalar $\mathcal{T}$ with the trace of the matter energy-momentum tensor $T$ \cite{Harko:2014aja}. The resulting $f(\mathcal{T},T)$ theory is a new modified gravity, since it is different from all of the existing torsion- or curvature-based constructions. Applied to a cosmological framework, it also leads to interesting phenomenology. In particular, one can obtain a unified description of the initial inflationary phase, the subsequent non-accelerating, matter-dominated expansion and, then, the transition to late-time accelerating phase. In the far future, the Universe results either in a de Sitter exponential expansion or in eternal power-law accelerated expansions. A similar analysis was investigated in \cite{Kiani:2013pba}, where using a perturbational approach, the stability of the solutions and, in particular, of the de Sitter phase was explored. Furthermore, the constraints imposed by the energy conditions were also considered. We refer the reader to \cite{Harko:2014aja,Kiani:2013pba} for more details.

The possibility that the behavior of the rotational velocities of test particles gravitating around galaxies can be explained in the framework of modified gravity models with nonminimal curvature-matter coupling was considered in \cite{Harko:2010vs}. Generally, the dynamics of test particles around galaxies, as well as the corresponding mass deficit, is explained by postulating the existence of dark matter. The extra terms in the gravitational field equations with curvature-matter coupling modify the equations of motion of test particles and induce a supplementary gravitational interaction. Starting from the variational principle describing the particle motion in the presence of the non-minimal coupling, the expression of the tangential velocity of a test particle, moving in the vacuum on a stable circular orbit in a spherically symmetric geometry, was derived. The tangential velocity depends on the metric tensor components, as well as on the coupling function between matter and geometry. The Doppler velocity shifts were also obtained in terms of the coupling function. If the tangential velocity profile is known, the coupling term between matter and curvature can be obtained explicitly in an analytical form. The functional form of this function was obtained in two cases, for a constant tangential velocity and for an empirical velocity profile obtained from astronomical observations, respectively. Therefore, these results open the possibility of directly testing the modified gravity models with a non-minimal curvature-matter coupling by using direct astronomical and astrophysical observations at the galactic or extra-galactic scale. These~issues will be reviewed in detail below.

It is rather important to discuss the theoretical motivations for these curvature-matter coupling theories more carefully. For instance, one may wonder if there is a fundamental theory or model that reduces to one of these theories in some limit. Indeed, for the specific case of the linear nonminimal curvature-matter coupling, one can show that this theory can be expressed as a scalar-tensor theory with two scalar fields \cite{Sotiriou:2008it}. These are reminiscent of the extensions of scalar-tensor gravity, which include similar couplings, such as the theories considered by Damour and Polyakov \cite{Damour:1994zq}, where it was shown that string-loop modifications of the low-energy matter couplings of the dilaton may provide a mechanism for fixing the vacuum expectation value of a massless dilaton. In addition to this, the results presented in \cite{Damour:1994zq} provide a new motivation for trying to improve by several orders of magnitude the various experimental tests of Einstein's equivalence principle, such as the universality of free fall and the constancy of the fundamental constants, amongst others. In addition to this, note that the modern revival of the Kaluza--Klein theory often leads to the introduction of several scalar fields, \textit{i.e.}, the ``compacton''~\cite{Damour:1992we}. Superstring theory also leads to several scalar fields coupled to the macroscopic distribution of energy; indeed, the ``dilaton'' is already present in ten dimensions, and the ``compacton'' comes from a dimensional reduction \cite{Damour:1992we}. Quantum-motivated, higher order generalizations of Einstein's General Relativity, under some conditions, are also equivalent to adding several scalar fields to the Einstein--Hilbert action \cite{Gottlober:1989ww}. 
In addition to this, we note that the nonminimal curvature-matter coupling also arises from one-loop vacuum-polarization effects in the formulation of quantum electrodynamics in curved spacetimes \cite{Drummond:1979pp}.
All of the above considerations make it natural to consider modified fields of gravity containing several scalar fields, which, in some cases, are equivalent, in the present context, to the presence of a nonminimal curvature-matter couplings. These considerations further motivate the analysis of the coupling between matter and curvature.

In this work, based on published literature, we review these generalized curvature-matter coupling-modified theories of gravity. The paper is outlined in the following manner: In Section~\ref{Sec:II}, we briefly introduce the linear curvature-matter coupling and present some of its interesting features. In Section~\ref{Sec:III}, we generalize the latter linear
curvature-matter coupling by considering the maximal extension of the Einstein--Hilbert action, which results in the $f\left(R,L_m\right)$ gravitational theory, and in Section~\ref{Sec:IV}, we extend the theory with the inclusion of general scalar field and kinetic term dependencies. In Section~\ref{Sec:V}, we consider another extension of general relativity, namely $f(R, T )$ modified theories of gravity, where the gravitational Lagrangian is given by an arbitrary function of the Ricci scalar $R$ and of the trace of the energy-momentum tensor $T$. In Section~\ref{Sec:VI}, we include an explicit invariant Ricci-energy-momentum tensor coupling and explore some of its astrophysical and cosmological phenomenology. This latter theory is motivated considering a traceless energy-momentum tensor, $T=0$; the gravitational field equations for the $f(R,T)$ theory reduce to that of the $f(R)$ gravity, and all non-minimal couplings of gravity to the matter field vanish, while the inclusion of the $R_{\mu\nu}T^{\mu\nu}$ term still allows a nonminimal coupling. The possibility of explaining dark matter as a consequence of the curvature-matter coupling is considered in Section \ref{dark}. Finally, in Section \ref{Sec:Concl}, we briefly discuss our results and conclude our review~paper.

\section{Gravity Theories with Linear Curvature-Matter Coupling}\label{Sec:II}

In the present Section, we review some of the basic features and properties of the simplest class of models involving a nonminimal coupling between matter and geometry, which is linear in the matter Lagrangian. For the sake of comparison of the different approaches for the description of the gravitational interaction, we also briefly introduce the $f(R)$ modified gravity theory.

\subsection{$f(R)$ Gravity}

One of the simplest extensions of standard general relativity, based on the Hilbert--Einstein action, is the $f(R)$-modified theory of gravity, whose action takes the form \cite{Bu70,Bar83}:
\begin{equation}\label{fRx}
S=\int \left[\frac{1}{16\pi G}f(R)+L_m\right] \sqrt{-g}\; d^{4}x
\end{equation}
where $f(R)$ is an arbitrary analytical function of the Ricci scalar $R$ and
$L_m$ is the Lagrangian density corresponding to matter.

Varying the action with respect to the metric $g_{\mu \nu }$ yields the
field equations of $f(R)$ gravity:
\begin{equation}\label{eqfR}
FR_{\mu \nu }-\frac{1}{2}fg_{\mu \nu }-\left(\nabla _{\mu }\nabla _{\nu
}-g_{\mu \nu }\square \right)F=8\pi G \,T_{\mu \nu}
\end{equation}
where we have denoted $F=df/d R$, and $T_{\mu \nu}$ is the standard, minimally
coupled, matter energy-momentum tensor. Note that the covariant derivative of
the field equations and of the matter energy-momentum tensor vanishes for
all $f(R)$ by means of generalized Bianchi identities~\cite{Bertolami:2007gv,Ko06}. By~contracting the field equations, Equation~(\ref{eqfR}), we obtain the useful relation:
\begin{equation} \label{trace}
3\square F+FR-2f=8\pi G\,T
\end{equation}
where $T$ is the trace of the energy-momentum tensor and from which one verifies that the Ricci scalar is now a fully dynamical
degree of freedom.

By introducing the Legendre transformation $\left\{ R,f\right\} \rightarrow
\left\{ \phi ,V\right\} $ defined as:
\begin{equation}
\phi \equiv F\left( R\right), \quad V\left( \phi \right) \equiv R\left( \phi
\right) F-f\left( R\left( \phi \right) \right)
\end{equation}
the field equations of $f(R)$ gravity can be reformulated as~\cite{equiv1,equiv2,equiv3,equiv4}:
\begin{equation}
R_{\mu \nu }-\frac{1}{2}g_{\mu \nu }R=8\pi \frac{G}{\phi }T_{\mu \nu
}+\theta _{\mu \nu } \label{fieldin}
\end{equation}
where:
\begin{equation}
\theta _{\mu \nu }=-\frac{1}{2}V\left( \phi \right) g_{\mu \nu }+\frac{1}{%
\phi }\left( \nabla _{\mu }\nabla _{\nu }-g_{\mu \nu }\square \right) \phi
\end{equation}

Using these variables,  Equation~(\ref{trace}) takes the following form:
\begin{equation} \label{trace1}
3 \square \phi + 2V(\phi) -\phi \frac{dV}{d\phi }=8\pi G\, T
\end{equation}

In this scalar-tensor representation, the field equations of $f(R)$ gravity can be derived
from a Brans--Dicke-type gravitational action, with parameter $\omega =0$, given by:
\begin{equation}\label{scalx}
S=\frac{1}{16\pi G}\int \left[ \phi R-V(\phi) +L_{m}\right] \sqrt{-g}\;
d^{4}x
\end{equation}

The only requirement for the $f(R)$ model equations to be expressed in the
form of a Brans--Dicke theory is that $F(R)$ be invertible, that is, $R(F)$
exists~\cite{Olmo07}. This condition is necessary for the construction of $%
V(\phi)$.

The modification of the standard Einstein--Hilbert action leads to the
appearance in the field equations of an effective gravitational constant $G_{%
\mathrm{eff}}=G/\phi$, which is a function of the curvature. Secondly, a new
source term for the gravitational field, given by the tensor $\theta_{\mu
\nu}$, is also induced. The tensor $\theta_{\mu \nu}$ is determined by the
trace of the energy-momentum tensor via Equation~(\ref{trace1}), which thus acts
as an independent physical parameter determining the metric of the
space-time.

\subsection{Linear Non-Minimal Curvature-Matter Coupling}

The action of the $f(R)$ modified gravity theory can be generalized by introducing in the action the linear nonminimal coupling between matter and geometry. The action of this modified gravity theory is given by \cite{Bertolami:2007gv}:
\begin{equation}
S=\int \left\{\frac{1}{2}f_1(R)+\left[1+\lambda f_2(R)\right]{
L}_{m}\right\} \sqrt{-g}\;d^{4}x~
 \label{actionlinear}
\end{equation}
where the factors $f_i(R)$ (with $i=1,2$) are arbitrary functions of the Ricci scalar $R$ and ${L}_{m}$ is the matter Lagrangian density, and the coupling constant $\lambda$ determines the strength of the interaction between $f_2(R)$ and the matter Lagrangian.

Now, varying the action with respect to the metric $g_{\mu\nu}$ provides the following field equations:
\bea
F_1(R)R_{\mu \nu }-\frac{1}{2}f_1(R)g_{\mu \nu }-\nabla_\mu
\nabla_\nu \,F_1(R)+g_{\mu\nu}\square F_1(R)
=-2\lambda F_2(R){L}_m R_{\mu\nu}
 \nonumber \\
+2\lambda(\nabla_\mu
\nabla_\nu-g_{\mu\nu}\square){ L}_m F_2(R) +[1+\lambda f_2(R)]T_{\mu \nu }^{(m)}~
\label{field1a}
\eea
where $F_i(R)=f'_i(R)$, and the prime represents the derivative with respect to the scalar curvature. The
matter energy-momentum tensor is defined as:
\begin{equation}
T_{\mu \nu
}^{(m)}=-\frac{2}{\sqrt{-g}}\frac{\delta(\sqrt{-g}\,{
L}_m)}{\delta(g^{\mu\nu})} ~ \label{EMTdef}
\end{equation}

Throughout this paper, we use the metric formalism. However, the field equations and the equations of motion were derived, for massive test particles in modified theories of gravity for a linear curvature-matter coupling, using the Palatini formalism in \cite{Harko:2010hw}.

A general property of these nonminimal curvature-matter coupling theories is the non-conservation of the energy-momentum tensor. This can be easily verified by taking into account the covariant derivative of the field Equation~(\ref{field1a}), the Bianchi identities, $\nabla^\mu G_{\mu\nu}=0$, and the following identity \mbox{$(\square\nabla_\nu -\nabla_\nu\square)F_i=R_{\mu\nu}\,\nabla^\mu F_i$}, which then imply the following relationship:
\begin{equation}
\nabla^\mu T_{\mu \nu }^{(m)}=\frac{\lambda F_2}{1+\lambda
f_2}\left[g_{\mu\nu}{L}_m- T_{\mu \nu
}^{(m)}\right]\nabla^\mu R ~ \label{cons1}
\end{equation}

Thus, the coupling between the matter and the higher derivative
curvature terms may be interpreted as an exchange of energy and momentum
between both. Analogous couplings arise after a conformal
transformation in the context of scalar-tensor theories of
gravity and also in string theory. In the absence of the
coupling, one verifies the conservation of the energy-momentum
tensor \cite{Ko06}, which can also be verified from the
diffeomorphism invariance of the matter part of the action. It is also interesting to
note that, from Equation~(\ref{cons1}),
the conservation of the energy-momentum tensor is verified if
$f_2(R)$ is a constant or the matter Lagrangian is not an explicit
function of the metric.

In order to test the motion in our model, we consider for the
energy-momentum tensor of matter a perfect fluid
$T_{\mu \nu }^{(m)}=\left( \rho +p\right) U_{\mu }U_{\nu
}+pg_{\mu \nu }$, where $\rho$ is the overall energy density and $p$, the
pressure, respectively. The four-velocity, $U_{\mu }$, satisfies
the conditions $U_{\mu }U^{\mu }=-1$ and $\nabla _{\nu }U^{\mu }U_{\mu }=0$.
We also introduce the projection operator $h_{\mu \lambda }=g_{\mu
\lambda }+U_{\mu }U_{\lambda }$ from which one obtains $h_{\mu
\lambda }U^{\mu }=0$.

By contracting Equation~(\ref{cons1}) with the projection operator
$h_{\mu \lambda }$, one deduces the following expression:
\bea
\left( \rho +p\right) g_{\mu \lambda }U^{\nu }\nabla_\nu
U^{\mu} -(\nabla_\nu p)(\delta_\lambda^\nu-U^\nu U_\lambda)
-\frac{\lambda F_2}{1+\lambda f_2}\left({
L}_m-p\right)(\nabla_\nu R)\,(\delta_\lambda^\nu-U^\nu
U_\lambda)=0 ~
\eea

Finally, contraction with $g^{\alpha \lambda }$ gives rise to the
equation of motion for a fluid element:
\begin{equation}
\frac{D U^{\alpha }}{ds} \equiv \frac{dU^{\alpha }}{ds}+\Gamma _{\mu
\nu }^{\alpha }U^{\mu }U^{\nu }=f^{\alpha }~ \label{eq1}
\end{equation}
where the extra force is given by:
\begin{eqnarray}
\label{force}
f^{\alpha }&=&\frac{1}{\rho +p}\Bigg[\frac{\lambda
F_2}{1+\lambda f_2}\left({L}_m -p\right)\nabla_\nu
R+\nabla_\nu p \Bigg] h^{\alpha \nu }\,
\end{eqnarray}

As one can immediately verify, the extra force $f^{\alpha }$ is
orthogonal to the four-velocity of the particle,
$f^{\alpha }U_{\alpha }=0$
which can be seen directly, from the properties of the projection
operator. This is consistent with the usual interpretation of the
force, according to which only the component of the four-force
that is orthogonal to the particle's four-velocity can influence
its trajectory.

It has also been shown that an $f(G)$-modified Gauss--Bonnet gravity with a non-minimal coupling to matter also induces an extra force, which is normal to their four-velocities and, as a result, moves along nongeodesic world-lines \cite{Mohseni:2009ns}.

A particularly intriguing feature is that the extra force depends on the form of the Lagrangian density. Note that considering the Lagrangian density $L_m = p$, where $p$ is the pressure, the extra-force vanishes~\cite{Bertolami:2008ab}. It has been argued that this is not the unique choice for the matter Lagrangian density and that more natural forms for $L_m$, such as $L_m = -\rho $, do not imply the vanishing of the extra-force. Indeed, in the presence of nonminimal coupling, they give rise to two distinct theories with different predictions~\cite{Faraoni:2009rk}, and this issue has been further investigated in different contexts \cite{Bertolami:2013raa,Minazzoli:2013bva}, including phantom energy~\cite{Bisabr:2013laa}. In this context, a matter Lagrangian density as an arbitrary function of the energy-density of the matter only was explored \cite{Harko:2008qz} (this possibility has also been explored in five dimensions \cite{Wu:2014upa} and in the braneworld context \cite{Olmo:2014sra}). It was also argued that the corresponding energy-momentum tensor of the matter in modified gravity models with non-minimal coupling is more general than the usual general-relativistic energy-momentum tensor for perfect fluids \cite{Harko:2010zi}, and it contains a supplementary, equation of state-dependent term, which could be related to the elastic stresses in the body or to other forms of internal energy. Therefore, the extra force induced by the coupling between matter and geometry never vanishes as a consequence of the thermodynamic properties of the system or for a specific choice of the matter Lagrangian, and it is non-zero in the case of a fluid of dust particles.
In the following subsection, we discuss in detail the problem of the matter Lagrangian and of the energy-momentum tensor in modified gravity theories with linear curvature-matter coupling \cite{Harko:2010zi}.

\subsection{The Matter Lagrangian and the Energy-Momentum Tensor in Modified Gravity with Non-Minimal Linear Coupling between Matter and Geometry}\label{section21}

It is an interesting and novel feature of the $f(R)$ gravity
with non-minimal curvature-matter coupling that the matter Lagrangian, and the
energy-momentum tensor obtained from it, are not model-independent quantities, but they are
completely and uniquely determined by the coupling function between
matter and geometry. This important result can be obtained by deriving first the
equations of motion of test particles (or test fluid) in the modified gravity model from a variational principle and then considering the
Newtonian limit of the particle action for a fluid obeying a barotropic equation of
state~\cite{Harko:2010zi}. The energy-momentum tensor of the matter obtained in this way is more general than the usual general-relativistic energy-momentum tensor for perfect fluids, and it
contains a supplementary term that may be related to the elastic stresses
in the test fluid or to other sources of internal energy. Since we assume that the matter obeys a barotropic equation of state, the matter Lagrangian can be expressed either in terms of the
density or in terms of the pressure, and in both representations, the
physical description of the system is equivalent. Therefore, the presence (or
absence) of the extra-force is independent of the specific form of the
matter Lagrangian, and it never vanishes, except in the particular case of (un)physical
matter systems with zero sound speed. In particular, in the case of dust particles, the extra-force is always non-zero.

We define the energy-momentum tensor of the matter as \cite{LaLi}:
\begin{equation}
T_{\mu \nu }=-\frac{2}{\sqrt{-g}}\left[ \frac{\partial \left( \sqrt{-g}%
L_{m}\right) }{\partial g^{\mu \nu }}-\frac{\partial }{\partial x^{\lambda }}%
\frac{\partial \left( \sqrt{-g}L_{m}\right) }{\partial \left( \partial
g^{\mu \nu }/\partial x^{\lambda }\right) }\right]
\end{equation}

By assuming that the Lagrangian density $L_{m}$ of the matter depends only
on the metric tensor components $g_{\mu \nu }$, and not on its derivatives,
we obtain:
\begin{equation}
T_{\mu \nu }=L_{m}g_{\mu \nu }-2\frac{\partial L_{m}}{\partial g^{\mu
\nu }} \label{SETdef}
\end{equation}

By taking into account the explicit form of the field equations Equation~(\ref{field1a}), one
obtains for the covariant divergence of the energy-momentum tensor the
equation:
\begin{equation}
\nabla ^{\mu }T_{\mu \nu }=2\left\{\nabla ^{\mu }\ln \left[ 1+\lambda f_{2}(R)%
\right] \right\}\frac{\partial L_{m}}{\partial g^{\mu \nu }} \label{cons1b}
\end{equation}

As a specific example of generalized gravity models with linear
curvature-matter coupling, we~consider the case in which matter, assumed
to be a perfect thermodynamic fluid, obeys a barotropic equation of state,
with the thermodynamic pressure $p$ being a function of the rest mass density of the matter (for short: matter density) $\rho $
only, so that $p=p\left( \rho \right)$. In this case, the matter Lagrangian
density, which, in the general case, could be a function of both density and
pressure, $L_{m}=L_{m}\left( \rho ,p\right)$, or of only one of the
thermodynamic parameters, becomes an arbitrary function of the density of
the matter $\rho $ only, so that $L_{m}=L_{m}\left( \rho \right) $. Then, the matter
energy-momentum tensor is obtained as \cite{Harko:2010zi}:
\begin{equation}
T^{\mu \nu }=\rho \frac{dL_{m}}{d\rho }U^{\mu }U^{\nu }+\left( L_{m}-\rho
\frac{dL_{m}}{d\rho }\right) g^{\mu \nu } \label{tens}
\end{equation}
where the four-velocity $U^{\mu }=dx^{\mu }/ds$ satisfies the condition $%
g^{\mu \nu }U_{\mu }U_{\nu }=-1$. To obtain Equation~(\ref{tens}), we have imposed
the condition of the conservation of the matter current:
\begin{equation}
\nabla _{\nu
}\left( \rho U^{\nu }\right) =0
\end{equation}
and we have used the relation:
\begin{equation}
\delta \rho
=\frac{ 1}{2} \rho \left( g_{\mu \nu }-U_{\mu } U_{\nu }\right) \delta
g^{\mu \nu }
\end{equation}
whose proof is given in the Appendix of \cite{Harko:2010zi}. With the use of the
identity $U^{\nu }\nabla _{\nu }U^{\mu }=d^{2}x^{\mu }/ds^{2}+\Gamma _{\nu
\alpha }^{\mu }U^{\nu }U^{\alpha }$, from Equations~(\ref{cons1}) and (\ref
{tens}), we obtain the equation of motion of a massive test particle, or~of a test fluid in the modified
gravity model with linear coupling between matter and geometry, as:
\begin{equation}
\frac{d U^{\mu }}{ds}+\Gamma _{\nu \alpha }^{\mu }U^{\nu }U^{\alpha
}=f^{\mu } \label{eqmota}
\end{equation}
where:
\begin{equation}
f^{\mu }=-\nabla _{\nu }\ln \left\{ \left[ 1+\lambda f_{2}(R)\right] \frac{%
dL_{m}\left( \rho \right) }{d\rho }\right\} \left( U^{\mu }U^{\nu }-g^{\mu
\nu }\right) 
\end{equation}

The extra-force $f^{\mu }$, generated due to the presence of the coupling
between matter and geometry, is perpendicular to the four-velocity, $f^{\mu
}U_{\mu }=0$.

The equation of motion Equation~(\ref{eqmota}) can be obtained from
the variational principle \cite{Harko:2010zi}:
\begin{equation}
\delta S_{p}=\delta \int L_{p}ds=\delta \int \sqrt{Q}\sqrt{g_{\mu \nu
}U^{\mu }U^{\nu }}ds=0 \label{actpart}
\end{equation}
where $S_{p}$ and $L_{p}=\sqrt{Q}\sqrt{g_{\mu \nu }U^{\mu }U^{\nu }}$ are
the action and the Lagrangian density for test particles (test fluid), respectively,
and:
\begin{equation}
\sqrt{Q}=\left[ 1+\lambda f_{2}(R)\right] \frac{dL_{m}\left( \rho \right) }{%
d\rho } \label{Q}
\end{equation}

To prove this result, we start with the Lagrange equations corresponding to
the action  Equation~(\ref{actpart}):
\begin{equation}
\frac{d}{ds}\left( \frac{\partial L_{p}}{\partial U^{\alpha }}\right) -%
\frac{\partial L_{p}}{\partial x^{\alpha }}=0
\end{equation}

Since:
\begin{equation}
\frac{\partial L_{p}}{\partial U^{\alpha }}=\sqrt{Q}u_{\alpha }
\end{equation}
 and:
\begin{equation}
\frac{\partial L_{p}}{\partial x^{\alpha }}=\frac{ 1}{2} \sqrt{Q}g_{\mu \nu
,\alpha }U^{\mu }U^{\nu }+\frac{1}{2}\frac{ Q_{,\alpha }}{Q}
\end{equation}
where a comma indicates the derivative with respect to $x^{\lambda }$, a straightforward calculation gives the equations of motion of the particle as:
\begin{equation}
\frac{d^{2}x^{\mu }}{ds^{2}}+\Gamma^{\mu }{}_{\nu \alpha }U^{\nu }U^{\alpha
}+\left( U^{\mu }U^{\nu }-g^{\mu \nu }\right) \nabla _{\nu }\ln \sqrt{Q}=0
\end{equation}

By simple identification with the equation of motion of the modified gravity
model with linear curvature-matter coupling, given by Equation~(\ref{eqmota}), we
obtain the explicit form of $\sqrt{Q}$, as given by Equation~(\ref{Q}). When $\sqrt{Q}\rightarrow 1$, we reobtain the standard general relativistic equation for geodesic~motion.

\subsubsection{The Newtonian Limit}

The variational principle  Equation~(\ref{actpart}) can be used to study the Newtonian
limit of the model. In the limit of the weak gravitational fields:
\begin{equation}
ds\approx \sqrt{1+2\phi -\vec{v}^{2}}dt\approx \left( 1+\phi -\frac{\vec{v}%
^{2}}{2}\right) dt
\end{equation}
where $\phi $ is the Newtonian potential and $\vec{v}$ is
the usual tridimensional velocity of the fluid. By representing the function
$\sqrt{Q}$ as:
\begin{equation}
\sqrt{Q}=\frac{dL_{m}\left( \rho \right) }{d\rho }+\lambda f_{2}(R)\frac{%
dL_{m}\left( \rho \right) }{d\rho }
\end{equation}
in the first order of approximation, the equations of motion of the fluid can
be obtained from the variational principle:
\begin{equation}
\delta \int \left[ \frac{dL_{m}\left( \rho \right) }{d\rho }+\lambda f_{2}(R)%
\frac{dL_{m}\left( \rho \right) }{d\rho }+\phi -\frac{\vec{v}^{2}}{2}\right]
dt=0
\end{equation}
and are given by:
\begin{equation}
\vec{a}=-\nabla \phi -\nabla \frac{dL_{m}\left( \rho \right) }{d\rho }%
-\nabla U_{E}=\vec{a}_{N}+\vec{a}_{H}+\vec{a}_{E}
\end{equation}
where $\vec{a}$ is the total acceleration of the system:
\begin{equation}
\vec{a}%
_{N}=-\nabla \phi
\end{equation}
 is the Newtonian gravitational acceleration, and:
 \begin{equation}
 \vec{a}%
_{E}=-\nabla U_{E}=-\lambda \nabla \left[ f_{2}(R)\frac{dL_{m}\left( \rho \right)
}{d\rho }\right]
\end{equation}
 is a supplementary acceleration induced due to the curvature-matter coupling. As for the term:
\begin{equation}
\vec{a}_{H}=-\nabla \left[
\frac{dL_{m}\left( \rho \right)} {d\rho }\right]
\end{equation}
it has to be identified with the
hydrodynamic acceleration term in the perfect fluid Euler equation:
\begin{equation}\label{eqlm}
\vec{a}_{H}=-\nabla \frac{dL_{m}\left( \rho \right) }{d\rho }=-\nabla
\int_{\rho _{0}}^{\rho }\frac{dp}{d\rho }\frac{d\rho }{\rho }
\end{equation}
where $\rho _{0}$, an integration constant, plays the role of a limiting
density.

\subsubsection{The Matter Lagrangian in Modified Gravity Theories with Curvature-Matter Coupling}

With the use of Equation~(\ref{eqlm}), the matter Lagrangian in modified gravity theories with curvature-matter coupling can be obtained by a simple integration as \cite{Harko:2010zi}:
\begin{equation}
L_{m}\left( \rho \right) =\rho \left[ 1+\Pi \left( \rho \right) \right]
-\int_{p_{0}}^{p}dp \label{Lm}
\end{equation}
where:
\begin{equation}
\Pi \left( \rho \right) =\int_{p_{0}}^{p}{\frac{dp}{\rho }}
\end{equation}
and we have normalized an arbitrary integration constant to one; $p_{0}$ is an
integration constant, or a limiting pressure. The corresponding
energy-momentum tensor of matter is given by:
\begin{equation}
T^{\mu \nu }=\left\{ \rho \left[ 1+\Phi \left( \rho \right) \right] +p\left(
\rho \right) \right\} U^{\mu } U^{\nu }-p\left( \rho \right) g^{\mu \nu }
\label{tens1}
\end{equation}
respectively, where:
\begin{equation}
\Phi \left( \rho \right) =\int_{\rho _{0}}^{\rho }\frac{p}{\rho ^{2}}d\rho
=\Pi \left( \rho \right) -\frac{p\left( \rho \right) }{\rho }
\end{equation}
and with all of the constant terms included in the definition of $p$. By introducing the energy density of the body according to the definition:
\begin{equation}
\varepsilon=\rho \left[ 1+\Phi \left( \rho \right) \right]
\end{equation}
the energy-momentum tensor of a test fluid can be written in the modified gravity models with curvature-matter coupling in a form similar to the standard general relativistic case:
\begin{equation}
T^{\mu \nu }=\left[\varepsilon \left(\rho \right)+p\left(\rho \right)\right] U^{\mu }U^{\nu }+p\left( \rho \right) g^{\mu \nu }
\end{equation}

From a physical point of view, $\Phi \left( \rho \right) $ can be interpreted
as the elastic (deformation) potential energy of the body, and therefore,
Equation~(\ref{tens1}) corresponds to the energy-momentum tensor of a
compressible elastic isotropic system \cite{Fock}. The matter Lagrangian can also be
written in the simpler form:
\begin{equation}
L_{m}\left( \rho \right) =\rho \Phi \left( \rho
\right)
\end{equation}

If the pressure does not have a thermodynamic or radiative component, one can
take $p_{0}=0$. If the pressure is a constant background quantity,
independent of the density, so that $p=p_{0}$, then $L_{m}\left( \rho
\right) =\rho $, and the matter energy-momentum tensor takes the form
corresponding to dust:
\begin{equation}
T^{\mu \nu }=\rho U^{\mu } U^{\nu }
\end{equation}

Since matter is supposed to
obey a barotropic equation of state, these results are independent of the
concrete representation of the matter Lagrangian in terms of the
thermodynamic quantities \cite{Harko:2010zi}. The~same results are obtained by assuming $%
L_{m}=L_{m}\left( p\right) $; due to the equation of state, $\rho $ and $p$
are freely interchangeable thermodynamic quantities, and the Lagrangians
expressed in terms of $\rho $ and $p$ only are completely equivalent. More
general situations, in which the density and pressure are functions of the
particle number and temperature, respectively, and the equation of state is
given in a parametric form, can be analyzed in a similar way.

The forms of
the matter Lagrangian and the energy-momentum tensor are strongly
dependent on the equation of state of the test fluid. For example, if the barotropic equation
of state is linear, $p=\left( \gamma -1\right) \rho $, $\gamma =$ constant, $%
1\leq \gamma \leq 2$, then:
\begin{equation}
L_{m}\left( \rho \right) =\rho \left\{ 1+\left(
\gamma -1\right) \left[ \ln \left( \frac{\rho }{\rho _{0}}\right) -1 %
\right] \right\}
\end{equation}
 and $\Phi \left( \rho \right) =\left( \gamma -1\right)
\ln \left( \rho /\rho _{0}\right) $, respectively. For the case of a
polytropic equation of state $p=K\rho ^{1+1/n}$, $K,n=$ constant, we have:
\begin{equation}
L_{m}\left( \rho \right) =\rho +K\left(\frac{n^{2}}{n+1}-1\right)
\rho ^{1+1/n}
\end{equation}
and $\Phi \left( \rho \right) =Kn\rho ^{1+1/n}=np\left( \rho
\right) $, respectively, where we have taken for simplicity $\rho
_{0}=p_{0}=0$. For a test fluid satisfying the ideal gas equation of state:
\begin{equation}
p=\frac{k_{B}}{\mu }\rho T
\end{equation}
where $k_{B}$ is Boltzmann's constant, $T$ is the
temperature and $\mu $ is the mean molecular weight, we obtain:
\begin{equation}
L_{m}\left( \rho \right) =\rho \left\{ 1+\frac{k_{B}T}{\mu }\left[
\ln \left( \frac{\rho }{\rho _{0}}\right) -1\right] \right\} +p_{0}
\end{equation}

In the case of a physical system satisfying the ideal gas equation of state, the
extra acceleration induced by the presence of the non-minimal coupling
between matter and geometry is given by:
\begin{equation}
\vec{a}_{E}\approx -\lambda \frac{k_{B}T}{\mu }\nabla \left[ f_{2}\left( R\right) \ln
\frac{\rho }{\rho _{0}}\right]
\end{equation}
and it is proportional to the temperature of the fluid. It is also
interesting to note that the limiting density and pressure $\rho _{0}$ and $%
p_{0}$ generate in the energy-momentum tensor some extra constant terms,
which may be interpreted as dark energy.

In conclusion, the extra-force induced by the coupling between matter and
geometry does not vanish for any specific choices of the matter Lagrangian.
In the case of the dust, with $p=0$, the extra force is given by:
\begin{equation}
f^{\mu }=-\nabla _{\nu }\ln \left[ 1+\lambda f_{2}(R)\right] \left( U^{\mu
}U^{\nu }-g^{\mu \nu }\right)
\end{equation}
and it is independent of the thermodynamic properties of the system, being
completely determined by the geometry, kinematics and coupling. In the limit of
small velocities and weak gravitational fields, the extra-acceleration of a
dust fluid is given by:
\begin{equation}
\vec{a}_{E}=-\lambda \nabla \left[ f_{2}(R)\right] %
\end{equation}

The thermodynamic condition for the vanishing of the extra-force is:
\begin{equation}
\frac{\partial L_{m}}{\partial g^{\mu \nu }}=\frac{ 1}{2} \left( \frac{\partial
L_{m}}{\partial \rho }\right) \rho \left( g_{\mu \nu }-U_{\mu }U_{\nu }\right)
=0
\end{equation}
only.  If the matter Lagrangian is written as a function of the pressure,
then:
\begin{equation}
\frac{\partial L_{m}}{\partial \rho }=\left( \frac{\partial L_{m}}{\partial p}\right)
\left( \frac{\partial p}{\partial \rho }\right)
\end{equation}
and for all physical systems
satisfying an equation of state (or, equivalently, for all systems with a
non-zero sound velocity), the extra-force is non-zero. Therefore, the
curvature-matter coupling is introduced in the generalized gravity models with a curvature-matter coupling in a consistent way. The coupling determines all of the physical properties of the system, including the extra-force, the matter Lagrangian and the energy-momentum tensor, respectively.

Hence, we have shown that in $f(R)$ gravity
with a non-minimal coupling, the matter Lagrangian and the corresponding
energy-momentum tensor are not model and thermodynamic parameters (independent quantities), but they are
completely and uniquely determined by the nature of the coupling between
matter and geometry, which, in the present model, is given by the function $f_2(R)$. We have obtained this result by deriving first the equations of motion in the modified gravity model with curvature-matter coupling from a variational principle and, then, by taking the
Newtonian limit of the particle action for a fluid obeying a barotropic equation of
state. The energy-momentum tensor of the matter obtained in this way is more
general than the usual general-relativistic energy-momentum tensor for perfect fluids, and it
contains a supplementary term that may be related to the elastic stresses
in the body, or to other sources of internal energy. The matter Lagrangian can be expressed either in terms of the density or in terms of the pressure, and in both representations, the
physical description of the system is equivalent. Therefore, the presence (or
absence) of the extra-force is independent of the specific form of the
matter Lagrangian, and it never vanishes, except in the case of (un)physical
systems with zero sound speed. In particular, in the case of dust particles, the extra-force is always non-zero.

\subsection{Equivalence of the Modified Gravity Theory with Linear Matter-Geometry Coupling with an Anomalous Scaler-Tensor Theory}

As we have shown in Section 2.1, pure $f(R)$ gravity, with action given by Equation~(\ref{fRx}), is
equivalent to a scalar-tensor theory, with action given by Equation~(\ref{scalx}). The equivalence between the modified gravity models with linear geometry-matter coupling was established by Faraoni \cite{Faraoni:2007sn}. In the following, we show that the action  Equation~(\ref{actionlinear})
is equivalent with a scalar-tensor Brans--Dicke-type theory, with a single scalar field,
a vanishing Brans--Dicke parameter $\omega$ and an unusual
coupling of the potential $U(\psi)$ of the theory to matter. 

By introducing a new field $\phi$, the action~Equation~(\ref{actionlinear}) becomes:
\begin{equation} \label{100}
S\int d^4x \sqrt{-g} \left\{ \frac{f_1(\phi)}{2} +\frac{1}{2} \,
\frac{df_1}{d\phi} \left( R-\phi \right) +\left[ 1+\lambda f_2(
\phi) \right] {\cal L}_m \right\}
\end{equation}

Next, we further introduce the field $
\psi(\phi) \equiv f_1'(\phi) $
(with a prime denoting the
differentiation with respect to $\phi$), and we obtain for the action the expression:
\begin{equation} \label{300}
S=\int d^4x \sqrt{-g} \left[ \frac{\psi R }{2} -V(\psi)\,
+U(\psi) {\cal L}_m \right] \;
\end{equation}
where:
\begin{eqnarray}
V(\psi) &=& \frac{\phi(\psi) f_1' \left[ \phi (\psi ) \right]
-f_1\left[ \phi( \psi ) \right] }{2} \; \label{400}\\
U( \psi) & =& 1+\lambda f_2\left[ \phi( \psi ) \right]
\label{500}
\end{eqnarray}

$\phi (\psi)$ must be obtained by inverting $ \psi(\phi) \equiv
f_1'(\phi) $. The
actions~Equations~(\ref{actionlinear}) and (\ref{300}) are equivalent when $f_1''(R)
\neq 0$ \cite{Faraoni:2007sn}. We can see this by setting $\phi=R$, and then Equation~(\ref{300}) reduces
trivially to Equation~(\ref{actionlinear}). On the other hand, the variation of Equation~(\ref{100})
with respect to $\phi$ gives:
\begin{equation} \label{600}
\left( R-\phi \right) f_1''(\phi)+2\lambda f_2'(\phi) {\cal
L}_m=0
\end{equation}

In a vacuum, we have ${\cal L}_m=0$, and Equation~(\ref{600}) gives $\phi=R$
whenever $f_1''\neq 0$ \cite{equiv1,equiv2,equiv3,equiv4}. However, in~the presence of
matter, there seem to be other possibilities, which,
however, can be excluded as~follows. 

When ${\cal L}_m\neq 0$, the
action~Equations~(\ref{actionlinear}) and (\ref{100}) are equivalent if:
\begin{equation}
\left(
R-\phi \right) f_1''(\phi) +2\lambda f_2''(\phi) {\cal L}_m \neq
0
\end{equation}

 When Equation~(\ref{600}) is satisfied, we have a pathological
case. It corresponds to:
\begin{equation} \label{700}
\lambda f_2(\phi) {\cal L}_m= \frac{ f_1'(\phi)}{2} \left(
\phi-R \right) -\frac{ f_1(\phi)}{2} \;
\end{equation}

However, if Equation~(\ref{700}) holds, then the action~Equation~(\ref{100})
reduces to the trivial case of 
pure matter without the gravity sector. Then, it follows that in the modified gravity theories with linear geometry-matter coupling, the
actions~Equations~(\ref{actionlinear}) and (\ref{300}) are equivalent when $f_1''(R)
\neq 0$, similar to the case of pure $f(R)$ gravity \cite{equiv1,equiv2,equiv3,equiv4}. 

\subsection{Further Theoretical Developments in Modified Gravity with Linear Curvature-Matter Coupling}

This linear nonminimal curvature-matter coupling has been extensively explored in a plethora of contexts in the literature. For instance, the equations of motion of test bodies for a theory with nonminimal coupling by means of a multipole method \cite{Puetzfeld:2008xu} was also studied, and it was shown that the propagation equations for pole-dipole particles allow for a systematic comparison with the equations of motion of general relativity and other gravity theories.

The consequences that a non-minimal coupling between curvature and matter can have on the dynamics of perfect fluids has also been investigated \cite{Bertolami:2011rb}. It was argued that the presence of a static, axially-symmetric, pressureless fluid does not imply an asymptotically Minkowski space-time, such as in General Relativity. This feature can be attributed to a pressure mimicking mechanism related to the non-minimal coupling. The case of a spherically symmetric black hole surrounded by fluid matter was analyzed, and it was shown that under equilibrium conditions, the total fluid mass is about twice that of the black hole \cite{Bertolami:2011rb}.

The consequences of the curvature-matter coupling on stellar equilibrium and constraints on the nonminimal coupling was considered, where particular attention was paid to the validity of the Newtonian regime and on the boundary and exterior matching conditions \cite{Bertolami:2007vu}. This explicit ``anomalous'' coupling of the Ricci curvature to matter has also raised the question of curvature instabilities, and in \cite{Faraoni:2007sn, Bertolami:2009cd, Wang:2010zzr, Wang:2010bh, Wang:2012mws}, constraints imposed by the energy condition and the conditions in order to avoid the notorious Dolgov--Kawasaki instability \cite{Dolgov:2003px} were obtained. It has also been claimed that the curvature-matter action leads to a theory of gravity that includes higher order derivatives of the matter fields without introducing more dynamics in the gravity sector and, therefore, cannot be a viable theory for gravitation \cite{Sotiriou:2008dh,Sotiriou:2008dh2}. However, we emphasize that the results of \cite{Sotiriou:2008dh,Sotiriou:2008dh2} only apply to the specific case of $f_2(R)=R$ and not to action Equation~(\ref{actionlinear}) in general. For more generic functional forms, the theory propagates extra degrees of freedom and the conclusions of \cite{Sotiriou:2008dh,Sotiriou:2008dh2} do not apply. The relation between these theories and ordinary scalar-tensor gravity was also analyzed, as well as its implications for the equivalence principle \cite{Sotiriou:2008it}.

In fact, the theoretical consistency of these nonminimal curvature-matter couplings was studied using a scalar field Lagrangian to model the matter content \cite{Tamanini:2013aca}. The conditions that the coupling does not introduce ghosts, classical instabilities or superluminal propagation of perturbations were derived. These~consistency conditions were then employed to rule out or severely restrict the forms of the non-minimal coupling functions \cite{Tamanini:2013aca}.
Lagrange--Noether methods have been used to derive the conservation laws for models in which matter interacts non-minimally with the gravitational field~\cite{Obukhov:2013ona}.
Furthermore, a covariant derivation of the equations of motion for test bodies for a wide class of gravitational theories with nonminimal coupling was presented in \cite{Puetzfeld:2013ir}, encompassing a general interaction via the complete set of nine parity-even curvature invariants. The equations of motion for spinning test bodies in such theories were explicitly derived by means of Synge's expansion technique. The authors' findings generalize previous results in the literature and allowed for a direct comparison to the general relativistic equations of motion of pole-dipole test bodies.
In \cite{Puetzfeld:2013sca}, the authors derived multipolar equations of motion for gravitational theories with general nonminimal coupling in spacetimes admitting torsion. Their general findings allow for a systematic testing of whole classes of theories by means of extended test bodies. One peculiar feature of certain subclasses of nonminimal theories turns out to be their sensitivity to post-Riemannian spacetime structures, even in experiments without microstructured test matter.
Weak field constraints have also been studied in detail \cite{Bertolami:2013qaa,Castel-Branco:2014exa}, and wormhole solutions were further explored, where it is the higher curvature coupling terms that support these exotic geometries \cite{Garcia:2010xb,MontelongoGarcia:2010xd, Bertolami:2012fz}.

In a cosmological context, the perturbation equation of matter on subhorizon scales was deduced, and specific bounds on the theory from weak lensing observations and the primordial nucleosynthesis were obtained in order to constrain the parameters of the model \cite{Nesseris:2008mq}. It was also shown that a non-minimal coupling between the scalar curvature and the matter Lagrangian density may account for the accelerated expansion of the Universe \cite{Bertolami:2010cw,Thakur:2010yv, Bisabr:2012tg} and provide, through mimicking, for a viable unification of dark energy and dark matter \cite{Bertolami:2010cw}. It was shown that a generalized
non-minimal coupling between curvature and matter is compatible with Starobinsky inflation and leads to a successful process of preheating \cite{Bertolami:2010ke}, and the problem of a cosmological constant was further explored \cite{Bertolami:2013uwl}.

The effects of the non-minimal curvature-matter coupling on the evolution of cosmological perturbations around a homogeneous and isotropic Universe and, hence, the formation of large-scale structure have also been analyzed \cite{Bertolami:2013kca}. This framework places constraints on the terms, which arise due to the coupling with matter and, in particular, on the modification in the growth of matter density perturbations. Approximate analytical solutions were obtained for the evolution of matter overdensities during the matter dominated era, and it was shown that these favor the presence of a coupling function that is compatible with the late-time cosmic acceleration.

The observations related to the growth of matter has also shown that there is a small, but finite window, where one can distinguish the non-minimally-coupled $f(R)$ models with the concordance $\Lambda$CDM \cite{Thakur:2013oya}.
The possibility that the behavior of the rotational velocities of test particles gravitating around galaxies can be explained in the framework of modified gravity models with nonminimal curvature-matter coupling has also been extensively explored \cite{Bertolami:2009ic,Harko:2010vs, Bertolami:2011ye}.

In fact, the literature is extremely vast, and rather than enumerate all of the features of these models, we~refer the reader to \cite{Bertolami:2008zh,Bertolami:2013xda} for a review on the topic of the linear nonminimal curvature-matter~coupling.

\newpage
\section{\boldmath{$f\left(R,L_m\right)$} Gravity}\label{Sec:III}

In this section, we generalize the $f(R)$-type gravity models by assuming that the gravitational Lagrangian is given by an arbitrary function of the Ricci scalar $R$ and of the matter Lagrangian $L_m$~\cite{Harko:2010mv}. This consists in a maximal extension of the Hilbert--Einstein action, and the action takes the following~form:
\begin{equation}
S=\int f\left(R,L_m\right) \sqrt{-g}\;d^{4}x
\end{equation}
where $f\left(R,L_m\right)$ is an arbitrary function of the Ricci scalar $R$, and of the Lagrangian density corresponding to matter, $L_{m}$. The
energy-momentum tensor of the matter is defined by Equation~(\ref{EMTdef}). Thus, by assuming that the Lagrangian density $L_{m}$ of the matter depends only
on the metric tensor components $g_{\mu \nu }$, and not on its derivatives,
we obtain $T_{\mu \nu }=g_{\mu \nu }L_{m}-2\;\partial L_{m}/\partial g^{\mu
\nu }$,
which will be useful~below.

Now, varying the action with respect to the metric yields the following field equation:
\begin{eqnarray}\label{field2a}
&&f_{R}\left( R,L_{m}\right) R_{\mu \nu }+\left( g_{\mu \nu }\Box -\nabla
_{\mu }\nabla _{\nu }\right) f_{R}\left( R,L_{m}\right)
\nonumber\\
&&-\frac{1}{2}\left[
f\left( R,L_{m}\right) -f_{L_{m}}\left( R,L_{m}\right)L_{m}\right] g_{\mu \nu }=
\frac{1}{2}%
f_{L_{m}}\left( R,L_{m}\right) T_{\mu \nu }
\end{eqnarray}

If $f\left( R,L_{m}\right) =R/2+L_{m}$ (the Hilbert--Einstein Lagrangian), we
recover the standard Einstein field equation of general relativity, $R_{\mu \nu
}-(1/2)g_{\mu \nu }R=T_{\mu \nu }$. For $f\left( R,L_{m}\right)
=f_{1}(R)+f_{2}(R)G\left( L_{m}\right) $, where $f_{1}$ and $f_{2}$ are
arbitrary functions of the Ricci scalar and $G$ a function of the matter Lagrangian
density, respectively, we reobtain the field equations of the modified
gravity with arbitrary curvature-matter coupling, considered in \cite{Harko:2008qz}.

The contraction of Equation~(\ref{field2a}) provides the following relation between the Ricci scalar $R$, the matter Lagrange density $L_{m}$ and the trace $T=T_{\mu }^{\mu }$ of the energy-momentum tensor:
\begin{eqnarray}
f_{R}\left( R,L_{m}\right) R+3\Box f_{R}\left( R,L_{m}\right) -2\left[
f\left( R,L_{m}\right) -f_{L_{m}}\left( R,L_{m}\right)L_{m}\right]
=\frac{1}{2}f_{L_{m}}\left(
R,L_{m}\right) T \label{contr2a}
\end{eqnarray}

By eliminating the term $\nabla _{\mu }\nabla ^{\mu } f_{R}\left( R,L_{m}\right) $ between Equations~(\ref{field2a}) and (\ref{contr2a}), we obtain another form of the gravitational
field equations as:
\begin{eqnarray}
&&f_{R}\left( R,L_{m}\right) \left( R_{\mu \nu }-\frac{1}{3}Rg_{\mu \nu
}\right) +
\frac{1}{6}\left[ f\left( R,L_{m}\right) -f_{L_{m}}\left( R,L_{m}\right)L_{m}\right] g_{\mu
\nu }=\nonumber\\
&&\frac{1}{2}f_{L_{m}}\left( R,L_{m}\right) \left( T_{\mu \nu }-\frac{1}{%
3}Tg_{\mu \nu }\right) +
\nabla _{\mu }\nabla _{\nu }f_{R}\left(
R,L_{m}\right)
\end{eqnarray}

By taking the covariant divergence of
Equation~(\ref{field2a}), with the use of the mathematical identity \cite{Ko06}:
\begin{eqnarray}
\nabla ^{\mu }\left[ f_R\left(R,L_m\right)R_{\mu \nu }-\frac{1}{2}f\left(R,L_m\right)g_{\mu \nu }+
\left(g_{\mu \nu }\Box -\nabla _{\mu }\nabla _{\nu }\right) f_R\left(R,L_m\right)\right] \equiv 0\,
\end{eqnarray}
we obtain for the divergence of the energy-momentum tensor $T_{\mu \nu}$, the following equation:
\begin{eqnarray}
\nabla ^{\mu }T_{\mu \nu } &=& \nabla ^{\mu }\ln \left[
f_{L_m}\left(R,L_m\right)\right] \left\{ L_{m}g_{\mu \nu
}-T_{\mu \nu }\right\}
 \nonumber \\
&=& 2\nabla
^{\mu }\ln \left[ f_{L_m}\left(R,L_m\right) \right] \frac{\partial L_{m}}{%
\partial g^{\mu \nu }}\, \label{noncons1}
\end{eqnarray}

The requirement of the conservation of the energy-momentum tensor
of matter, $\nabla ^{\mu }T_{\mu \nu }=0$, gives an effective
functional relation between the matter Lagrangian density and the function $f_{L_m}\left(R,L_m\right)$:
\begin{equation}
\nabla
^{\mu }\ln \left[ f_{L_m}\left(R,L_m\right) \right] \frac{\partial L_{m}}{%
\partial g^{\mu \nu }}=0\,
\end{equation}

Thus, once the matter Lagrangian density is known, by an
appropriate choice of the function $f\left( R,L_{m}\right) $, one can construct, at least in principle, conservative models with arbitrary curvature-matter~dependence.

Now, assuming that the matter Lagrangian is a function of the rest mass density $\rho $ of the matter only, from Equation~(\ref{noncons1}), we obtain explicitly the equation of motion of the test particles in the $f\left(R,L_m\right)$ gravity model as:
\begin{equation}
\frac{D^{2}x^{\mu }}{ds^{2}}=U^{\nu }\nabla _{\nu }U^{\mu }=\frac{d^{2}x^{\mu }}{ds^{2}}+\Gamma _{\nu \lambda }^{\mu }U^{\nu }U^{\lambda
}=f^{\mu }
\label{eqgeod}
\end{equation}
where the world-line parameter $s$ is taken as the proper time, $U^{\mu
}=dx^{\mu }/ds$ is the four-velocity of the particle, $\Gamma
_{\sigma \beta }^{\nu }$ are the Christoffel symbols associated with the
metric and the extra-force $f^{\mu}$ is defined as:
\begin{equation}
f^{\mu }=-\nabla _{\nu }\ln \left[ f_{L_m}\left(R,L_m\right) \frac{%
dL_{m}\left( \rho \right) }{d\rho }\right] \left( U^{\mu }U^{\nu }-g^{\mu
\nu }\right)
\label{extra}
\end{equation}

Note that as in the linear curvature-matter coupling, the extra-force $f^{\mu }$, generated by the
curvature-matter coupling, is perpendicular to the four-velocity, $f^{\mu}U_{\mu }=0$. Due to the presence of the extra-force $f^{\mu }$, the motion of the test particles in modified theories of gravity with an arbitrary coupling between matter and curvature is non-geodesic. From the relation $U_{\mu }\nabla _{\nu }U^{\mu }\equiv 0$, it follows that the force $f^{\mu }$ is always perpendicular to the velocity, so that $U_{\mu }f^{\mu }=0$.

Building on this analysis, the geodesic deviation equation, describing the relative accelerations of nearby particles, and the Raychaudhury equation, giving the evolution of the kinematical quantities associated with deformations (expansion, shear and rotation) were extensively considered in this framework of modified theories of gravity with an arbitrary curvature-matter coupling, by taking into account the effects of the extra force
\cite{Harko:2012ve,Harko:2012ve2}. As a physical application of the geodesic deviation equation, the modifications of the tidal forces due to the supplementary curvature-matter coupling were obtained in the weak field approximation. The tidal motion of test particles is directly influenced not only by the gradient of the extra force, which is basically determined by the gradient of the Ricci scalar, but also by an explicit coupling between the velocity and the Riemann curvature tensor. As a specific example, the expression of the Roche limit (the orbital distance at which a satellite will begin to be tidally torn apart by the body it is orbiting) was also obtained for this class of models. These aspects will be presented~below.

Furthermore, the energy conditions and cosmological applications were explored \cite{Wang:2012rw}. In \cite{Huang:2013dca}, the Wheeler--DeWitt equation of $f(R,L_m)$ gravity was analysed in a flat FRW (Friedmann-Robertson-Walker) 
 universe, which is the first step of the study of quantum $f(R,L_m)$ cosmology. In the minisuperspace spanned by the FRW scale factor and the Ricci scalar, the equivalence of the reduced action was examined, and the canonical quantization of the $f(R,L_m)$ model was undertaken and the corresponding Wheeler--DeWitt equation derived. The~introduction of the invariant contractions of the Ricci and Riemann tensors were further considered, and applications to black hole and wormhole physics were analyzed \cite{Tian:2014mta}.

\subsection{Solar System Tests of $f\left(R,L_m\right)$ Gravity}

One of the basic predictions of the modified gravity theories with a curvature-matter coupling is the existence of an extra force, which makes the motion of the test particles non-geodesic. The existence of this force can also be tested at the level of the Solar System, by estimating its effects on the orbital parameters of the motion of the
planets around the Sun. The impact on the planetary motion of the extra force can be obtained in a simple way by using the
properties of the Runge--Lenz vector, defined as:
\begin{equation}
\vec{A}=\vec{v}\times
\vec{L}-\alpha \vec{e}_{r}
\end{equation}
where $\vec{v}$ is the velocity of the planet of mass $m$ relative to the Sun, with mass $M_{\odot}$, $\vec{r}=r\vec{e}_{r}$ the two-body position vector, $\vec{p}=\mu \vec{v}$ the relative
momentum, $\mu =mM_{\odot }/\left( m+M_{\odot}\right) $ the reduced~mass:
\begin{equation}
 \vec{L}=\vec{r}\times \vec{p}=\mu r^{2}\dot{\theta}\vec{k}
 \end{equation}
 is the angular momentum, and $\alpha =GmM_{\odot}$, respectively (see \cite{Harko:2008qz} and the references therein). For an elliptical orbit of
eccentricity $e$, major semi-axis $a$ and period $T$, the equation of the orbit is given by $\left( L^{2}/\mu
\alpha \right) r^{-1}=1+e\cos \theta $. The Runge--Lenz vector can
be expressed as:
\begin{equation}
\vec{A}=\left( \frac{\vec{L}^{2}}{\mu r}-\alpha \right) \vec{e}_{r}-%
\dot{r}L\vec{e}_{\theta }
\end{equation}
and its derivative with respect to the polar
angle $\theta $ is given by $d\vec{A}/d\theta =r^{2}\left[ dV(r)/dr-\alpha
/r^{2}\right] \vec{e}_{\theta }$, where~$%
V(r)$ is the potential of the central force \cite{Harko:2008qz}.

The gravitational potential term acting on a planet consists of the Post--Newtonian potential:
\begin{equation}
V_{PN}(r)=-\frac{\alpha }{r}-3\frac{\alpha ^{2}}{mr^{2}}
\end{equation}
plus the gravitational term induced by the general coupling between matter and geometry.
Thus, we~have:
\begin{equation}
\frac{d\vec{A}}{d\theta }=r^{2}\left[ 6\frac{\alpha ^{2}}{mr^{3}}+m\vec{a}%
_{E}(r)\right] \vec{e}_{\theta }
\end{equation}
where we have also assumed that $\mu \approx m$. Then, we obtain the change of $\Delta \phi $ of the perihelion with a change of $\theta $ of $2\pi $ as:
\begin{equation}
\Delta \phi =\left( \frac{1}{\alpha e}\right) \int_{0}^{2\pi }\left\vert \dot{\vec{L}}\times d%
\frac{\vec{A}}{d\theta }\right\vert d\theta
\end{equation}
which can be explicitly calculated as:
\begin{equation}\label{prec}
\Delta \phi =24\pi ^{3}\left( \frac{a}{T}\right) ^{2}\frac{1}{1-e^{2}}+\frac{%
L}{8\pi ^{3}me}\frac{\left( 1-e^{2}\right) ^{3/2}}{\left( a/T\right) ^{3}}%
\int_{0}^{2\pi }\frac{a_{E}\left[ L^{2}\left( 1+e\cos \theta \right)
^{-1}/m\alpha \right] }{\left( 1+e\cos \theta \right) ^{2}}\cos \theta
d\theta
\end{equation}%
where we have used the relation $\alpha /L=2\pi \left( a/T\right) /\sqrt{%
1-e^{2}}$. The first term of Equation~(\ref{prec}) gives the expression of the standard general relativistic precession of the perihelion of the planets, while the second term gives the contribution to the perihelion precession due to the presence of the extra force generated by the coupling between matter and curvature.

As an example of the application of Equation~(\ref{prec}), we consider the simple case for which the extra force may be considered as a constant, $a_E\approx$ constant, an approximation that could be valid for small regions of space-time. This case may also correspond to a MOND-type acceleration \mbox{$a_E\approx \sqrt{a_0a_N}=\sqrt{GM_{\odot}a_0}/r$}, where $a_0$ is a constant acceleration, which was proposed phenomenologically as a dynamical model for dark matter \cite{Milgrom1, Milgrom2}. In the Newtonian limit, the extra-acceleration generated by the curvature-matter coupling can be expressed in a similar form \cite{Bertolami:2007gv}. With the use of Equation~(\ref{prec}), one finds for the perihelion precession of a planet in the Solar System the~expression:
\begin{equation}\label{prec1}
\Delta \phi =\frac{6\pi GM_{\odot}}{a\left( 1-e^{2}\right) }+\frac{2\pi a^{2}%
\sqrt{1-e^{2}}}{GM_{\odot}}a_{E}
\end{equation}
where we have also used Kepler's third law, $T^2=4\pi ^2a^3/GM_{\odot}$. For the planet Mercury \mbox{$a=57.91\times 10^{11}$~cm}, and $e=0.205615$,
respectively, while $M_{\odot }=1.989\times 10^{33}$ {\it g} \cite{Will:2014xja}. With the use of these
numerical values, the first term in Equation~(\ref{prec1}) gives the standard
general relativistic value for the precession angle, $\left( \Delta \phi
\right) _{GR}=42.962$ arcsec per century. On the other hand, the observed value of the precession of the perihelion of Mercury is $\left(\Delta \phi \right)_{obs}=43.11\pm0.21$ arcsec per century \cite{Will:2014xja}. Therefore, the difference $\left(\Delta \phi \right)_{E}=\left(\Delta \phi \right)_{obs}-\left( \Delta \phi
\right) _{GR}=0.17$ arcsec per century can be attributed to other physical effects. Hence, the observational constraint on the perihelion precession of the planet Mercury requires
that the value of the constant acceleration $a_E$ induced by the curvature-matter coupling at the scale of the Solar System must satisfy the condition $a_E\leq 1.28\times 10^{-9}$ cm/s$^2$. This value of $a_E$, obtained from the
high precision Solar System observations, is somewhat smaller than the value of the extra acceleration $%
a_{0}\approx 10^{-8}$ cm/s$^{2}$, necessary to explain
the ``dark matter'' properties, as well as the Pioneer anomaly \citep{Bertolami:2007gv}. However, this value does not rule out the possibility of the presence of some extra curvature-matter coupling-type gravitational effects, acting at both the Solar System and galactic levels, since the assumption of a constant extra acceleration may not be correct on larger astronomical scales.

\subsection{The Geodesic Deviation Equation and the Raychaudhury Equation in $f\left(R,L_m\right)$ Gravity}

In the present subsection, we present the equation of the geodesic deviation and the Raychaudhury equation in $f\left(R,L_m\right)$ gravity, which explicitly contain the effects of the curvature-matter coupling and of the extra force. Some of the physical implications of the geodesic deviation equation, namely, the problem of the tidal forces in $f\left(R,L_m\right)$ gravity is also considered, and the generalization of the Roche limit will be considered in the next subsection. In our presentation, we closely follow \cite{Harko:2012ve,Harko:2012ve2}.

Consider a one-parameter congruence of curves $x^{\mu }\left( s;\lambda
\right) $, so that for each $\lambda =\lambda _{0}=$ constant, $x^{\mu
}\left( s,\lambda _{0}\right) $ satisfies Equation~(\ref{eqgeod}). We suppose the
parametrization to be smooth, and hence, we can introduce the tangent vector
fields along the trajectories of the particles as $U^{\mu }=\partial x^{\mu
}\left( s;\lambda \right) /\partial s$ and $n^{\mu }=\partial x^{\mu }\left(
s;\lambda \right) /\partial \lambda $, respectively. We also introduce the
four-vector:
\begin{equation}
\eta ^{\mu }=\left[ \frac{\partial x^{\mu }\left( s;\lambda \right)
}{\partial \lambda }\right] \delta \lambda \equiv n^{\mu }\delta \lambda
\end{equation}
joining points on infinitely close geodesics, corresponding to parameter
values $\lambda $ and $\lambda +\delta \lambda $, which have the same value
of $s$. By taking into account Equation~(\ref{eqgeod}), we obtain the geodesic deviation equation
(Jacobi equation), giving the second order derivative with respect to the parameter $s$ of the deviation vector $\eta ^{\mu }$ as \cite{LaLi,Haw}:
\begin{equation}
\frac{D^{2}\eta ^{\mu }}{ds^{2}}=R_{\nu \alpha \beta }^{\mu }\eta ^{\alpha
}U^{\beta }U^{\nu }+\eta ^{\alpha }\nabla _{\alpha }f^{\mu } \label{eq3}
\end{equation}

In the case $f^{\mu }\equiv 0$, we reobtain the standard Jacobi equation,
corresponding to the geodesic motion of test particles in standard general relativity. The interest in the
deviation vector $\eta ^{\mu }$ derives from the fact that if $x_{0}^{\mu
}(s)=x^{\mu }\left( s;\lambda _{0}\right) $ is a solution of Equation~(\ref{eq3}%
), then to first order $x_{1}^{\mu }(s)=x_{0}^{\mu }(s)+\eta ^{\mu }$ is a
solution of the geodesic equation, as well, since $x^{\mu }\left( s;\lambda _{1}\right) \approx x^{\mu
}\left( s;\lambda _{0}\right) +n^{\mu }\left( s;\lambda _{0}\right) \delta
\lambda \approx x^{\mu }\left( s;\lambda _{0}+\delta \lambda \right) $.

By taking into account the explicit form of the extra force
given by Equation~(\ref{extra}), in $f\left(R,L_m\right)$-modified theories of gravity with a curvature-matter coupling, the geodesic deviation equation can be written as:
\bea
\frac{D^{2}\eta ^{\mu }}{ds^{2}}=R_{\nu \alpha \beta }^{\mu }\eta ^{\alpha
}U^{\beta }U^{\nu }+\eta ^{\alpha }\nabla _{\alpha }\left\{ \nabla _{\nu
}\ln \left[ f_{L_m}\left( R, L_m\right)\frac{%
dL_{m}\left( \rho \right) }{d\rho } \right] \left(
L_{m}g^{\mu \nu }-T^{\mu \nu}\right) \right\}
\eea

Explicitly, the geodesic deviation equation becomes:
\bea
\frac{D^{2}\eta ^{\mu }}{ds^{2}}&=&R_{\nu \alpha \beta }^{\mu }\eta ^{\alpha
}U^{\beta }U^{\nu }
 +\eta ^{\alpha }\left\{\nabla _{\alpha }\nabla _{\nu }\ln %
\left[f_{L_m}\left( R,L_m\right)\frac{%
dL_{m}\left( \rho \right) }{d\rho } \right]\right\} \left(
L_{m}g^{\mu \nu }-T^{\mu \nu }\right)
 \nonumber\\
&&+\eta ^{\alpha }\nabla _{\nu }\ln \left[ f_{L_m}\left( R,L_m\right)\frac{%
dL_{m}\left( \rho \right) }{d\rho } \right]\left( g^{\mu \nu } \nabla _{\alpha }L_{m} -\nabla
_{\alpha }T^{\mu \nu }\right) 
\eea

In the presence of an extra force, the Raychaudhury equation is obtained as \cite{Haw}:
\begin{equation}
\dot{\theta}+\frac{1}{3}\theta ^{2}+\left( \sigma ^{2}-\omega ^{2}\right)
=\nabla _{\mu }f^{\mu }+R_{\mu \nu }U^{\mu }U^{\nu }
\end{equation}
where $\theta =\nabla _{\nu }U^{\nu }$ is the expansion
of the congruence of particles, $\sigma ^{2}=\sigma _{\mu \nu }\sigma ^{\mu \nu }$ and \mbox{$\omega^{2}=\omega _{\mu \nu }\omega ^{\mu \nu }$},~respectively.

With the explicit use of the field equation Equation~(\ref{field2a}) and the
expression of the extra force given by Equation~(\ref{extra}), in $f\left(R,L_m\right)$-modified theories of gravity with an arbitrary coupling between curvature and matter, the Raychaudhury equation assumes the following generalised form:
\begin{eqnarray}
\dot{\theta} &=&-\frac{1}{3}\theta ^{2}-\left( \sigma ^{2}-\omega
^{2}\right) + \Lambda \left( R,L_{m}\right)
 +\nabla _{\mu }\left\{ \nabla _{\nu }\ln \left[ f_{L_m}\left( R,L_m\right)\frac{%
dL_{m}\left( \rho \right) }{d\rho } \right] \left( L_{m}g^{\mu \nu }-T^{\mu
\nu }\right) \right\}
 \nonumber \\
&&+\frac{1}{f_R\left( R,L_{m}\right) }U^{\mu
}U^{\nu }\nabla _{\mu }\nabla _{\nu }f_R\left( R,L_{m}\right)+\Phi \left(
R,L_{m}\right) \left( T_{\mu \nu }U^{\mu }U^{\nu }-\frac{1}{3}T\right)
\end{eqnarray}
where we have denoted:
\begin{equation}
\Lambda \left( R,L_{m}\right) =\frac{2f_R\left( R,L_{m}\right)R
-f\left( R,L_{m}\right) +f_{L_m}\left(R,
L_{m}\right) L_{m}}{6f_R\left( R,L_{m}\right) }
\label{lambda}
\end{equation}
and:
\begin{equation}
\Phi \left( R,L_{m}\right) =\frac{f_{L_m}\left(R, L_{m}\right)
}{f_R\left( R,L_{m}\right) }
\end{equation}
respectively.

\subsection{Tidal Forces and the Roche Limit}

Tides are common astrophysical phenomena, and they are due to the presence of a gradient of the gravitational force field
induced by a mass above an extended body or a system of particles. In the Solar System, tidal perturbations act on compact celestial bodies, such as planets,
moons and comets. On larger scales than the Solar System, as, for example, in a galactic
or cosmological context, one can observe tidal deformations or disruptions
of a stellar cluster by a galaxy or in galaxy encounters \cite{mas1,mas2, mas3}. In~the
relativistic theories of gravitation, as well as in Newtonian gravity, a
local system of coordinates can be chosen, which is inertial, except for the
presence of the tidal forces. In strong gravitational fields, relativistic
tidal effects can lead to important physical phenomena, such as the emission of
tidal gravitational waves \cite{mas1,mas2, mas3}.

In the following, we denote by a prime the reference frame in which all of the Christoffel symbols vanish. In
such a system, one can always take the deviation vector component $\eta ^{\prime 0}=0$, which means that the
particle accelerations are compared at equal times. $\eta ^{\prime i}$ is
then the displacement of the particle from the origin \cite{Oh}. Moreover,
in the static/stationary case, in which the metric, the Ricci scalar and the
thermodynamic parameters of the matter do not depend on time, we have $f^{\prime 0}=0$. With the use of the equation of motion, this condition implies $U^{\prime 0}=$ constant $=1$. Therefore, with these assumptions, in a static or stationary space-time, the equation of the geodesic deviation (the Jacobi equation) takes the form \cite{Harko:2012ve}:
\begin{equation}
\frac{d^{2}\eta ^{\prime i}}{dt^{\prime 2}}=R_{0l0}^{\prime i}\eta ^{\prime
l}+R_{jlm}^{\prime i}\eta ^{\prime l}U^{\prime j}U^{^{\prime }m}+\eta
^{\prime l}\frac{\partial f^{\prime i}}{\partial x^{\prime l}}
\label{geodin}
\end{equation}

Equation~(\ref{geodin}) can be reformulated as:
\begin{equation}
F^{\prime i}=\frac{d^{2}\eta ^{\prime i}}{dt^{\prime 2}}=K_{j}^{i}\eta
^{\prime j}
\end{equation}
where $F^{\prime i}$ is the tidal force, and we have also introduced the
generalized tidal matrix $K_{l}^{i}$ \cite{mas1,mas2, mas3}, which is defined as:
\begin{equation}
K_{j}^{i}=R_{0j0}^{\prime i}+R_{kjm}^{\prime i}U^{\prime k}U^{^{\prime }m}+%
\frac{\partial f^{\prime i}}{\partial x^{\prime j}}
\end{equation}

The tidal force has the property:
\begin{equation}
\frac{\partial F^{\prime i}}{\partial \eta
^{\prime j}}=K_{j}^{i}
\end{equation}
and its divergence is given by:
\begin{equation}
\frac{\partial F^{\prime
i}}{\partial \eta ^{\prime i}}=K
\end{equation}
where the trace $K$ of the tidal matrix is:
\begin{equation}
K=K_{i}^{i}=R_{00}^{\prime }+R_{km}^{\prime }U^{\prime k}U^{^{\prime }m}+%
\frac{\partial f^{^{\prime }j}}{\partial x^{\prime j}}
\end{equation}

With the use of the gravitational field equations Equation~(\ref{field2a}), we can
express $K$ as:
\begin{eqnarray}
K &=&\Lambda \left( R^{\prime },L_{m}^{\prime }\right) \eta _{00}+\frac{1}{%
f_R\left( R^{\prime },L_{m}\right) }\frac{\partial ^{2}}{\partial t^{\prime 2}}%
f_R\left( R^{\prime },L_{m}^{\prime }\right) +\Phi \left( R^{\prime
},L_{m}^{\prime }\right) \left( T_{00}^{\prime }-\frac{1}{3}T^{\prime }\eta
_{00}\right) \nonumber\\
&&+\Lambda \left( R^{\prime },L_{m}^{\prime }\right) \eta _{km}U^{\prime
k}U^{^{\prime }m}
+\frac{1}{f_R\left( R^{\prime },L_{m}^{\prime }\right) }%
U^{\prime k}U^{\prime m}\frac{\partial ^{2}}{\partial x^{\prime k}\partial
x^{\prime m}}f_R\left( R^{\prime },L_{m}^{\prime }\right) \nonumber\\
&&+\phi \left( R^{\prime },L_{m}^{\prime }\right) \left( T_{km}^{\prime }-%
\frac{1}{3}T\eta _{km}\right) U^{\prime k}U^{\prime m}
 \nonumber \\
&&+\frac{\partial }{\partial x^{\prime k}}\left\{ \frac{\partial }{\partial
x^{\prime m}}\ln \left[ f_{L_m}\left( R^{\prime
},L_{m}^{\prime } \right) \frac{%
dL_{m}^{\prime }\left( \rho \right) }{d\rho }\right] \left( L_{m}^{\prime }\eta ^{km}-T^{\prime km}\right)
\right\}
\end{eqnarray}

Then, since in the Newtonian limit, one can omit all time derivatives, we
obtain for $R_{0l0}^{i}$ the expression~\cite{Oh}:
\begin{equation}
R_{0l0}^{i}=\frac{\partial ^{2}\phi}
{\partial x^{i}\partial x^{l}}
\end{equation}

For simplicity, in the
following, we omit the primes for the geometrical and physical
quantities in the Newtonian approximation. Therefore, in $f\left(R,L_m\right)$-modified gravity with a
curvature-matter coupling, we obtain for the tidal acceleration of the test
particles the expression:
\begin{equation}
\frac{d^{2}\eta ^{i}}{dt^{2}}=F^{i}\approx \frac{\partial ^{2}\phi }{%
\partial x^{i}\partial x^{l}}\eta ^{l}+R_{jlm}^{i}\eta ^{l}V^{j}V^{m}+\eta
^{l}\frac{\partial f^{i}}{\partial x^{l}} \label{newgeod}
\end{equation}
where $V^{j}$ and $V^{m}$ are the Newtonian three-dimensional velocities. In
the Newtonian approximation, in modified theories of gravity with a curvature-matter coupling, the tidal force tensor is defined as:
\begin{equation}
\frac{\partial F^{i}}{\partial \eta ^{l}}=\frac{\partial ^{2}\phi }{\partial
x^{i}\partial x^{l}}+R_{jlm}^{i}V^{j}V^{m}+\frac{\partial f^{i}}{\partial
x^{l}}
\end{equation}
and its trace gives the generalized Poisson equation:
\begin{equation}
\frac{\partial F^{i}}{\partial \eta ^{i}}=\Delta \phi +R_{jm}V^{j}V^{m}+%
\frac{\partial f^{i}}{\partial x^{i}}
\end{equation}

In Newtonian gravity:
\begin{equation}
\chi_{il}=-\frac{\partial ^{2}\phi }{\partial x^{i}\partial x^{l}}
\end{equation}
 represents the Newtonian tidal tensor \cite{Oh}. In the Newtonian approximation, the spherical potential of a given particle with
mass $M$ is $\phi (r)=-M/8\pi r$. By choosing a frame of reference, so that
the $x$-axis passes through the particle's position, corresponding to the
radial spherical coordinate, that is, $(x=r,y=0,z=0)$, the Newtonian tidal
tensor is diagonal and has the only non-zero components:
\begin{equation}
\chi _{ii}=\mathrm{%
diag}\left( \frac{2M}{8\pi r^{3}},-\frac{M}{8\pi r^{3}},-\frac{M}{8\pi r^{3}}\right)
\end{equation}

The Newtonian tidal force $\vec{F}_{t}$ can be written as $F_{tx}=2M\Delta
x/8\pi r^{3}$, $F_{ty}=-GM\Delta y/8\pi r^{3}$ and $F_{tz}=-M\Delta z/8\pi
r^{3}$, respectively \cite{Oh}. These results can be used to derive the
generalization of the Roche limit in modified gravity with an arbitrary coupling between matter and curvature.

The Roche limit is the closest distance $r_{Roche}$ that a celestial object
with mass $m$, radius $R_{m}$ and density $\rho _{m}$, held together only by
its own gravity, can come to a massive body of mass $M$, radius $R_{M}$ and
density $\rho _{M}$, respectively, without being pulled apart by the massive
object's tidal (gravitational) force \cite{mas1,mas2,mas3}. For simplicity, we will
consider $M \gg m$, so that the center of mass of the system nearly coincides
with the geometrical center of the mass $M$.

The elementary Newtonian theory of this process is as follows. Consider a small mass $\Delta m$ located at the surface of the small object of mass $m$. There are two forces acting on $\Delta m$, the gravitational attraction of the mass $m$, given by:
\be
F_{G}=\frac{m\Delta m}{8\pi R_{m}^{2}}
\ee
and the tidal force exerted by the massive object, which is given by:
\be
F_{tr}=\frac{M\Delta mR}{8\pi r^{3}}
\ee
where $r$ is the distance between the centers of the two celestial bodies. The Roche limit is reached at the distance $r=r_{Roche}$, when the gravitational force and the tidal force exactly balance each other, $F_{G}=F_{tr}$, thus giving \cite{mas1,mas2,mas3}:
\be
r_{Roche}=R_{m}\left(\frac{ M}{m}\right) ^{1/3}=2^{1/3}R_{M}\left( \frac{\rho _{M}}{\rho
_{m}}\right) ^{1/3}
\ee

In modified gravity with a curvature-matter coupling, the equilibrium
between gravitational and tidal forces occurs at a distance $r_{Roche}$ given by the equation:
\begin{equation}
\left( \frac{M}{8\pi r_{Roche}^{3}}+R_{jrm}^{r}V^{j}V^{m}+\frac{\partial
f^{r}}{\partial r}\right) R_{m}=\frac{m}{8\pi R_{m}^{2}}+f^{r}
\end{equation}
where $f^{r}$ is the radial component of the extra-force, which modifies the
Newtonian gravitational force, and the curvature tensor $R_{jrm}^{r}$ (no
summation upon $r$) must be evaluated in the coordinate system in which the
Newtonian tidal tensor is diagonal. Hence, we obtain the generalized Roche
limit in the presence of arbitrary curvature-matter coupling as \cite{Harko:2012ve}:
\begin{eqnarray}
r_{Roche}&\approx & R_{m}\left( \frac{M}{m}\right) ^{1/3} \left[ 1+\frac{8\pi
R_{m}^{3}}{3m}\left( R_{jrm}^{r}V^{j}V^{m}+\frac{\partial f^{r}}{\partial r}%
\right) -\frac{8\pi R_{m}^{2}}{3m}f^{r}\right]
\end{eqnarray}
where we have assumed that the gravitational effects due to the coupling
between matter and curvature are small as compared to the Newtonian ones. Relativistic corrections to the
Newtonian tidal accelerations caused by a massive rotating source, such as,
for example, the Earth, could be determined experimentally, at least in
principle, thus leading to the possibility of testing relativistic theories
of gravitation by measuring such effects in a laboratory.

\section{Extended \boldmath{$f\left(R,L_m \right)$} Gravity with the Generalized Scalar Field and Kinetic Term Dependencies}\label{Sec:IV}\vspace{-24pt}

\subsection{Action and Field Equations}\label{sect3_1}

We generalize $f\left(R,L_m \right)$ gravity, outlined in the previous section, by considering a novel gravitational model with an action given by an arbitrary function of the Ricci scalar, the matter Lagrangian density, a~scalar field and a kinetic term constructed from the gradients of the scalar field, respectively \cite{Harko:2012hm}. The~action is given by:
\begin{equation}\label{1}
S=\int f\left(R,L_m, \phi ,g^{\mu \nu}\nabla _{\mu }\phi \nabla _{\nu }\phi \right) \sqrt{-g}\;d^{4}x
\end{equation}
where $\sqrt{-g}$ is the determinant of the metric tensor $g_{\mu \nu}$, and $f\left(R,L_m, \phi, g^{\mu \nu}\nabla _{\mu }\phi \nabla _{\nu }\phi \right)$ is an arbitrary function of the Ricci scalar $R$, the matter Lagrangian density, $L_{m}$, a scalar field $\phi $ and the gradients constructed from the scalar field, respectively. The only restriction on the function $f$ is to be an analytical function of $R$, $L_m$, $\phi $, and of the scalar field kinetic energy, respectively, that is, $f$ must possess a Taylor series expansion about any point.

Note that one may motivate the introduction of the action  Equation~(\ref{1}) by the extensive interest in the literature between couplings between the scalar and the matter fields \cite{gasperiniPRD02,damourPRD96,dilaton2EP,dilaton2EP2,dilaton2EP3,armendarizPRD02,EP,EP2,int,int2,int3,int4,NMC,min2, Cota}. Indeed, such couplings generically appear in Kaluza--Klein theories with compactified dimensions \cite{KK} or in the low energy effective limit of string theories \cite{gasperiniPRD02,damourPRD96,dilaton2EP,dilaton2EP2,dilaton2EP3}, where the dilaton has been proposed to be a good candidate for quintessence \cite{gasperiniPRD02} or the inflaton \cite{damourPRD96}. The theory represented by action Equation~(\ref{1}) offers a generalization of the above theories in a single theoretical framework.

Here, we introduce first the ``reduced'' energy-momentum tensor $\tau _{\mu \nu }$ of the matter, which is defined as \cite{LaLi},
\begin{equation}
\tau _{\mu \nu }=-\frac{2}{\sqrt{-g}}\frac{\delta \left( \sqrt{-g}L_{m}\right)}{
\delta g^{\mu \nu }}=g_{\mu \nu }L_{m}-2\frac{\delta L_{m}}{\delta g^{\mu
\nu }}
\end{equation}

In addition to this, we assume that the scalar field $\phi $ is independent of the metric, \textit{i.e.}, $\delta \phi /\delta g^{\mu \nu }\equiv 0$, and in the following, we define, for simplicity,
$\left(\nabla \phi\right)^2=g^{\mu \nu}\nabla _{\mu }\phi \nabla _{\nu }\phi $.

By varying the action $S$ with respect to the metric tensor and to $\phi $, we obtain the following field~equation:
\begin{eqnarray}\label{field3a}
f_{R} R_{\mu \nu }+\left( g_{\mu \nu }\nabla _{\lambda }\nabla^{\lambda }
-\nabla
_{\mu }\nabla _{\nu }\right)f_R
- \frac{1}{2}\left(
f -f_{L_{m}}L_m \right)g_{\mu \nu }
=\frac{1}{2}%
f_{L_{m}} \tau _{\mu \nu }-f_{\left(\nabla \phi \right)^2 }\nabla _{\mu }\phi \nabla _{\nu }\phi
\end{eqnarray}
and the evolution equation for the scalar field:
\begin{equation}
\square _{\left( \nabla \phi \right) ^{2}}\phi =\frac{1}{2}f_{\phi }
\end{equation}
respectively. The subscript of $f$ denotes a partial derivative with respect to the arguments, \textit{i.e.}, \mbox{$f_R=\partial f/\partial R$, $f_{L_m}=\partial f/\partial L_m$}, $f_{\left(\nabla \phi \right)^2 }=\partial f/\partial \left(\nabla \phi \right)^2$, $f_\phi=\partial f/\partial \phi$, and:
\begin{equation}\label{2}
\square _{\left( \nabla \phi \right) ^{2}}=\frac{1}{\sqrt{-g}}\frac{\partial
}{\partial x^{\mu }}\left[ f_{\left( \nabla \phi \right) ^{2}}\sqrt{-g}%
g^{\mu \nu }\frac{\partial }{\partial x^{\nu }}\right]
\end{equation}
is the generalized d'Alembert operator of $f\left(R,L_m, \phi, \nabla _{\mu }\phi \nabla ^{\mu }\phi \right)$ gravity.

The contraction of Equation~(\ref{field3a}) provides the following relation between the Ricci scalar $R$, the matter Lagrangian density $L_{m}$, the derivatives of the scalar field and the trace $\tau =\tau _{\mu }^{\mu }$ of the ``reduced'' energy-momentum tensor:
\begin{eqnarray}
f_{R} R+3\nabla _{\mu }\nabla ^{\mu } f_{R} -2\left(
f-f_{L_{m}}L_{m}\right)=
\frac{1}{2}f_{L_{m}} \tau -f_{\left( \nabla \phi \right) ^{2}}\nabla _{\mu }\phi \nabla ^{\mu }\phi \label{contrb}
\end{eqnarray}

By taking the covariant divergence of Equation~(\ref{field3a}), we obtain for the covariant divergence of the ``reduced'' energy-momentum tensor the following expression:
\begin{eqnarray}
\frac{1}{2} \nabla^\sigma \left(f_{L_m} \tau_{\mu \sigma}\right) = \frac{1}{2} \left( L_m \nabla_\mu f_{L_m}-f_\phi \nabla_\mu \phi \right)+f_{\left(\nabla \phi \right)^2}\nabla_\mu \phi \nabla_\sigma \nabla^\sigma \phi+
\nabla_\mu \phi \nabla^\sigma \phi \nabla_\sigma f_{\left(\nabla \phi \right)^2}
\end{eqnarray}

This relationship was deduced by taking into account the following mathematical identities:
\bea
\nabla ^{\mu }R_{\mu \nu }=\frac{1}{2} \nabla _{\nu }R \,, \qquad
\left( \nabla _{\nu }\square
-\square \nabla _{\nu }\right) f_{R}=-\left( \nabla ^{\mu }f_{R}\right)R_{\mu \nu }
\eea
and considering torsion-free space-times, such that
$\left[ \nabla_\sigma \nabla_\epsilon - \nabla_\epsilon \nabla_\sigma \right] \psi =0$,
where $\psi$ is any scalar-field. Now, using Equation~(\ref{2}), we get:
\begin{eqnarray}
\nabla^\sigma \left(f_{L_m} \tau_{\mu \sigma} \right) = L_m \nabla_\mu f_{L_m}
\end{eqnarray}

For $\phi \equiv 0$, Equation~(\ref{field3a}) reduces to the field equations of the $f\left(R,L_m\right)$ model considered in the previous section \cite{Harko:2010mv}. For $\phi \neq 0$, one recovers the conservation equations for either General Relativity and Brans--Dicke-like scalar-tensor theories (with and without scalar/matter coupling \cite{Min}). For instance, the total Lagrangian of the simplest matter-scalar field-gravitational field theory, with a scalar field kinetic term and a self-interacting potential $V(\phi)$, corresponds to the choice:
\begin{equation}
f=\frac{R}{2}+L_{m}+\frac{\lambda }{2} g^{\mu \nu }\nabla _{\mu }\phi \nabla _{\nu }\phi +V(\phi )
\end{equation}
where $\lambda $ is a constant. The corresponding field equations can be immediately obtained from Equation~(\ref{field3a}) as:
 \bea
 R_{\mu \nu }-\frac{1}{2}Rg_{\mu \nu }=\tau _{\mu \nu }-\lambda \nabla _{\mu }\phi \nabla _{\nu }\phi
 + \left[\frac{\lambda}{2} g^{\alpha \beta }\nabla _{\alpha }\phi \nabla _{\beta }\phi +V(\phi)\right]g_{\mu \nu }
 \eea

The scalar field satisfies the evolution equation:
\begin{equation}
\frac{1}{\sqrt{-g}}\frac{\partial }{\partial x^{\mu}}\left[\sqrt{-g}g^{\mu \nu }\frac{\partial \phi }{\partial x^{\nu }}\right]=\frac{1}{\lambda }\frac{dV(\phi )}{d\phi }
\end{equation}
while the energy-momentum tensor obeys the conservation equation $\nabla ^{\sigma }\tau _{\mu \sigma }=0$.

\subsection{Models with Nonminimal Matter-Scalar Field Coupling}\label{sect3_2}

As an example of the application of the formalism developed in the previous section, we consider a simple phenomenological model, in which a scalar field is non-minimally coupled to pressureless matter with rest mass density $\rho $. For the action of the system, we consider:
\be
S=\int{\left[\frac{R}{2}-F(\phi )\rho +\lambda g^{\mu \nu }\nabla _{\mu }\phi \nabla _{\nu }\phi \right]\sqrt{-g}d^4x}
\ee
where $F(\phi )$ is an arbitrary function of the scalar field that couples non-minimally to ordinary matter. The field equations for this model are given by:
\bea
R_{\mu \nu }-\frac{1}{2}Rg_{\mu \nu }&=&F(\phi )\rho U_{\mu } U_{\nu }
 +2\lambda \left[\nabla _{\mu }\phi \nabla _{\nu }\phi -\frac{1}{2}g_{\mu \nu }\nabla _{\alpha }\phi \nabla ^{\alpha }\phi \right]
\eea
where $U^{\mu }$ is the four-velocity of the matter fluid. The scalar field satisfies the evolution equation:
\be \label{square1}
\square \phi =-\frac{1}{2\lambda }\frac{dF(\phi)}{d\phi }\rho
\ee
where $\square $ is the usual d'Alembert operator defined in a curved space. The total energy-momentum tensor of the scalar field-matter system is given by:
\be
T_{\mu \nu }=F(\phi )\rho U_{\mu } U_{\nu }+ 2\lambda \left[\nabla _{\mu }\phi \nabla _{\nu }\phi -\frac{1}{2}g_{\mu \nu }\nabla _{\alpha }\phi \nabla ^{\alpha }\phi \right]
\ee

The Bianchi identities imply that $\nabla _{\nu }T^{\mu \nu }=0$, and in the following, we assume that the mass density current is conserved, \textit{i.e.},
$\nabla _{\nu }\left(\rho U^{\nu }\right)=0$. Using the latter condition and the mathematical identity given by $\left[ \nabla_\sigma \nabla_\epsilon - \nabla_\epsilon \nabla_\sigma \right] \psi =0$, we have:
\be
F(\phi )\rho \left( U^{\nu }\nabla _{\nu } U^{\mu}\right)+\rho U^{\mu} U^{\nu }\frac{dF(\phi )}{d\phi }\nabla _{\nu }\phi +2\lambda \left(\nabla ^{\mu }\phi \right)\square \phi =0
\ee
Then, by eliminating the term $\square \phi $ with the help of Equation~(\ref{square1}), we obtain:
\be
U^{\nu }\nabla _{\nu } U^{\mu }+\left[\frac{d}{d\phi }\ln F(\phi )\right]\left(U^{\mu } U^{\nu }\nabla _{\nu }\phi -\nabla ^{\mu }\phi \right)=0
\ee

Using the identity $U^{\nu }\nabla _{\nu } U^{\mu }\equiv \frac{d^2x^{\mu }}{ds^2}+\Gamma _{\alpha \beta }^{\mu }u^{\alpha }u^{\beta },$
where $\Gamma _{\alpha \beta }^{\mu }$ are the Christoffel symbols corresponding to the metric, the equation of motion of the test particles
non-minimally coupled to an arbitrary scalar field takes the form:
\be\label{20}
\frac{d^2x^{\mu }}{ds^2}+\Gamma _{\alpha \beta }^{\mu } U^{\alpha } U^{\beta }+ \left[\frac{d}{d\phi }\ln F(\phi )\right]\left( U^{\mu } U^{\nu }\nabla _{\nu }\phi -\nabla ^{\mu }\phi \right)=0
\ee

A particular model can be obtained by assuming that $F(\phi )$ is given by a linear function:
\be
F(\phi )=\frac{\Lambda +1}{2}\left[1+\frac{1}{2}\left(\Lambda -1\right)\phi \right]
\ee
where $\Lambda $ is a constant. Then, the equation of motion becomes:
\be
\frac{d^2x^{\mu }}{ds^2}+\Gamma _{\alpha \beta }^{\mu }U^{\alpha }U^{\beta }+\left(U^{\mu }U^{\nu }-g^{\mu \nu }\right)\nabla _{\nu }\ln\left[1+\frac{\Lambda -1}{2}\phi \right]=0
\ee

In order to simplify the field equations, we adopt for $\lambda $ the value $\lambda =-\left(\Lambda ^2-1\right)/8$. Then, Equation~(\ref{square1}), determining the scalar field, takes the simple form
$\square \phi =\rho$.

The gravitational field equations take the form:
\be
R_{\mu \nu}-\frac{1}{2}g_{\mu \nu }R=\frac{\Lambda +1}{2}T_{\mu \nu }
\ee
with the total energy-momentum tensor given by:
\bea
T_{\mu \nu }=\left[1+\frac{\Lambda -1}{2}\phi \right]\rho U_{\mu} U_{\nu}
 -\frac{\Lambda -1}{2}\left[\nabla _{\mu }\phi \nabla _{\nu }\phi -\frac{1}{2}g_{\mu \nu }\nabla _{\alpha }\phi \nabla ^{\alpha }\phi \right]
\eea

For $\Lambda =1$, we reobtain the general relativistic model for dust. Other possible choices of the function $F(\phi)$, such as $F(\phi )=\exp (\phi )$, can be discussed in a similar way.

A more general model can be obtained by adopting for the matter Lagrangian the general expression~\cite{Harko:2010zi,Fock,Min}:
\be\label{lagr}
L_m=-\left[\rho +\rho \int{\frac{dp(\rho )}{\rho }}-p(\rho )\right]
\ee
where $\rho $ is the rest-mass energy density and $p$ is the thermodynamic pressure, which, by assumption, satisfies a barotropic equation of state, $p=p\left(\rho \right)$. By assuming that the matter Lagrangian does not depend on the derivatives of the metric and that the particle matter fluid current is conserved \mbox{($\nabla _{\nu }\left(\rho u^{\nu }\right)=0$)}, the~Lagrangian given by Equation~(\ref{lagr}) is the unique matter Lagrangian that can be constructed from the thermodynamic parameters of the fluid, and it is valid for all gravitational theories satisfying the two previously mentioned conditions \cite{Min}.

The gravitational field equations and the equation describing the
matter-scalar field coupling are given~by:
\be
R_{\mu \nu }-\frac{1}{2}g_{\mu \nu }R=F(\phi ) ~\epsilon ~U_{\mu }U_{\nu }-p g_{\mu \nu }+\lambda Q_{\mu \nu}
\ee
 \be
\square \phi =\frac{1}{2\lambda }\frac{dF(\phi )}{d\phi }~\epsilon
 \ee
where
$Q_{\mu \nu}=\nabla _{\mu }\phi \nabla _{\nu }\phi -\frac{1}{2}\nabla _{\lambda }\phi \nabla ^{\lambda }\phi g_{\mu \nu }$ and where the total energy density is \mbox{$\epsilon =\rho+\rho \int{dp/\rho }-p$}~\mbox{\cite{Fock,Min}}. With the use of the conservation equation $\nabla _{\nu }\left(\rho U^{\nu }\right)=0$, one~obtains the equation of motion of massive test particles as:
\be\label{eqmot}
\frac{d^2x^{\mu }}{ds^2}+\Gamma _{\alpha \beta }^{\mu }U^{\alpha } U^{\beta }+\left(U^{\mu }U^{\nu }-g^{\mu \nu }\right)\nabla _{\nu }\ln\left[1+\int{\frac{dp}{\rho }} \right]=0
\ee

Models with scalar field-matter coupling were considered in the framework of the Brans--Dicke theory~\cite{NMC}, with the action of the model given by \mbox{$S=\int{\left[\phi R/2+\left(\omega /\phi \right)\nabla _{\mu }\phi \nabla ^{\mu }\phi +F(\phi )L_m\right]\sqrt{-g}d^4x}$}. Such models can give rise to a late-time accelerated expansion of the Universe for very high values of the Brans--Dicke parameter $\omega $. Other~models with interacting scalar field and matter have been considered in \cite{int}. We emphasize that the gravitational theory considered in this work generalizes all of the above models.

In conclusion, the general formalism outlined in this
work can be extremely useful in a variety of scenarios,
such as, in describing the interaction between dark
energy, modeled as a scalar field, and dark matter, or
ordinary matter (neutrinos), with or without pressure,
matter-scalar field interactions in inflation, as well as in
the study of the interactions of the scalar field (representing
dark matter and/or dark energy) and the electromagnetic
component in the very early Universe. Moreover, they can provide a realistic description of the late expansion
of the Universe, where a possible interaction between
ordinary matter and dark energy cannot be excluded \textit{a
priori}.

\section{\boldmath{$f(R,T)$} Gravity}\label{Sec:V}\vspace{-12pt}

\subsection{Action and Gravitational Field Equations}\label{sec5_1}

In this section, we consider another extension of standard General Relativity, namely, $f(R, T )$-modified theories of gravity, where the gravitational Lagrangian is given by an arbitrary function of the Ricci scalar $R$ and of the trace of the energy-momentum tensor $T$ \cite{Harko:2011kv}. The dependence from $T$ may be induced by exotic imperfect fluids or quantum effects (conformal anomaly). The action takes the following form:
\begin{equation}
S=\frac{1}{16\pi}\int
f\left(R,T\right)\sqrt{-g}\;d^{4}x+\int{L_{m}\sqrt{-g}\;d^{4}x}\,
\end{equation}
where $f\left(R,T\right)$ is an arbitrary function of the Ricci
scalar, $R$, and of the trace $T$ of the energy-momentum tensor of
the matter, $T_{\mu \nu}$. For the matter content, we assume that it consists of a fluid that can be characterized by two thermodynamic parameters only, the energy density and the pressure, respectively. With the use of   Equation~(\ref{SETdef}), it follows that the trace of the energy-momentum tensor $T$ can be expressed as a function of the matter Lagrangian as:
\begin{equation}
T=g^{\mu \nu}T_{\mu \nu}=4L_m-2g^{\mu \nu}\frac{\delta L_m}{\delta g^{\mu \nu}}
\end{equation}

Hence, $f(R,T)$ gravity theories can be interpreted as extensions of the $f\left(R,L_m\right)$-type gravity theories, with the gravitational action depending not only on the matter Lagrangian only, but also on its variation with respect to the metric.

By varying the action $S$ of the gravitational field with respect
to the metric tensor components $g^{\mu \nu }$,
we obtain the field equations of the $f\left( R,T\right) $ gravity model as:
\begin{eqnarray}\label{field}
f_{R}\left( R,T\right) R_{\mu \nu } - \frac{1}{2}
f\left( R,T\right) g_{\mu \nu }
+\left( g_{\mu \nu }\square -\nabla_{\mu }\nabla _{\nu }\right)
f_{R}\left( R,T\right)
 \nonumber \\
=8\pi T_{\mu \nu}-f_{T}\left( R,T\right)
T_{\mu \nu }-f_T\left( R,T\right)\Theta _{\mu \nu}\,
\end{eqnarray}

We have defined the variation of $T$ with respect to the metric tensor as:
\begin{equation}
\frac{\delta \left(g^{\alpha \beta }T_{\alpha \beta }\right)}{\delta g^{\mu
\nu}}
=T_{\mu\nu}+\Theta _{\mu \nu}\,
\end{equation}
where:
\begin{equation}
\Theta_{\mu \nu}\equiv g^{\alpha \beta }\frac{\delta T_{\alpha \beta
}}{\delta g^{\mu \nu}}\,
\end{equation}

Note that when $f(R,T)\equiv f(R)$, from Equation~(\ref{field}), we
obtain the field equations of $f(R)$ gravity.

Contracting Equation~(\ref{field}) gives the following relation
between the Ricci scalar $R$ and the trace $T$ of the
energy-momentum tensor:
\begin{eqnarray}
&&f_{R}\left( R,T\right) R+3\square f_{R}\left( R,T\right) -2
f\left( R,T\right) =8\pi T-f_{T}\left(R,T\right)
T-f_{T}\left(R,T\right)\Theta\,
\label{contr}
\end{eqnarray}
where we have denoted $\Theta =\Theta^{\ \mu}_{\mu}$.

By eliminating the term $\square f_{R}\left( R,T\right)$ between
Equations~(\ref{field}) and (\ref{contr}), the gravitational field
equations can be written in the form:
\begin{eqnarray}
f_{R}\left( R,T\right) \left( R_{\mu \nu }-\frac{1}{3}Rg_{\mu \nu
}\right) +\frac{1}{6} f\left( R,T\right) g_{\mu\nu }
&=&8\pi \left(T_{\mu \nu}-\frac{1}{3}T
g_{\mu \nu}\right)-f_{T}\left( R,T\right)
\left( T_{\mu \nu }-\frac{1}{3}T g_{\mu \nu }\right)
\nonumber \\
&&-f_T\left(R,T\right)\left(\Theta_{\mu \nu}-\frac{1}{3}
\Theta g_{\mu\nu}\right)
+ \nabla _{\mu }\nabla _{\nu }f_{R}\left(R,T\right) \nonumber\\
\end{eqnarray}

Taking into account the covariant divergence of Equation~(\ref{field}),
with the use of the following mathematical identity \cite{Ko06}:
\begin{eqnarray}
\nabla ^{\mu }\left[ f_R\left(R,T\right)
R_{\mu\nu}-\frac{1}{2}f\left(R,T\right)g_{\mu\nu}
+\left(g_{\mu \nu }\square -\nabla_{\mu }\nabla_{\nu}\right)
f_R\left(R,T\right)\right] \equiv -\frac{1}{2}g_{\mu \nu}f_T(R,T)\nabla ^{\mu }T\,
\end{eqnarray}
we obtain, for the divergence of the energy-momentum tensor $T_{\mu
\nu}$, the equation:
\begin{equation}\label{noncons}
\nabla ^{\mu }T_{\mu \nu }
=\frac{f_{T}\left( R,T\right) }{8\pi -f_{T}\left(R,T\right) }
\left[ \left( T_{\mu \nu }+\Theta _{\mu \nu }\right) \nabla^{\mu }
\ln f_{T}\left( R,T\right) +\nabla ^{\mu }\Theta _{\mu \nu }-\frac{1}{2}g_{\mu \nu}\nabla ^{\mu }T\right]\,
\end{equation}

Once the matter Lagrangian is known, the calculation of the tensor $\Theta _{\mu
\nu}$ yields (we refer the reader to~\cite{Harko:2011kv} for more details):
\begin{equation}\label{var}
\Theta _{\mu \nu}=-2T_{\mu \nu}+g_{\mu \nu }L_{m}-2g^{\alpha \beta }
\frac{\partial ^2L_{m}}{\partial g^{\mu \nu }\partial g^{\alpha \beta
}}\,
\end{equation}

More specifically, in the case of the electromagnetic field, the matter Lagrangian is
given by:
\begin{equation}
L_{m}=-\frac{1}{16\pi }F_{\alpha \beta }F_{\gamma \sigma }
g^{\alpha \gamma }g^{\beta \sigma }\,
\end{equation}
where $F_{\alpha \beta }$ is the electromagnetic field tensor, we obtain $\Theta _{\mu\nu}=-T_{\mu \nu }$. For a massless scalar field $\phi $ with Lagrangian
$L_{m}=g^{\alpha \beta }\nabla_{\alpha }\phi \nabla _{\beta
}\phi $, we obtain $\Theta _{\mu \nu}=-T_{\mu \nu}+(1/2)Tg_{\mu
\nu}$.

The problem of the perfect fluids, described by an energy
density $\rho $, pressure $p$ and four-velocity $U^{\mu}$, is more
subtle, since there is no unique definition of the matter
Lagrangian. However, in the present study we assume that the
energy-momentum tensor of the matter is given by
$T_{\mu \nu}=\left(\rho +p\right)U_{\mu }U_{\nu}+pg_{\mu \nu}$,
and the matter Lagrangian can be taken as $L_{m}=p$. The
four-velocity $U_{\mu }$ satisfies the conditions
$U_{\mu} U^{\mu}=-1$ and $U^{\mu }\nabla _{\nu } U_{\mu }=0$,
respectively. Then, with the use of Equation~(\ref{var}), we obtain, for
the variation of the energy-momentum of a perfect fluid, the
expression:
\begin{equation}
\Theta _{\mu \nu }=-2T_{\mu \nu }+pg_{\mu \nu }\,
\end{equation}

The choice $L_m=-\rho $ leads to a different form for $\Theta _{\mu \nu }$, $\Theta _{\mu \nu }=-2T_{\mu \nu}+\rho g_{\mu \nu}$. With this form of the matter Lagrangian, the gravitational field equations take a different form. The nature of the qualitative and quantitative differences between the two different versions of the $f(R,T)$ gravity, corresponding to different matter Lagrangians, is still an open question.

\subsection{Specific Cosmological Solution}\label{sec5_3}

In the present Section, we consider a particular class of
$f(R,T)$-modified gravity models, obtained by explicitly
specifying the functional form of $f$. Generally, the field
equations also depend, through the tensor $\Theta _{\mu \nu }$, on
the physical nature of the matter field. Hence, in the case of
$f(R,T)$ gravity, depending on the nature of the matter source,
for each choice of $f$, we can obtain several theoretical models,
corresponding to different matter models \cite{Harko:2012ar}.


As a specific case of a $f(R,T)$ model, consider a correction to the Einstein--Hilbert action given by
$f\left(R,T\right)=R+2f(T)$, where $f(T)$ is an arbitrary function
of the trace of the energy-momentum tensor of matter. The
gravitational field equations immediately follow from
Equation~(\ref{field}), and is given by:
\begin{equation}
R_{\mu\nu}-\frac{1}{2}Rg_{\mu\nu}
=8\pi T_{\mu \nu }-2f'\left(T\right)T_{\mu\nu}-2f'(T)
\Theta _{\mu \nu}+f(T)g_{\mu \nu }\,
\end{equation}
where the prime denotes a derivative with respect to the argument.

Consider a perfect fluid, so that $\Theta _{\mu\nu}=-2T_{\mu\nu}-pg_{\mu \nu}$, and the field equations become:
\begin{equation}
R_{\mu\nu}-\frac{1}{2}Rg_{\mu\nu}=8\pi T_{\mu\nu}
+2f'\left(T\right)T_{\mu\nu}+\left[2pf'(T)+f(T)\right]g_{\mu \nu }\,
\end{equation}

In the case of dust with $p=0$, the gravitational field equations
are given by:
\begin{equation}
R_{\mu\nu}-\frac{1}{2}Rg_{\mu\nu}
=8\pi T_{\mu\nu}+2f'(T)T_{\mu\nu}+f(T)g_{\mu\nu}\,
\label{Ein1}
\end{equation}

These field equations were proposed in \cite{Poplawski:2006ey} to
solve the cosmological constant problem. The~simplest cosmological
model can be obtained by assuming a dust universe ($p=0$, $T=-\rho
$) and by choosing the function $f(T)$, so that $f(T)=\lambda T$,
where $\lambda $ is a constant. By assuming that the metric of the
universe is given by the flat Robertson--Walker metric:
\be
ds^2=-dt^2+a^2(t)\left(dx^2+dy^2+dz^2\right)\,
\ee
the gravitational field equations are given by:
\begin{eqnarray}
3\frac{\dot{a}^2}{a^2}&=&\left(8\pi +3\lambda \right)\rho\, \\
2\frac{\ddot{a}}{a}+\frac{\dot{a}^2}{a^2}&=&\lambda \rho\,
\end{eqnarray}
respectively. Thus, this $f(R,T)$ gravity model is equivalent to a
cosmological model with an effective dark energy density
$\Lambda _\mathrm{eff}\propto H^2$, where $H=\dot{a}/a$ is the
Hubble function \cite{Poplawski:2006ey}. It is also interesting to
note that, generally, for this choice of $f(R,T)$, the gravitational
coupling becomes an effective and time-dependent coupling, of the
form $G_\mathrm{eff}=G\pm 2f'(T)$. Thus, the term $2f(T)$ in the
gravitational action modifies the gravitational interaction
between matter and curvature, replacing $G$ by a running
gravitational coupling parameter.

The field equations reduce to a single equation for $H$:
\be
2\dot{H}+3\frac{8\pi +2\lambda }{8\pi +3\lambda }H^2=0\,
\ee
with the general solution given by:
\be H(t)=\frac{2\left(8\pi
+3\lambda \right)}{3\left(8\pi +2\lambda \right)}\frac{1}{t}\,
\ee

The scale factor evolves according to $a(t)=t^{\alpha }$, with
$\alpha ={2\left(8\pi +3\lambda \right)/3\left(8\pi +2\lambda
\right)}$.

\subsection{Further Applications}\label{sec5_4}

The $f(R,T)$ gravitational model analysed above has been given a great amount of recent attention. We will briefly outline a few cosmological applications. In \cite{Houndjo:2011tu}, the cosmological reconstruction of $f(R, T)$ gravity describing matter-dominated and accelerated phases was analysed. Special attention was paid to the specific case of $f (R, T) = f_1 (R) + f_2 (T)$. The use of an auxiliary scalar field was considered with two known examples for the scale factor corresponding to an expanding Universe. In~the first example, where ordinary matter is usually neglected for obtaining the unification of matter-dominated and accelerated phases with $f(R)$ gravity, it was shown that this unification can be obtained in the presence of ordinary matter. In the second example, as in $f(R)$ gravity, the model of $f(R, T)$ gravity with a transition of the matter-dominated phase to the acceleration phase was obtained.

In \cite{Momeni:2011am}, it was shown that the dust fluid reproduces $\Lambda$CDM, he phantom-non-phantom era and phantom cosmology. Furthermore, different cosmological models were reconstructed, including the Chaplygin gas and the scalar field with some specific forms of $f(R,T)$. The numerical simulation for the Hubble parameter shows good agreement with the  baryon acoustic oscillations (BAO) 
 observational data for low redshifts $z<2$.
In \cite{Houndjo:2011fb}, through a numerical reconstruction, it was shown that specific $f(R,T)$ models are able to reproduce the same expansion history generated, in standard General Relativity, by dark matter and holographic dark energy. It was further shown that these theories are able to reproduce the four known types of future finite-time singularities \cite{Houndjo:2012ij}.
In \cite{Sharif:2012zzd}, a non-equilibrium picture of thermodynamics was discussed at the apparent horizon of an FRW universe in $f(R,T)$ gravity, where the validity of the first and second law of thermodynamics in this scenario were checked. It was shown that the Friedmann equations can be expressed in the form of the first law of thermodynamics and that the second law of thermodynamics holds both in the phantom and non-phantom phases.

The energy conditions have also been extensively explored in $f(R,T)$ gravity, for instance, in an FRW universe with perfect fluid \cite{Sharif:2012gz}, and in the context of exact power-law solutions \cite{Sharif:2012ce}, where certain constraints were found that have to be satisfied to ensure that power law solutions may be stable and match the bounds prescribed by the energy conditions.
In \cite{Alvarenga:2012bt}, it was also shown that the energy conditions were satisfied for specific models. Furthermore, an analysis of the perturbations and stabilities of de Sitter solutions and power-law solutions was performed, and it was shown that for those models in which the energy conditions are satisfied, de Sitter solutions and power-law solutions may be~stable.

Solutions of the G\"{o}del universe in the framework of $f(R,T)$-modified theories of gravity were also obtained \cite{Santos:2013bfa}. Algorithms were derived for constructing five-dimensional Kaluza--Klein cosmological space-times in the presence of a perfect fluid source in $f(R,T)$ gravity \cite{Ram:2013kya}.
Spatially homogeneous and anisotropic Bianchi type-V cosmological models in a scalar-tensor theory of gravitation \cite{Naidu:2013aga} were also explored. The Bianchi type I model with perfect fluid as a matter content in $f(R,T)$ gravity was analysed, and the physical and kinematical properties of the model were also discussed \cite{Sharif:2014cpa}.
In \cite{Chakraborty:2012kj}, the specific case of the conservation of the energy-momentum tensor was considered, and cosmological solutions were obtained for a homogeneous and isotropic model of the Universe.

In \cite{Alvarenga:2013syu}, the evolution of scalar cosmological perturbations in the metric formalism were analysed. According to restrictions on the background evolution, a specific model within these theories was assumed in order to guarantee the standard continuity equation. Using a completely general procedure, the complete set of differential equations for the matter density perturbations was found. In the case of sub-Hubble modes, the density contrast evolution reduces to a second-order equation, and it was shown that for well-motivated $f(R,T)$ Lagrangians, the quasistatic approximation provide different results from the ones derived in the frame of the concordance $\Lambda$CDM model, which constrains severely the viability of such theories.
In \cite{Sharif:2014jpa}, the Ricci and modified Ricci dark energy models were considered in the context of $f(R,T)$ gravity, and it was found that specific models can reproduce the expansion history of the Universe in accordance with the present observational data. The Dolgov--Kawasaki stability condition were also obtained for the specific reconstructed $f(R,T)$ functions.

\newpage

\section{\boldmath{$f\left(R,T,R_{\mu\nu}T^{\mu\nu}\right)$} Gravity}\label{Sec:VI}\vspace{-12pt}

\subsection{ Action and Field Equations}\label{sec6_1}

For the specific case of a traceless energy-momentum tensor, $T=0$, for instance, when the electromagnetic field is involved, the gravitational field equations for the $f(R,T)$ theory \cite{Harko:2011kv} reduce to that of the field equations for $f(R)$ gravity, and all non-minimal couplings of gravity to the matter field vanish. This motivates a further generalization of $f(R,T)$ gravity that consists of including an explicit first order coupling between the matter energy-momentum $T_{\mu \nu}$ and the Ricci tensor \cite{Haghani:2013oma, Odintsov:2013iba}.
In contrast to $f(R,T)$ gravity, for $T=0$, this extra coupling still has a non-minimal coupling to the electromagnetic field via the $R_{\mu\nu}T^{\mu\nu}$ coupling term in the action, which is non-zero in general. As in the previous section, we consider the matter content as consisting of a perfect fluid characterized by its energy density and thermodynamic pressure only.

The action, is given by:
\be\label{eq200}
S=\f{1}{16\pi G}\int d^4x\sqrt{-g}f\left(R,T,R_{\mu\nu}T^{\mu\nu}\right)+\int
d^4x\sqrt{-g}L_m \,
\ee

The only requirement imposed on the function $f\left(R,T,R_{\mu\nu}T^{\mu\nu}\right)$ is that it is an arbitrary analytical function in all arguments.
The gravitational field equations take the following form:
\bea \label{eq203}
(f_R-f_{RT}L_m)G_{\mu\nu}+\left[\Box
f_R+\f{1}{2}Rf_R-\f{1}{2}f+f_TL_m+\f{1}{2}\nabla_\alpha\nabla_\beta\left(f_{RT}T^{
\alpha\beta}\right)\right]g_{\mu\nu}
 \nonumber\\
-\nabla_\mu\nabla_\nu f_R +\f{1}{2}\Box\left(f_{RT}T_{\mu\nu}\right)
 +2f_{RT}R_{\alpha(\mu}T_{\nu)}^{~\alpha}-\nabla_\alpha\nabla_{(\mu}\left[T^\alpha_{
~\nu)}f_{RT}\right]
 \nonumber\\
-\left(f_T+\f{1}{2}f_{RT}R+8\pi
G\right)T_{\mu\nu} -2\left(f_Tg^{\alpha\beta}+f_{RT}R^{\alpha\beta}\right)\f{\partial^2
L_m}{\partial g^{\mu\nu}\partial g^{\alpha\beta}}=0
\eea

The trace of the gravitational field equation, Equation~\eqref{eq203}, is obtained as:
\begin{eqnarray}\label{trace2}
3\Box f_R+\f{1}{2}\Box\left(f_{RT}T\right)+\nabla_\alpha\nabla_\beta\left(f_{RT}T^{\alpha\beta}\right)+Rf_R-Tf_T -\f{1}{2}RTf_{RT}+2R_{\alpha\beta}T^{\alpha\beta}f_{RT}
 \nonumber\\
+Rf_{RT}L_m +4f_TL_m -2f
 -8\pi GT -2g^{\mu\nu}\left(g^{\alpha\beta}f_T+R^{\alpha\beta}f_{RT}\right)\f{\partial^2 L_m}{\partial g^{\mu\nu}\partial g^{\alpha\beta}}=0
\end{eqnarray}

The second derivative of the matter Lagrangian with respect to the metric is non-zero if the matter Lagrangian is of second or higher order in the metric. Thus, for a perfect fluid with $L_m=-\rho$, or a scalar field with
$L_m=-\partial_\mu\phi\partial^\mu\phi /2$,
this term can be dropped. However, for instance, considering the Maxwell field, we have
$L_m=-F_{\mu\nu}F^{\mu\nu}/4$; this term results in $\partial^2 L_m/\partial g^{\mu\nu}\partial
g^{\alpha\beta}=-F_{\mu\alpha}F_{\nu\beta}/2$,
thus giving a non-zero contribution to the field equations.

In analogy with the standard Einstein field equation, one can write the field Equation~\eqref{eq203} as:
\be\label{eq204}
G_{\mu\nu}=8\pi G_{\rm eff}T_{\mu\nu}-\Lambda_{\rm eff}g_{\mu\nu}+T^{\rm eff}_{\mu\nu}
\ee
where we have defined the effective gravitational coupling $G_{\rm eff}$, the effective cosmological constant $\Lambda _{\rm eff}$ and an effective energy-momentum tensor $T^{\rm eff}_{\mu\nu}$ as:
\be\label{eq204-1}
G_{\rm eff}=\f{G+\f{1}{8\pi}\big(f_T+\f{1}{2}f_{RT}R-\f{1}{2}\Box f_{RT}\big)}{f_R-f_{RT}L_m}
\ee
\be
\Lambda_{\tiny{\rm eff}}=\f{2\Box
f_R+Rf_R-f+2f_TL_m+\nabla_\alpha\nabla_\beta(f_{RT}T^{
\alpha\beta})}{2(f_R-f_{RT}L_m)}
\ee
and:
\bl\label{eq204-2}
T^{\rm eff}_{\mu\nu}=\f{1}{f_R-f_{RT}L_m}\Bigg\{\nabla_\mu\nabla_\nu f_R-\nabla_\alpha f_{RT}\nabla^\alpha T_{\mu\nu}
 -\f{1}{2}f_{RT}\Box T_{\mu\nu}-
2f_{RT}R_{\alpha(\mu}T_{\nu)}^{~\alpha}
 \nonumber\\
+\nabla_\alpha\nabla_{(\mu}\left[T^\alpha_{
~\nu)}f_{RT}\right]+2\left(f_Tg^{\alpha\beta}+f_{RT}R^{\alpha\beta}\right)
\f{\partial^2
L_m}{\partial g^{\mu\nu}\partial g^{\alpha\beta}}\Bigg\}
\end{align}
respectively. Note that, in general, $G_{\rm eff}$ and $\Lambda_{\rm eff}$ are not constants and depend on the specific model~considered.

\subsection{ Equation of Motion of the Massive Test Particles in the $f\left(R,T,R_{\mu\nu}T^{\mu\nu}\right)$ Gravity Theory}\label{sec6_2a}

The covariant divergence of the energy-momentum tensor can be obtained by taking the divergence of the gravitational field equation, Equation~\eqref{eq203}, which takes the following form:
\begin{align}\label{eq401}
\nabla^\mu T_{\mu\nu}=\f{2}{\left(1+Rf_{TR}+2f_T\right)}\Bigg\{
\nabla_\mu\left(f_{RT}R^{\sigma\mu}T_{\sigma\nu}
\right)
 +\nabla_\nu\left(L_mf_T\right)-\f{1}{2}\bigg(f_{RT}R_{\rho\sigma}+f_T
g_{\rho\sigma}\bigg)
\nabla_\nu T^{\rho\sigma}
 \nonumber\\
 -G_{\mu\nu}\nabla^\mu \left(f_{RT}L_m\right)-
\f{1}{2}\left[\nabla^\mu\left(Rf_{RT}\right)+2\nabla^\mu f_T\right]T_{\mu\nu}\Bigg\}
\end{align}
where we have assumed that $\partial^2
L_m/\partial g^{\mu\nu}\partial g^{\alpha\beta}=0$ and have used the mathematical identities:
\bea
\nabla^\mu \left(f_R R_{\mu\nu}+\Box f_R
g_{\mu\nu}-\f{1}{2}fg_{\mu\nu}-\nabla_\mu\nabla_\nu
f_R\right)
=-\f{1}{2}\bigg[f_T \nabla_\nu
T+f_{RT}\nabla_\nu\left(R_{\rho\sigma}T^{\rho\sigma}\right)\bigg]
\eea
\be
2T_{\mu\tau;\delta[;\rho;\sigma]}=T_{\mu\tau;\alpha}R^\alpha_{~\delta\rho\sigma}
+T_{\alpha\tau;\delta}R^\alpha_{~\mu\rho\sigma}+T_{\mu\alpha;\delta}R^\alpha_{
~\tau\rho\sigma}
\ee
and $\left[\Box,\nabla_\nu\right]T=R_{\mu\nu}\nabla^\mu T$, respectively.

In order to find the equation of motion for a massive test particle, as in the previous sections, we~consider the energy-momentum tensor of a perfect fluid. Following the procedure outlined above, we obtain the equation of motion for a massive test particle, considering the matter Lagrangian $L_m=p$, as:
\be\label{eq404}
\f{d^2x^\lambda}{ds^2}+\Gamma^\lambda_{~\mu\nu} U^\mu U^\nu=f^\lambda
\ee
where the extra force acting on the test particles is given by:
\bea\label{eq405}
f^\lambda&=&\f{1}{\rho+p}\Big[\left(f_T+Rf_{RT}\right)\nabla_\nu\rho-
\left(1+3f_T\right)\nabla_\nu p
-(\rho+p)f_{RT}R^{\sigma\rho}\left(\nabla_\nu
h_{\sigma\rho}-2\nabla_\rho h_{\sigma\nu}\right)\nonumber\\
&& - f_{RT}
R_{\sigma\rho}h^{\sigma\rho}\nabla_\nu\left(\rho+p\right)\Big]
\f{h^{\lambda\nu}}{1+2f_T+Rf_{RT}}
\eea

Contrary to the nonminimal coupling presented in \cite{Bertolami:2007gv} and as can be seen from the above equations, the extra force does not vanish, even with the Lagrangian $L_m=p$.

The extra-force is perpendicular to the four-velocity, satisfying the relation $f^{\mu} U_{\mu}=0$. In the absence of any coupling between matter and geometry, with $f_T=f_{RT}=0$, the extra-force takes the usual form of the standard general relativistic fluid motion, \textit{i.e.}, $f^{\lambda }=-h^{\lambda \nu}\nabla _{\nu }p/\left(\rho +p\right)$. In the case of $f\left(R,T,R_{\mu\nu}T^{\mu\nu}\right)$ gravity, there is an explicit dependence of the extra-force on the Ricci tensor $R_{\sigma \rho }$, which makes the deviation from the geodesic motion more important for regions with strong curvatures.

\subsection{Cosmological Applications: Specific Case of $f=R+\alpha R_{\mu\nu}T^{\mu\nu}$}\label{sec6_3}

Let us now consider some examples of cosmological solutions of the theory. In order to obtain explicit results and as the first step, one has to fix the functional form of the function $f\left(R,T,R_{\mu\nu}T^{\mu\nu}\right)$. We analyze the evolution and dynamics of the Universe, assuming that the Universe is isotropic and homogeneous, with the matter content described by the energy density $\rho $ and thermodynamic pressure $p$ with the matter Lagrangian chosen as $L_m=-\rho$. Consider the Friedmann--Lemaitre--Robertson--Walker (FLRW) metric, the Hubble parameter $H=\dot{a}/a$ and the deceleration parameter $q$, defined as:
\be
q=\frac{d}{dt}\frac{1}{H}-1
\ee

Note that if $q<0$, the expansion of the Universe is accelerating, while positive values of $q> 0$ describe decelerating evolutions.

For simplicity, consider the case in which the interaction between matter and geometry takes place only via the coupling between the energy-momentum and Ricci tensors, \textit{i.e.},
\begin{equation}\label{113}
f=R+\alpha R_{\mu\nu}T^{\mu\nu} \,
\end{equation}

In order to pass the Solar System and the other astrophysical tests, the correction term in Equation~(\ref{113}) must be small, which implies that $\alpha $ is a small parameter. This simple case serves as an example to show the main differences of the present theory with $f(R,T)$ gravity, considered before~\cite{Harko:2011kv}.
The gravitational field equations for this form of $f$ are given by:
\begin{eqnarray}\label{eq203-1}
G_{\mu\nu}+\alpha\Bigg[2R_{\sigma(\mu}T^\sigma_{~\nu)}-
\f{1}{2}R_{
\rho\sigma}T^{\rho\sigma}g_{\mu\nu}-\f{1}{2}RT_{\mu\nu}
-\f{1}{2}\left(2 \nabla_\sigma\nabla_{(\nu}
T^\sigma_{~\mu)}-\Box
T_{\mu\nu}-\nabla_\alpha\nabla_\beta T^{\alpha\beta}g_{\mu\nu}\right)
 \nonumber\\
-G_{\mu\nu}L_m-2R^{\alpha\beta}\f{\partial^2 L_m}{\partial g^{\mu\nu}\partial
g^{\alpha\beta}}\Bigg]-
8\pi GT_{\mu\nu}=0
\end{eqnarray}

For the case of the FLRW metric, the modified Friedmann equations are:
\be\label{c1}
3H^2=\frac{\kappa}{1-\alpha \rho }\rho +\frac{3}{2}\frac{\alpha }{1-\alpha \rho }H\left(\dot{\rho }-\dot{p}\right)
\ee
and:
\be\label{c2}
2\dot{H}+3H^2=\frac{2\alpha }{1+\alpha p}H\dot{\rho }-\frac{\kappa p}{1+\alpha p}+\frac{1}{2}\frac{\alpha }{1+\alpha p}\left(\ddot{\rho }-\ddot{p}\right)
\ee
respectively, where we have denoted $\kappa =8\pi G$ for simplicity. When $\alpha =0$, we recover the standard Friedmann equations. To remove the under determinacy of the field equations, we impose an equation of state for the cosmological matter, $p=p(\rho)$. A standard form of the cosmological matter equation of state is $p=\omega \rho $, where $\omega ={\rm constant}$ and $0\leq \omega \leq 1$.


\subsubsection{ High Cosmological Density Limit of the Field Equations}

First, we consider the high energy density limit of the system of modified cosmological equations Equations~(\ref{c1}) and (\ref{c2}) and assume that the constant $\alpha $ is small, so that $\alpha \rho \ll 1$ and $\alpha p \ll 1$, respectively. In the high-energy limit, $\rho =p$, then Equations~(\ref{c1}) and (\ref{c2}) take the approximate form:
\bea
3H^2&=&\kappa \rho
 \\
2\dot{H}+3H^2&=&-\kappa \rho +2\alpha H\dot{\rho }
\eea

The time evolution of the Hubble parameter is described by the equation:
\be
\left(1-\frac{6\alpha}{\kappa }\right)H^2+3H^2=0
\ee
and, hence, for this model, the evolution of the Hubble parameter is given by:
\be
H(t)= \frac{\sqrt{\left(C_1+3 \kappa t\right){}^2-24 \alpha \kappa }+C_1+3 \kappa t}{12
 \alpha }
\ee
where $C_1>0$ is an arbitrary integration constant. The scale factor of the Universe is given by:
\be
a(t)=C_2\frac{\exp \left[\frac{\left(C_1+3 \kappa t\right) \sqrt{\left(C_1+3 \kappa
 t\right){}^2-24 \alpha \kappa }+9 \kappa t^2+6 \kappa C_1 t}{72 \alpha
 \kappa }\right]}{\sqrt[3]{\sqrt{\left(C_1+3 \kappa t\right){}^2-24 \alpha \kappa
 }+C_1+3 \kappa t}}
 \ee
where $C_2>0$ is an arbitrary integration constant. In order to have a physical solution, the integration constant $C_1$ must satisfy the constraint $C_1\geq \sqrt{24 \kappa \alpha}$.

The values of the integration constant can be determined from the condition $H(0)=H_0$, and \mbox{$a(0)=a_0$}, where $H_0$ and $a_0$ are the initial values of the Hubble parameter and of the scale factor of the Universe, respectively. This condition immediately provides for $C_1$ the following value
\mbox{$C_1=(6 \alpha H_0^2+\kappa)/H_0$}.
 For the integration constant $C_2$, we obtain:
 \bea
 C_2=a_0\sqrt[3]{\sqrt{\frac{\left(\kappa -6 \alpha H_0^2\right)^2}{H_0^2}}+6 \alpha
 H_0+\frac{\kappa }{H_0}} \;\;
 \exp \left[-\frac{\sqrt{\frac{\left(\kappa -6 \alpha
 H_0^2\right)^2}{H_0^2}} \left(6 \alpha H_0^2+\kappa \right)}{72 \alpha H_0 \kappa
 }\right]
 \eea

In the small time limit, the scale factor is approximated by:
\be
a(t)\approx a_0\left(1+\frac{\kappa }{6H_0\alpha}t\right)
\ee

The deceleration parameter is obtained as:
\be
q(t)=-\frac{36 \alpha H_0 \kappa }{\sqrt{\frac{\left(6 \alpha H_0^2+3 H_0 \kappa t+\kappa
 \right)^2}{H_0^2}-24 \alpha \kappa } \left[6 \alpha H_0^2+H_0 \sqrt{\frac{\left(6
 \alpha H_0^2+3 H_0 \kappa t+\kappa \right)^2}{H_0^2}-24 \alpha \kappa }+3 H_0 \kappa
 t+\kappa \right]}-1
 \ee
 and it can be represented in a form of a power series as:
 \be
 q(t)\approx -1-\frac{18 \alpha H_0^2
 }{\kappa -6
 \alpha H_0^2}+
 \frac{6 H_0 \kappa ^2 }{\left(\kappa -6 \alpha H_0^2\right)^3}t
 \ee

For small values of time, if $24\alpha H_0^2 \ll \kappa$, $q\approx -1$, and the Universe starts its expansion from a de Sitter-like phase, entering, after a finite time interval, into a decelerating phase. On the other hand, if~$\kappa >6 \alpha H_0^2$, $q<-1$, and the non-singular Universe experiences an initial super-accelerating phase.

\subsubsection{The Case of Dust}

Next, we consider the case of low density cosmological matter, with $p=0$. Moreover, we assume again that the condition $\alpha \rho \ll 1$ holds. Then, the gravitational field equations, Equations~(\ref{c1}) and~(\ref{c2}), corresponding to a FLRW universe, take the approximate form:
 \bea
 3H^2&=&\kappa \rho +\frac{3}{2}\alpha H\dot{\rho } \label{c3}
 \\
 2\dot{H}+3H^2&=&2\alpha H\dot{\rho }+\frac{1}{2}\alpha \ddot{\rho } \label{c4}
 \eea

First, we consider the matter dominated phase of the model, in which the non-accelerating expansion of the Universe can be described by a power law form of the scale factor, so that $a=t^m$, $m={\rm constant}$ and $H=m/t$, respectively. The deceleration parameter is given by $q=1/m-1$. Therefore, Equation~(\ref{c3}) gives, for the time evolution of the density, the equation:
 \be
 \frac{3\alpha m}{2t}\dot{\rho }+\kappa \rho -3\frac{m^2}{t^2}=0
 \ee
 with the general solution given by:
 \be
 \rho (t)=\frac{e^{-\frac{\kappa t^2}{3 \alpha }} \left[3 \rho _0 \alpha e^{\frac{\kappa
 t_0^2}{3 \alpha }}+\text{Ei}\left(\frac{t^2 \kappa }{3 \alpha
 }\right)-\text{Ei}\left(\frac{t_0^2 \kappa }{3 \alpha }\right)\right]}{3
 \alpha }
 \ee
where $\text{Ei}(z)=-\int_{-z}^{\infty}{e^{-t}dt/t}$ is the exponential integral function, and we have used the initial condition $\rho \left(t_0\right)=\rho _0$. By substituting the expressions of the density and of the Hubble parameter into Equation~(\ref{c4}), in the first order of approximation, we obtain the following constraint on $m$,
$ (9 m^2-10 m+1)/3t^2+O\left(t^2\right)\approx 0$,
which is (approximately) satisfied if $m$ is given by the algebraic equation $9m^2-10m+1=0$, having the solutions $m_1=1$ and $m_2=1/9$, respectively. The~deceleration parameters corresponding to these solutions are $q_1=0$ and $q_2=8$, respectively. Since a value of the deceleration parameter of the order of $q=8$ seems to be ruled out by the observations, the physical solution has a scale factor $a=t$, and $q=0$. The cosmological solutions with zero value of the deceleration parameter are called marginally accelerating, and they describe the pre-accelerating phase of the cosmic expansion.

Now, we look for a de Sitter-type solution of the field equations for the pressureless matter, Equations~(\ref{c3}) and (\ref{c4}), by taking $H=H_0={\rm constant}$. Then, it follows that, in order to have an accelerated expansion, the matter density must satisfy the equation:
 \be
 \ddot{\rho }-H_0\dot{\rho }+\frac{2\kappa }{\alpha }\rho =0
 \ee
with the general solution given by:
 \bea\label{r1}
 \rho (t)&=&e^{\frac{1}{2} H_0 \left(t-t_0\right)} \Bigg\{\frac{\sqrt{\alpha } \left(2 \rho
 _{01}-H_0 \rho _0\right) }{\sqrt{\alpha H_0^2-8
 \kappa }}\sinh \left[\frac{ \sqrt{\alpha
 H_0^2-8 \kappa }}{2 \sqrt{\alpha }}\left(t-t_0\right)\right]
 \nonumber\\
 &&+\cosh \left[\frac{ \sqrt{\alpha H_0^2-8 \kappa }}{2
 \sqrt{\alpha }}\left(t-t_0\right)\right]\Bigg\}
 \eea
where we have used the initial conditions $\rho \left(t_0\right)=\rho _0$ and $\dot{\rho }\left(t_0\right)=\rho _{01}$, respectively.
 Therefore, in the presence of a non-trivial curvature-matter coupling, once the evolution of the matter density is given by Equation~(\ref{r1}), the time evolution of the Universe is of the de Sitter-type.

\section{Dark Matter As a Curvature-Matter Coupling Effect}\label{dark}

According to
Newton's gravitation theory, the rotation of hydrogen clouds around galaxies should show a Keplerian decrease with distance $r$ of
the orbital rotational speed $v_{tg}$ outside the luminous baryonic matter, $%
v_{tg}^2\propto M(r)/r$, where $M(r)$ is the dynamical mass of the galaxy. However, observations show
that the rotation curves are rather flat \cite{dm1, dm2,dm3,BT08}.
The rotational velocities increase near the center of
the galaxy, and then, they remain nearly constant at a value of $v_{tg\infty }\sim$
200--300 km/s. Hence, the observations give a general mass profile $M(r)\approx
rv_{tg\infty }^2/G$ \cite{dm1,dm2,dm3, BT08}. Consequently, even at large
distances, where very little luminous matter can be detected, the mass within a distance $%
r$ from the center of the galaxy increases linearly with the radial distance $r$.

This unusual behavior of the galactic rotation curves is usually explained by postulating
the existence of some dark (invisible) matter, distributed in a spherical
halo around the galaxies. The rotation curves are obtained
by measuring the frequency shifts $z$ of the 21-cm radiation emission from
the neutral hydrogen gas clouds. Usually, astronomers report the
resulting $z$ in terms of a velocity field $v_{tg}$ \cite{dm1,dm2,dm3}. In the standard cosmological model, dark matter is assumed to be a cold,
pressureless medium. Many possible candidates for dark matter have been proposed, the
most popular ones being the weakly interacting massive particles (WIMP) (for
a review of the particle physics aspects of dark matter, see %
\cite{OvWe04}). The interaction cross-section with normal baryonic
matter is expected to be small, but~non-zero, and we expect
that dark matter particles can be detected directly. However, after more than 20 years of intense experimental and observational effort, presently, no convincing non-gravitational evidence for dark matter exists.

In the present section, we will briefly analyze, following \cite{Harko:2010vs}, the possibility that the dynamic behavior of hydrogen clouds rotating around the galactic center can be explained due to the presence of a curvature-matter coupling. We will restrict our analysis to the case of modified gravity models with linear curvature-matter coupling. The extra-terms in the gravitational field equations modify the equations of motion of test particles and induce a supplementary gravitational interaction, which can account for the observed behavior of the galactic rotation curves.

Note that the modified gravity approach has been explored, as an alternative to dark matter, in~different contexts \cite{Boehmer:2007kx,Bohmer:2007fh,Capozziello:2012qt, Capozziello:2013yha}, such as in $f(R)$ gravity and the recently proposed hybrid metric-Palatini gravity~\cite{Harko:2011nh,Capozziello:2012ny, Capozziello:2013uya, Capozziello:2013gza}.

\subsection{Stable Circular Orbits and Frequency Shifts in Modified Gravity with Linear Curvature-Matter~Coupling}

In the following, we assume that gas clouds behave like massive test particles, moving in a static and spherically-symmetric space-time outside the galactic baryonic mass distribution. In the galactic space-time, we consider two observers $O_{E}$ and $O_{\infty }$, with four-velocities $u_{E}^{\mu }$ and $u_{\infty }^{\mu}$, respectively. Observer $O_{E}$ is the light emitter (\textit{i.e.}, the gas clouds placed at a point $P_{E}$ of the space-time), and $O_{\infty } $ represents the detector at point $P_{\infty }$, located far from the emitter, at ``spatial infinity'' \cite{Nuc01}.

Without any loss of generality, we can assume that the gas clouds move in the
galactic plane $\theta =\pi /2$, so that $u_{E}^{\mu }=\left( \dot{t},\dot{r}%
,0,\dot{\phi}\right) _{E}$, where the dot stands for derivation with respect
to the affine parameter $s$. On the other hand, we suppose that the detector
is static (\textit{i.e.}, $O_{\infty }$'s four-velocity is tangent to the static
Killing field $\partial /\partial t$), and in the chosen coordinate system,
its four-velocity is $u_{\infty }^{\mu }=\left( \dot{t},0,0,0\right)
_{\infty }$ \citep{Nuc01}. In~this section, we use the $(+,-,-,-)$ signature.

The static spherically symmetric metric outside the galactic baryonic mass
distribution is given by:
\begin{equation}
ds^{2}=e^{\nu \left( r\right) }dt^{2}-e^{\lambda \left( r\right)
}dr^{2}-r^{2}\left( d\theta ^{2}+\sin ^{2}\theta d\phi ^{2}\right)
\label{line}
\end{equation}
where the metric coefficients $\nu $ and $\lambda $ are functions of the radial coordinate $r$
only. The motion of a test particle in the gravitational field in modified
gravity with a linear curvature-matter coupling is described by
the Lagrangian given by Equation~(\ref{actpart}):
\begin{equation}
L=Q\left[ e^{\nu \left( r\right) }\left( \frac{dt}{ds}\right)
^{2}-e^{\lambda \left( r\right) }\left( \frac{dr}{ds}\right)
^{2}-r^{2}\left( \frac{d\Omega }{ds}\right) ^{2}\right]
\end{equation}
where $d\Omega ^{2}=d\theta ^{2}+\sin ^{2}\theta d\phi ^{2}$, and the function $Q$ is given by:
\begin{equation}\label{Q1}
Q=\left[1+\zeta f_2(R)\right]\frac{dL_m(\rho )}{d\rho }
\end{equation}
where $\zeta $ is a constant.

For $\theta
=\pi /2$, $d\Omega ^{2}=d\phi ^{2}$. From the Lagrange equations, it follows
that we have two constants of motion, the energy $E$:
\begin{equation}
E=Qe^{\nu (r)}\dot{t}
\end{equation}
and the angular momentum $l$, given by:
\begin{equation}
l=Qr^{2}\dot{\phi}
\end{equation}
where a dot denotes the derivative with respect to the affine parameter $s$. The condition of the normalization of the four-velocity $U^{\mu }U_{\mu }=1$ gives:
\begin{equation}
1=e^{\nu \left( r\right) }\dot{t}%
^{2}-e^{\lambda (r)}\dot{r}^{2}-r^{2}\dot{\phi}^{2}
\end{equation}
from which, with the
use of the constants of motion, we obtain:
\begin{equation}
E^{2}=Q^{2}e^{\nu +\lambda }\dot{r}^{2}+e^{\nu }\left( \frac{l^{2}}{r^{2}}%
+Q^{2}\right) \label{energy}
\end{equation}

Equation (\ref{energy}) shows that the radial motion of the particles in modified gravity with linear matter-geometry coupling is the same as
that of a particle in ordinary Newtonian mechanics, with velocity $\dot{r}$,
position-dependent mass $m_{eff}=2Q^{2}e^{\nu +\lambda }$ and energy $E^{2}$%
, respectively, moving in the effective potential:
\begin{equation}
V_{eff}\left( r\right) =e^{\nu (r)}\left( \frac{l^{2}}{r^{2}}+Q^{2}\right)
\end{equation}

The conditions for stable circular orbits $\partial V_{eff}/\partial r=0$ and $\dot{%
r}=0$ determine the energy and the angular momentum of the particle as:
\begin{equation}\label{cons1c}
l^{2}=\frac{1}{2}\frac{r^{3}Q\left( \nu ^{\prime }Q+2Q^{\prime }\right) }{%
1-r\nu ^{\prime }/2}
\end{equation}
and:
\begin{equation}\label{cons2}
E^{2}=\frac{e^{\nu }Q\left( rQ^{\prime }+Q\right) }{1-r\nu ^{\prime }/2}
\end{equation}
respectively.

The line element given by Equation~(\ref{line}) can be rewritten in terms of the
spatial components of the velocity, normalized with the speed of light,
measured by an inertial observer far from the source, as:
\begin{equation}
ds^{2}=dt^{2}\left( 1-v^{2}\right)
\end{equation}
where:
\begin{equation}
v^{2}=e^{-\nu }\left[ e^{\lambda }\left( \frac{dr}{dt}\right)
^{2}+r^{2}\left( \frac{d\Omega }{dt}\right) ^{2}\right]
\end{equation}

For a stable circular orbit $dr/dt=0$, and the tangential velocity of the
test particle is obtained as:
\begin{equation}
v_{tg}^{2}=e^{-\nu }r^{2}\left( \frac{d\Omega }{dt}\right) ^{2}
\end{equation}
In terms of the conserved quantities $E$ and $l$, the angular velocity is given, for $%
\theta =\pi /2$, by:
\begin{equation}
v_{tg}^{2}=\frac{e^{\nu }}{r^{2}}\frac{l^{2}}{E^{2}}
\end{equation}
With the use of Equations~(\ref{cons1c}) and (\ref{cons2}), we obtain:
\begin{equation}
v_{tg}^{2}=\frac{1}{2}\frac{r\left( \nu ^{\prime }Q+2Q^{\prime }\right) }{%
rQ^{\prime }+Q} \label{vtg}
\end{equation}

Thus, the rotational velocity of the test body in modified gravity with
linear coupling between matter and geometry is determined by the metric
coefficient $\exp \left( \nu \right) $, and by the function $Q$ and its
derivative with respect to the radial coordinate $r$. In the standard general relativistic limit $\zeta =0$, $Q=1$, and we obtain $v_{tg}^2=r\nu '/2$.

\subsection{The Effect of the Curvature-Matter Coupling on the Light Shifts}

The velocities of the hydrogen clouds rotating around galactic centers are determined from the red and blue
shifts of the radiation emitted by the gas moving on circular orbits on
both sides of the central region. The radiation travels on null geodesics
with tangent $k^{\mu }$. We may restrict without any loss of generality $k^{\mu }$ to lie in the equatorial
plane $\theta =\pi /2$ and evaluate the frequency shift for a light signal
emitted from $O_{E}$, in motion, in a circular orbit, and detected by $O_{\infty }$ at infinity. The
frequency shift associated with the emission and detection of the light signal
is given by:
\begin{equation}
z=1-\frac{\omega _{E}}{\omega _{\infty }}
\end{equation}
where $\omega _{I}=-k_{\mu }u_{I}^{\mu }$, and the index $I$ refers to
emission ($I=E$) or detection ($I=\infty $) at the corresponding space-time
point \cite{Nuc01,La03}. Two frequency shifts, corresponding to the frequency shifts of a receding or approaching gas cloud, respectively, are associated with light propagation in the same and
opposite direction of motion of the emitter, respectively. In
terms of the tetrads $e_{(0)}=e^{-\nu /2}\partial /\partial t$, $%
e_{(1)}=e^{-\lambda /2}\partial /\partial r$, $e_{(2)}=r^{-1}\partial
/\partial \theta $, $e_{(3)}=\left( r\sin \theta \right) ^{-1}\partial
/\partial \phi $, the frequency shifts take the form \cite{Nuc01}:
\begin{equation}
z_{\pm }=1-e^{\left[ \nu _{\infty }-\nu \left( r\right) \right] /2}\left(
1\mp v\right) \Gamma
\end{equation}
where $v=\left[ \sum_{i=1}^{3}\left( u_{(i)}/u_{(0)}\right) ^{2}\right]
^{1/2}$, with $u_{(i)}$ the components of the particle's four velocity along
the tetrad (\textit{i.e.}, the velocity measured by an Eulerian observer, whose world-line is tangent to the static Killing field). $\Gamma =\left( 1-v^{2}\right)
^{-1/2}$ is the usual Lorentz factor, and $\exp \left( \nu _{\infty }\right)
$ is the value of $\exp \left[ \nu \left( \left( r\right) \right) \right] $
for $r\rightarrow \infty $. In the case of circular orbits in the $\theta
=\pi /2$ plane, we obtain:
\begin{equation}
z_{\pm }=1-e^{\left[ \nu _{\infty }-\nu \left( r\right) \right] /2}\frac{%
1\mp \sqrt{r\left( \nu ^{\prime }Q+2Q^{\prime }\right) /2\left( rQ^{\prime
}+Q\right) }}{\sqrt{1-r\left( \nu ^{\prime }Q+2Q^{\prime }\right) /2\left(
rQ^{\prime }+Q\right) }}
\end{equation}

It is convenient to define two other quantities, $z_{D}=\left(
z_{+}-z_{-}\right) /2$ and $z_{A}=\left( z_{+}+z_{-}\right) /2$, respectively \cite{Nuc01}. In the modified gravity model with a linear curvature-matter coupling, the redshift factors are given by:
\begin{equation}
z_{D}\left( r\right) =e^{\left[ \nu _{\infty }-\nu \left( r\right) \right]
/2}\frac{\sqrt{r\left( \nu ^{\prime }Q+2Q^{\prime }\right) /2\left(
rQ^{\prime }+Q\right) }}{\sqrt{1-r\left( \nu ^{\prime }Q+2Q^{\prime }\right)
/2\left( rQ^{\prime }+Q\right) }}
\end{equation}
and:
\begin{equation}
z_{A}\left( r\right) =1-\frac{e^{\left[ \nu _{\infty }-\nu \left( r\right) %
\right] /2}}{\sqrt{1-r\left( \nu ^{\prime }Q+2Q^{\prime }\right) /2\left(
rQ^{\prime }+Q\right) }}
\end{equation}
respectively, which can be easily related to the galactic observations \citep{Nuc01}.
$z_{A}$ and $z_{D}$ satisfy the relation:
\begin{equation}
\left( z_{A}-1\right)
^{2}-z_{D}^{2}=\exp \left[ 2\left( \nu _{\infty }-\nu \left( r\right)
\right) \right]
\end{equation}
and thus, in principle, by assuming that the metric tensor
component $\exp \left[ \nu \left( \left( r\right) \right) \right] $ is
known, $Q$ and $Q^{\prime }$ can be obtained directly from the astrophysical
observations. Hence, the observations of the red- and blue-shifts of the radiation emitted by hydrogen clouds rotating around the galactic center could provide a direct observational test of the galactic
geometry and, implicitly, of the modified gravity models with linear
coupling between matter and geometry.

\subsection{Galactic Rotation Curves and the Curvature-Matter Coupling}

The tangential velocities $v_{tg}$ of gas clouds, considered as massive test particles moving around the centers of the galaxies, cannot be measured directly. Instead, they can
be inferred from the redshift $z_{\infty }$ of the radiation observed at spatial infinity,
for which:
\begin{equation}
1+z_{\infty }=\exp \left[ \left( \nu _{\infty }-\nu \right) /2%
\right] \left( 1\pm v_{tg}\right) /\sqrt{1-v_{tg}^{2}}
\end{equation}
Since the velocities of the gas clouds are
non-relativistic, with $v_{tg}\leq \left(
4/3\right) \times 10^{-3}$, we observe $v_{tg}\approx z_{\infty }$ (as the
first term of a geometric series), with the consequence that the lapse
function $\exp \left( \nu \right) $ necessarily tends at infinity to unity,
\textit{i.e.}, $e^{\nu }\approx e^{\nu _{\infty }}/\left( 1-v_{tg}^{2}\right)
\approx e^{\nu _{\infty }}\rightarrow 1$. The observations show that at
distances large enough from the galactic center, $v_{tg}\approx $ constant.

In the following, we use this observational constraint to reconstruct the
coupling term between curvature and matter in the ``dark matter''-dominated
region, far away from the baryonic matter distribution. By assuming that $%
v_{tg}=$ constant, Equation~(\ref{vtg}) can be written as:
\begin{equation}
v_{tg}^{2}\frac{1}{rQ}\frac{d}{dr}\left( rQ\right) =\frac{\nu ^{\prime }}{2}+%
\frac{Q^{\prime }}{Q} \label{Q0}
\end{equation}
and can be immediately integrated to give:
\begin{equation}
Q(r)=\left( \frac{r}{r_{0}}\right) ^{v_{tg}^{2}/\left( 1-v_{tg}^{2}\right)
}\exp \left[ -\frac{\nu }{2\left( 1-v_{tg}^{2}\right) }\right] \label{Q2}
\end{equation}
where $r_{0}$ is an arbitrary constant of integration. Since
hydrogen clouds are a pressureless dust ($p=0$) that can be characterized by
their density $\rho $ only, the Lagrangian of the matter (gas cloud) is
given by $L_{mat}\left( \rho \right) =\rho $. Therefore, from Equations~(\ref{Q1})
and (\ref{Q2}), we obtain:
\begin{eqnarray}\label{coup}
\zeta f_{2}\left( R\right) &=&\left( \frac{r}{r_{0}}\right)
^{v_{tg}^{2}/2\left( 1-v_{tg}^{2}\right) }\exp \left[ -\frac{\nu }{4\left(
1-v_{tg}^{2}\right) }\right] -1\approx
\frac{v_{tg}^{2}}{2\left( 1-v_{tg}^{2}\right) }\ln \frac{r}{r_{0}}-\frac{%
\nu }{4\left( 1-v_{tg}^{2}\right) }-\nonumber\\
&&\frac{v_{tg}^{2}}{8\left(
1-2v_{tg}^{2}\right) }\nu \ln \frac{r}{r_{0}}
\end{eqnarray}

For an arbitrary velocity profile $v_{tg}=v_{tg}(r)$, the general solution of
Equation~(\ref{Q0}) is given by:
\begin{equation}
\sqrt{Q(r)}=1+\zeta f_{2}(R)=\sqrt{Q_{0}}\exp \left[ \frac{1}{2}\int \frac{%
v_{tg}^{2}(r)/r-\nu ^{\prime }/2}{1-v_{tg}^{2}(r)}dr\right] \label{fin}
\end{equation}
where $Q_{0}$ is an arbitrary constant of integration.

For $v_{tg}^{2}$, we assume the simple empirical dark halo
rotational velocity law \cite{per}:
\begin{equation}
v_{tg}^{2}=\frac{v_{0}^{2}x^{2}}{a^{2}+x^{2}}
\end{equation}
where $x=r/r_{opt}$, $r_{opt}$ is the optical radius containing 83$\%$ of
the galactic luminosity. The parameters $a$, the ratio of the halo core
radius and $r_{opt}$ and the terminal velocity $v_{0}$ are all functions of the
galactic luminosity $L$. For spiral galaxies $a=1.5\left( L/L_{\ast }\right)
^{1/5}$ and $v_{0}^{2}=v_{opt}^{2}\left( 1-\beta _{\ast }\right) \left(
1+a^{2}\right) $, where \mbox{$v_{opt}=v_{tg}\left( r_{opt}\right) $}, and $\beta _{\ast
}=0.72+0.44\log _{10}\left( L/L_{\ast }\right) $, with $L_{\ast
}=10^{10.4}L_{\odot}$.

One can assume that the coupling between the neutral hydrogen clouds and the geometry is small, $\zeta f_{2}(R)L_{mat}\ll 1$, and consequently, the galactic geometry is not significantly modified in the vacuum outside the baryonic mass distribution with mass $M_{B}$, corresponding to $L_{mat}\approx 0$. Moreover, we assume, for simplicity, that outside the baryonic matter distribution, the galactic metric is given by the Schwarzschild metric (which is also a solution of the vacuum field equations of the $f(R)$-modified gravity \cite{pun} and still gives the dominant contribution to the total metric, even if the curvature-matter coupling is not small), written as:
\begin{equation}
e^{\nu }=e^{-\lambda }=1-\frac{2R_{0}}{x}
\end{equation}
where $R_{0}=GM_{B}/r_{opt}$, from Equation~(\ref{fin}), we obtain:
\begin{eqnarray}
1+\zeta f_{2}(R)=\exp \left[ \alpha \times{\rm arctanh}\left( \frac{\sqrt{1-v_{0}^{2}%
}}{a}x\right) \right] \times
\left( 1-\frac{2R_{0}}{x}\right) ^{\beta }\frac{\left[
\left( 1-v_{0}^{2}\right) x^{2}+a^{2}\right] ^{\gamma }}{Q_0^{-1/2}x^{1/4}}
\end{eqnarray}
where:
\begin{equation}
\alpha =-\frac{aR_{0}v_{0}^{2}}{2\sqrt{1-v_{0}^{2}}\left[ a^{2}+4\left(
1-v_{0}^{2}\right) R_{0}^{2}\right] }
\end{equation}
\begin{equation}
\beta =\frac{a^{2}-4R_{0}^{2}}{4\left[ a^{2}+4\left( 1-v_{0}^{2}\right)
R_{0}^{2}\right] }
\end{equation}
and:
\begin{equation}
\gamma =-\frac{v_{0}^{2}\left[ a^{2}+6\left( 1-v_{0}^{2}\right) R_{0}^{2}%
\right] }{4\left( 1-v_{0}^{2}\right) \left[ a^{2}+4\left( 1-v_{0}^{2}\right)
R_{0}^{2}\right] }
\end{equation}
respectively. Thus, the geometric part of the curvature-matter coupling can be completely reconstructed from the observational data on the galactic rotation curves.

As one can see from Equation~(%
\ref{vtg}), in the limit of large $r$, when $\nu '\rightarrow 0$ (in the case of the Schwarzschild metric $\nu '\approx 2M_B/r^2$), the tangential velocity of test particles at infinity is given by:
\begin{equation}\label{vtg1}
v_{tg\infty}^{2}=\frac{r }{%
r+Q/Q^{\prime }}
\end{equation}
which, due to the presence of the curvature-matter coupling, does not decay to zero at large distances from the galactic center, a behavior that is perfectly consistent with the observational data and that is usually attributed to the existence of dark matter.

By using the simple observational fact of the constancy of the galactic rotation curves, the curvature-matter coupling function can be completely reconstructed, without any supplementary assumption.
If, for simplicity, we consider again that the metric in the vacuum outside the
galaxy can be approximated by the Schwarzschild metric, with $\exp \left(
\nu \right) =1-2GM_{B}/r$, where $M_{B}$ is the mass of the baryonic matter
of the galaxy, then in the limit of large $r$, we have $\nu \rightarrow 0$.
Therefore,from Equation~(\ref{coup}), we obtain:
\begin{equation}
\zeta \lim_{r\rightarrow \infty }f_{2}\left( R\right) \approx \frac{%
v_{tg}^{2}}{2\left( 1-v_{tg}^{2}\right) }\ln \frac{r}{r_{0}}
\end{equation}
If the galactic rotation velocity profiles and the galactic metric are known, the coupling function can be reconstructed exactly over the entire mass distribution of the galaxy.

One can formally associate an approximate ``dark matter'' mass profile $M_{DM}(r)$ with the tangential velocity profile, which is determined by the non-minimal curvature-matter coupling and is given by:
\begin{equation}\label{darkmass}
M_{DM}(r)\approx \frac{1}{2G}\frac{r^{2}\left( \nu ^{\prime }Q+2Q^{\prime
}\right) }{rQ^{\prime }+Q}
\end{equation}

The corresponding ``dark matter'' density profile $\rho _{DM}\left( r\right) $ can be
obtained as:
\begin{eqnarray}
\rho _{DM}\left( r\right) =\frac{1}{4\pi r^{2}}\frac{dM}{dr}=\frac{1}{4\pi G}\times
\left[ \frac{\nu ^{\prime }Q+2Q^{\prime }}{r\left( rQ^{\prime }+Q\right) }+%
\frac{\nu ^{\prime \prime }Q+\nu Q^{\prime }+2Q^{\prime \prime }}{2\left(
rQ^{\prime }+Q\right) }\right. -
\left.\frac{\left( \nu ^{\prime }Q+2Q^{\prime }\right)
\left( rQ^{\prime \prime }+2Q^{\prime }\right) }{2\left( rQ^{\prime
}+Q\right) ^{2}}\right]\nonumber\\
\end{eqnarray}

\subsection{Constraining the Curvature-Matter Coupling with Galactic Stellar Distributions}

Other observational constraints on $M_{DM}$ and $\rho _{DM}$ can be obtained from the study of the galactic stellar populations. We assume that each galaxy consists of a single, pressure-supported stellar population
that is in dynamic equilibrium and traces an underlying gravitational potential resulting from the non-minimal curvature-matter coupling. In spherical symmetry, the equivalent mass profile
induced by the geometry-matter coupling (the mass profile of the ``dark matter'' halo)
relates to the moments of the stellar distribution function via the Jeans
equation \cite{BT08}:
\begin{equation}
\frac{d}{dr}\left[ \rho _{s}\left\langle v_{r}^{2}\right\rangle \right] +%
\frac{2\rho _{s}\left( r\right) \beta }{r}=-\frac{G\rho _{s}M_{DM}(r)}{r^{2}}%
\end{equation}%
where $\rho _{s}(r)$, $\left\langle v_{r}^{2}\right\rangle $ and $\beta
(r)=1-\left\langle v_{\theta }^{2}\right\rangle /\left\langle
v_{r}^{2}\right\rangle $ describe the three-dimensional density, radial
velocity dispersion and orbital anisotropy of the stellar component, where $\left\langle v_{\theta }^{2}\right\rangle$ is the tangential velocity dispersion. By
assuming that the anisotropy is a constant, the Jeans equation has the
solution \cite{MaLo05}:
\begin{equation}
\rho _{s}\left\langle v_{r}^{2}\right\rangle =Gr^{-2\beta
}\int_{r}^{\infty }s^{2(1-\beta )}\rho _{s}\left( s\right) M_{DM}\left(
s\right) ds
\end{equation}

With the use of Equation~(\ref{darkmass}), we obtain for the stellar velocity dispersion equation:
\begin{equation}\label{integral}
\rho _{s}\left\langle v_{r}^{2}\right\rangle \approx \frac{1}{2}r^{-2\beta }\int_{r}^{\infty }s^{2(2-\beta )}\rho
_{s}\left( s\right) \frac{\nu ^{\prime }(s)Q(s)+2Q^{\prime }(s)}{sQ^{\prime
}(s)+Q(s)}ds
\end{equation}

After projection along the line of sight, the ``dark matter'' mass profile can be related to two observable profiles, the projected stellar density $I(R)$, and to the stellar velocity dispersion $\sigma _p(R)$, according to the relation \cite{BT08}:
\begin{equation}
\sigma _{P}^{2}(R)=\frac{2}{I(R)}\int_{R}^{\infty }\left( 1-\beta \frac{R^{2}%
}{r^{2}}\right) \frac{\rho _{s}\left\langle v_{r}^{2}\right\rangle r}{\sqrt{%
r^{2}-R^{2}}}dr
\end{equation}

Given a projected stellar density model $I(R)$, one recovers the
three-dimensional stellar density from~\cite{BT08}:
\begin{equation}
\rho _{s}(r)=-(1/\pi
)\int_{r}^{\infty }\left( \frac{dI}{dR}\right) \left( R^{2}-r^{2}\right) ^{-1/2}dR
\end{equation}

 Therefore, once the stellar density profile $I(R)$, the stellar velocity dispersion $\left\langle v_{r}^{2}\right\rangle$ and the galactic metric
 are known, with the use of the integral Equation~(\ref{integral}), one can obtain the explicit form of the curvature-matter coupling function $Q$ and the equivalent mass profile induced by the non-minimal
coupling between matter and curvature. The simplest analytic projected density profile is the Plummer profile \cite{BT08}, given by:
\begin{equation}
I(R)={\cal L}\left(\pi r_{half}^2\right)^{-1}\left(1+R^2/r_{half}^2\right)^{-2}
\end{equation}
where ${\cal L}$ is the total luminosity and $r_{half}$ is the projected half-light radius (the radius of the cylinder that encloses half of the total luminosity).

\subsection{Stability of the Stable Circular Orbits in Modified Gravity with Curvature-Matter Coupling}

An important physical requirement for the circular orbits of the test
particles moving around galaxies is that they must be stable. Let $r_{0}$ be a
circular orbit and consider a perturbation of it of the form $r=r_{0}+\delta
$, where $\delta \ll r_{0}$ \citep{La03}. Taking expansions of $V_{eff}\left(
r\right) $, $\exp \left( \nu +\lambda \right) $ and $Q^{2}(r)$ about $%
r=r_{0} $, it~follows from Equation~(\ref{energy}) that:
\begin{equation}
\ddot{\delta}+\frac{1}{2}Q^{2}\left( r_{0}\right) e^{\nu \left( r_{0}\right)
+\lambda \left( r_{0}\right) }V_{eff}^{\prime \prime }\left( r_{0}\right)
\delta =0
\end{equation}

The condition for the stability of the simple circular orbits requires $%
V_{eff}^{\prime \prime }\left( r_{0}\right) >0$ \citep{La03}. This gives for the coupling function $Q$ the
constraint:
\begin{eqnarray}
\left.\frac{d^{2}Q}{dr^{2}}\right| _{r=r_{0}}<
\left.\left[ \nu ^{\prime \prime
}\left( \frac{l^{2}}{r^{2}}+Q^{2}\right) +\nu ^{\prime }\left( -\frac{2l^{2}%
}{r^{2}}+2QQ^{\prime }\right) -\frac{6l^{2}}{r^{4}}\right] \right|
_{r=r_{0}}
\end{eqnarray}
a condition that must be satisfied at any point $r_0$ of the galactic space-time.

From the observational, as well as from the theoretical point of view, an important problem is to estimate an upper bound for the cutoff of the constancy of the tangential velocities. If in the large $r$ limit, the coupling function satisfies the condition $Q'/Q\rightarrow 0$, then $Q/Q'\rightarrow\infty$, and in this limit, the tangential velocity decays to zero. If the exact functional form of $Q$ is known, the value of $r$ at which the rotational velocity becomes zero can be accurately estimated.

\section{Discussion and Conclusions}\label{Sec:Concl}

In the present paper, we have reviewed the cosmological and astrophysical applications of generalized curvature-matter couplings in $f(R)$-type gravity models, and we have extensively analysed some of their applications. Specific models were explored and presented in detail. The gravitational field equations in the metric formalism, in the presence of a nonminimal coupling between curvature and matter, were presented, as well as the equations of motion for test particles, which follow from the covariant divergence of the energy-momentum tensor. Generally, the motion is non-geodesic and takes place in the presence of an extra force orthogonal to the four-velocity. The Newtonian limit of the equation of motion was also described, and a procedure for obtaining the energy-momentum tensor of the matter in the framework of these gravity models was presented. On the other hand, it was shown that the gravitational field equations are equivalent to the effective Einstein equations of the $f(R)$ model in empty spacetime, but differ from them, as well as from standard General Relativity, in the presence of matter. Therefore, the predictions of these gravity theories could lead to some major differences, as compared to the predictions of standard General Relativity or to its extensions that ignore the role of matter, in several problems of current interest, such as cosmology, dark matter, dark energy, gravitational collapse or the generation of gravitational waves.

We have also reviewed, based on the already existing literature, the implications of the curvature-matter coupling models on the galactic dynamics, and we have shown that the behavior of the neutral hydrogen gas clouds outside of the galaxies and their flat rotation curves can be explained in terms of a non-minimal coupling between matter and curvature. We have also shown that in the curvature-matter coupling ``dark matter'' model, all of the relevant physical quantities, including the ``dark~mass'' associated with the coupling, and which plays the role of dark matter, its corresponding density profile, as well as the curvature-matter coupling function can be expressed in terms of observable parameters: the tangential velocity, the baryonic (luminous) mass and the Doppler frequency shifts of test particles moving around the galaxy. Therefore, this opens the possibility of directly testing the modified gravity models with non-minimal coupling between matter and geometry by using direct astronomical and astrophysical observations at the galactic or extra-galactic scale. Since the observations on the galactic rotation curves are obtained from the Doppler frequency shifts, we have generalized the expression of the frequency shifts by including the effect of the curvature-matter coupling. Thus, at least in principle, the coupling function can be obtained directly from astronomical observations. We~have also presented two other possibilities for testing these classes of gravity theories, at the level of the Solar System, by using the perihelion precession of Mercury, and at astrophysical scales, from possible observations of the gravitational tidal effects.

As future avenues of research, one should aim to characterize as much as possible the phenomenology predicted by these theories with curvature-matter couplings in order to find constraints arising from observations. The study of these phenomena may also provide some specific signatures and effects, which could distinguish and discriminate between the various theories of modified gravity. We also propose to use a background metric to analyse the dynamic system for specific curvature-matter coupling models and to use the data of  SNIa (Type Ia supernovae), BAO and CMB shift parameter to obtain restrictions for the respective models and to explore in detail the analysis of structure formation.

\acknowledgements{Acknowledgments}

 Francisco S.N. Lobo  acknowledges financial support of the Funda\c{c}\~{a}o para a
Ci\^{e}ncia e Tecnologia through an Investigador  FCT (Funda\c{c}\~{a}o para a
Ci\^{e}ncia e Tecnologia) 
 Research contract, with
Reference IF/00859/2012, funded by  FCT/MCTES (Funda\c{c}\~{a}o para a
Ci\^{e}ncia e Tecnologia e Ministerio da Ciencia, Tecnologia e Ensino Superior) 
 (Portugal) and grants
CERN/FP/123618/2011 and EXPL/FIS-AST/1608/2013.


\authorcontributions{Author Contributions}

 Both authors have contributed to the writing of this manuscript through their joint published work over the past few years. 


\conflictofinterests{Conflicts of Interest}

 The authors declare no conflict of interest. 

\newpage

\bibliographystyle{mdpi}
\makeatletter
\renewcommand\@biblabel[1]{#1. }
\makeatother


\begin{thebibliography}{-------}

\bibitem{1}
Perlmutter, S.; Aldering, G.; Goldhaber, G.; Knop, R.A.; Nugent, P.; Castro, P.G.; Deustua, S.; Fabbro, S.; Goobar, A.; Groom, D.E.; {\it et al}. Measurements of $\Omega$ and $\Lambda$ from 42 High-Redshift Supernovae. {\em Astrophys. J.} {\bf 1999}, {\em 517}, 565--586.

\bibitem{2}
Riess, A.G.; Filippenko, A.V.; Challis, P.; Clocchiattia, A.; Diercks, A.; Garnavich, P.M.; Gilliland, R.L.; Hogan, C.J.; Jha, S.; Kirshner, R.P.; {\it et al}. Observational Evidence from Supernovae for an Accelerating Universe and a Cosmological Constant. {\em Astron. J.} {\bf 1998}, {\em 116}, 1009--1038.

\bibitem{3}
Riess, A.G.; Strolger, L.-G.; Tonry, J.; Casertano, S.; Ferguson, H.C.; Mobasher, B.; Challis, P.; Filippenko, A.V.; Jha, S.; Li, W.; {\it et al}. Type Ia Supernova Discoveries at $z$ > 1 from the
Hubble Space Telescope: Evidence for Past Deceleration and Constraints on Dark Energy Evolution. {\em Astrophys. J.} {\bf 2004}, {\em 607}, 665--687.

\bibitem{4}
Maartens, R.; Koyama, K. Brane-world gravity. {\em Living Rev. Relativ.} {\bf 2004}, {\em 7},  7.

\bibitem{5}
Lobo, F.S.N. The Dark side of gravity: Modified theories of gravity. {\em ArXiv E-Prints}, {\bf 2009}, \href{http://xxx.lanl.gov/abs/0807.1640}{{arXiv:0807.1640}}.

\bibitem{7}
Nojiri, S.; Odintsov, S.D. Unified cosmic history in modified gravity: From $F(R)$ theory to Lorentz non-invariant models. \textit{Phys. Rep.} {\bf 2011}, \textit{505}, 59--144. 

\bibitem{8}
Nojiri, S.; Odintsov, S.D.; Sasaki, M. Gauss-Bonnet dark energy. \textit{Phys. Rev. D} {\bf 2005}, \textit{71},  doi:10.1103/PhysRevD.71.123509.

\bibitem{6}
Carroll, S.M.; Duvvuri, V.; Trodden, M.; Turner, M.S. Is cosmic speed-up due to new gravitational physics? \textit{Phys. Rev. D} {\bf 2004}, \textit{70}, doi:10.1103/PhysRevD.70.043528.

\bibitem{Bertolami:2007gv}
Bertolami, O.; Boehmer, C.G.; Harko, T.; Lobo, F.S.N. Extra force in $f(R)$ modified theories of gravity. \textit{Phys. Rev. D} {\bf 2007}, \textit{75},  doi:10.1103/PhysRevD.75.104016.

\bibitem{Nojiri:2004bi}
Nojiri, S.; Odintsov, S.D. Gravity assisted dark energy dominance and cosmic
acceleration. \textit{Phys. Lett. B} {\bf 2004}, \textit{599}, 137--142. 
 %

\bibitem{Nojiri:2004bi2}
Nojiri, S.; Odintsov, S.D. Dark energy and cosmic speed-up from consistent modified gravity. \textit{PoS WC} {\bf 2004}, \textit{2004}, 024. 
%

\bibitem{Nojiri:2004bi3}
Allemandi, G.; Borowiec, A.; Francaviglia, M.; Odintsov, S.D. Dark energy dominance and cosmic acceleration in first order formalism. \textit{Phys. Rev. D} {\bf 2005}, \textit{72}, doi:10.1103/\linebreak PhysRevD.72.063505.

\bibitem{Bamba:2008ja}
Bamba, K.; Odintsov, S.D. Inflation and late-time cosmic acceleration in non-minimal Maxwell-$F(R)$ gravity and the generation of large-scalemagnetic fields. \textit{J. Cosmol. Astropart. Phys.} {\bf 2008}, {\em 2008}, doi:10.1088/1475-7516/2008/04/024.

\bibitem{Bamba:2008xa}
Bamba, K.; Nojiri, S.; Odintsov, S.D. Inflationary cosmology and the late-time accelerated expansion of the universe in non-minimal Yang-Mills-$F(R)$ gravity and non-minimal vector-$F(R)$ gravity. \textit{Phys. Rev. D} {\bf 2008}, \textit{77}, doi:10.1103/PhysRevD.77.123532. 

\bibitem{Harko:2010mv}
Harko, T.; Lobo, F.S.N. $f(R,L_{m}$) gravity. \textit{Eur. Phys. J. C} {\bf 2010}, \textit{70}, 373--379.

\bibitem{Harko:2011kv}
Harko, T.; Lobo, F.S.N.; Nojiri, S.; Odintsov, S.D. $f(R,T)$ gravity. \textit{Phys. Rev. D} {\bf 2011}, \textit{84}, doi:10.1103/PhysRevD.84.024020.

\bibitem{Faraoni}
Faraoni, V. {\it Cosmology in Scalar-Tensor Gravity};  Springer: Berlin, Germany,  2004.

\bibitem{BPT06}
Bertolami, O.; Paramos, J.; Turyshev, S. General theory of relativity: Will it survive the next decade? In {\em Lasers, Clocks, and Drag-Free Control}; Springer: Berlin, Germany, 2006; pp. 27--67.

\bibitem{Will:2014xja}
Will, C.M. The Confrontation between General Relativity and Experiment.
{\em ArXiv E-Prints}, {\bf 2005}, \href{http://xxx.lanl.gov/abs/gr-qc/0510072}{{arXiv:gr-qc/0510072}}.


\bibitem{BPL07}
Bertolami, O.; Pedro, F.G.; le Delliou, M. Dark Energy-Dark Matter Interaction and the Violation of the Equivalence Principle from the Abell Cluster A586. \textit{Phys. Lett. B} {\bf 2007}, \textit{654},  165--169. 

\bibitem{Damour:2001fn}
Damour, T.  Questioning the equivalence principle. 
{\em ArXiv E-Prints}, {\bf 2001}, \href{http://xxx.lanl.gov/abs/gr-qc/0109063}{{arXiv:gr-qc/0109063}}.

\bibitem{Damour:1996xt}
Damour, T. Testing the equivalence principle: Why and how? \textit{Class. Quantum Gravity} {\bf 1996}, \textit{13},   doi:10.1088/0264-9381/13/11A/005.

\bibitem{Damour:2010rp}
Damour, T.; Donoghue, J.F. Equivalence Principle Violations and Couplings of a Light Dilaton. \textit{Phys. Rev. D} {\bf 2010}, \textit{82}, doi:10.1103/PhysRevD.82.084033.  

\bibitem{Harko:2012hm}
Harko, T.; Lobo, F.S.N.; Minazzoli, O. Extended $f(R,L_m)$ gravity with generalized scalar
field and kinetic term dependences. \textit{Phys. Rev. D} {\bf 2013}, \textit{87}, doi:10.1103/PhysRevD.87.047501. 

\bibitem{Haghani:2013oma}
Haghani, Z.; Harko, T.; Lobo, F.S.N.; Sepangi, H.R.; Shahidi, S. Further matters in space-time geometry: $f(R,T,R_{\mu\nu}T^{\mu\nu})$ gravity. \textit{Phys. Rev. D} {\bf 2013}, \textit{88}, doi:10.1103/PhysRevD.88.044023. 

\bibitem{Odintsov:2013iba}
Odintsov, S.D.; S\'{a}ez-G\'{o}mez, D. $f(R, T, R_\mu\nu T^\mu\nu)$ gravity phenomenology and $\Lambda$CDM universe. \textit{Phys. Lett. B} {\bf 2013}, \textit{725},  437--444.

\bibitem{Haghani:2014ina}
Haghani, Z.; Harko, T.; Sepangi, H.R.; Shahidi, S. Matter may matter.
{\em ArXiv E-Prints}, {\bf 2014}, \href{http://xxx.lanl.gov/abs/1405.3771}{{arXiv:1405.3771}}.

\bibitem{deser}
Deser, S.; Gibbons, G.W. Born-Infeld-Einstein actions? \textit{Class. Quantum Gravity} {\bf 1998}, \textit{15}, doi:10.1088/0264-9381/15/5/001.

\bibitem{Dolgov:2003px}
Dolgov, A.D.; Kawasaki, M. Can modified gravity explain accelerated cosmic expansion?
\textit{Phys. Lett. B} {\bf 2003}, \textit{573},  1--4. 

\bibitem{Unzicker:2005in}
Unzicker, A.; Case, T.
 Translation of Einstein's attempt of a unified field
theory with teleparallelism. 
{\em ArXiv E-Prints}, {\bf 2005}, \href{http://xxx.lanl.gov/abs/physics/0503046}{{arXiv:physics/0503046}}.


\bibitem{TEGR1}
M\"{o}ller, C. Conservation laws and absolute parallelism in general relativity. {\em Math.  Phys. Skr. Danske Vid. Selsk.} {\bf 1961}, {\em 1}, 3--50.

\bibitem{TEGR2}
Pellegrini, C.; Plebanski, J. Tetrad fields and gravitational fields. {\em Math. Phys. Skr. Danske Vid. Selsk.} {\bf 1963}, {\em 2}, 1--39.


\bibitem{Hayashi:1979qx}
Hayashi, K.; Shirafuji, T. New general relativity. \textit{Phys. Rev. D} {\bf 1979}, \textit{19}, doi:10.1103/\linebreak PhysRevD.19.3524.

\bibitem{Maluf:2013gaa}
Maluf, J.W. The teleparallel equivalent of general relativity. \textit{Ann. Phys.} {\bf 2013}, \textit{525},  339--357.


\bibitem{JGPereira}
Aldrovandi, R.; Pereira, J.G. {\it Teleparallel
Gravity: An Introduction}; Springer: Dordrecht, The~Netherlands, 2013.


\bibitem{Ferraro:2006jd1}
Ferraro, R.; Fiorini, F. Modified teleparallel gravity: Inflation without an inflaton. \textit{Phys. Rev. D} {\bf 2007}, \textit{75}, doi:10.1103/PhysRevD.75.084031.

\bibitem{Ferraro:2006jd2}
Bengochea, G.R.; Ferraro, R. Dark torsion as the cosmic speed-up. \textit{Phys. Rev. D} {\bf 2009}, \textit{79}, 124019. doi:10.1103/PhysRevD.79.124019.

\bibitem{Linder:2010py}
Linder, E.V. Einstein's other Gravity and the Acceleration of the Universe. \textit{Phys. Rev. D} {\bf 2010}, \textit{81}, doi:10.1103/PhysRevD.81.127301 .


\bibitem{Harko:2014sja}
Harko, T.; Lobo, F.S.N.; Otalora, G.; Saridakis, E.N. Non-minimal torsion-matter coupling
extension of $f(T)$ gravity. \textit{Phys. Rev. D} {\bf 2014}, \textit{89}, doi:10.1103/PhysRevD.89.124036.

\bibitem{Harko:2014aja}
Harko, T.; Lobo, F.S.N.; Otalora, G.; Saridakis, E.N. $f(T,\mathcal{T})$ gravity and cosmology.  {\em ArXiv~E-Prints}, {\bf 2014}, \href{http://xxx.lanl.gov/abs/1405.0519}{{ 	arXiv:1405.0519}}.

\bibitem{Kiani:2013pba}
Kiani, F.; Nozari, K. Energy conditions in $F(T,\Theta)$ gravity and compatibility with a stable de
Sitter solution. \textit{Phys. Lett. B} {\bf 2014}, \textit{728},  554--561. 




\bibitem{Harko:2010vs}
Harko, T. Galactic rotation curves in modified gravity with non-minimal coupling between matter and geometry. \textit{Phys. Rev. D} {\bf 2010}, \textit{81}, doi:10.1103/PhysRevD.81.084050 

\bibitem{Sotiriou:2008it}
Sotiriou, T.P.; Faraoni, V. Modified gravity with \textit{R}-matter couplings and (non-)geodesic
motion. \textit{Class. Quantum Gravity} {\bf 2008}, \textit{25}, doi:10.1088/0264-9381/25/20/205002.


\bibitem{Damour:1994zq}
Damour, T.; Polyakov, A.M. The String dilaton and a least coupling principle. \textit{Nucl. Phys.
B} {\bf 1994}, \textit{423},  532--558.


\bibitem{Damour:1992we}
Damour, T.; Esposito-Farese, G. Tensor-multi-scalar theories of gravitation. \textit{Class. Quantum Gravity} {\bf 1992}, \textit{9}, doi:10.1088/0264-9381/9/9/015.


\bibitem{Gottlober:1989ww}
Gottlober, S.; Schmidt, H.J.; Starobinsky, A.A. Sixth-order gravity and conformal
transformations. \textit{Class. Quantum Gravity} {\bf 1990}, \textit{7}, doi:10.1088/0264-9381/7/5/018.


\bibitem{Drummond:1979pp} 
Drummond, I.T.; Hathrell, S.J. QED Vacuum Polarization in a Background Gravitational Field and Its Effect on the Velocity of Photons. \textit{Phys. Rev. D} {\bf 1980}, \textit{22}, doi:10.1103/PhysRevD.22.343.


\bibitem{Bu70} 
Buchdahl, H.A. Non-linear Lagrangians and cosmological theory. \textit{Mon. Not. R. Astron. Soc.} {\bf 1970}, \textit{150}, 1--8.

\bibitem{Bar83} 
Barrow, J.D.; Ottewill, A.C. The stability of general relativistic cosmological theory. \textit{J. Phys. A Math. Gen.} {\bf 1983}, \textit{16}, 2757--2776.

\bibitem{Ko06} 
Koivisto, T. A note on covariant conservation of energy-momentum in modified gravities. \textit{Class. Quantum Gravity} {\bf 2006}, \textit{23}, doi:10.1088/0264-9381/23/12/N01.

\bibitem{equiv1} 
Teyssandier, P.; Tourrenc, Ph. The Cauchy problem for the $R+R^2$ theories of gravity without torsion. \textit{J. Math. Phys.} {\bf 1983}, \textit{24},  doi:10.1063/1.525659.

\bibitem{equiv2} 
Whitt, B. Fourth-order gravity as general relativity plus matter. \textit{Phys. Lett. B} {\bf 1984}, \textit{145},  176--178. 

\bibitem{equiv3} 
Wands, D. Extended gravity theories and the Einstein--Hilbert action. \textit{Class. Quantum Gravity} {\bf 1994}, \textit{11}, doi:10.1088/0264-9381/11/1/025.

\bibitem{equiv4} 
Faraoni, V. De Sitter space and the equivalence between $f(R)$ and scalar-tensor gravity. \textit{Phys.~Rev.~D} {\bf 2007}, \textit{75}, doi:10.1103/PhysRevD.75.067302.

\bibitem{Olmo07} 
Olmo, G.J. Limit to general relativity in $f(R)$ theories of gravity. \textit{Phys. Rev. D} {\bf 2007}, \textit{75}, doi:10.1103/PhysRevD.75.023511.

\bibitem{Harko:2010hw}
Harko, T.; Koivisto, T.S.; Lobo, F.S.N. Palatini formulation of modified gravity with a~non-minimal curvature-matter coupling. \textit{Mod. Phys. Lett. A} {\bf 2011}, \textit{26}, doi:10.1142/\linebreak S0217732311035869.

\bibitem{Mohseni:2009ns}
Mohseni, M. Non-geodesic motion in f(G) gravity with non-minimal coupling. \textit{Phys. Lett. B} {\bf 2009}, \textit{682}, 89--92. 


\bibitem{Bertolami:2008ab}
Bertolami, O.; Lobo, F.S.N.; P\'{a}ramos, J. Nonminimum coupling of perfect fluids to curvature. \textit{Phys. Rev. D} {\bf 2008}, \textit{78}, doi:10.1103/PhysRevD.78.064036.

\bibitem{Faraoni:2009rk}
Faraoni, V.  Lagrangian description of perfect fluids and modified gravity with an extra
force. \textit{Phys. Rev. D} {\bf 2009}, \textit{80}, doi:10.1103/PhysRevD.80.124040.


\bibitem{Bertolami:2013raa}
Bertolami, O.; P\'{a}ramos, J.  Homogeneous spherically symmetric bodies with a nonminimal coupling between curvature and matter: The choice of the Lagrangian density for matter. {\em ArXiv~E-Prints}, {\bf 2013}, 
\href{http://xxx.lanl.gov/abs/1306.1177}{{arXiv:1306.1177}}.

\bibitem{Minazzoli:2013bva}
Minazzoli, O. Conservation laws in theories with universal gravity/matter coupling. \textit{Phys.~ Rev.~D} {\bf 2013}, \textit{88}, doi:10.1103/PhysRevD.88.027506. 

\bibitem{Bisabr:2013laa}
Bisabr, Y. Non-minimal Gravitational Coupling of Phantom and Big Rip Singularity. \textit{Gen.~Relativ.~Gravit.} {\bf 2013}, \textit{45}, 1559--1566. 

\bibitem{Harko:2008qz}
Harko, T. Modified gravity with arbitrary coupling between matter and geometry. \textit{Phys. Lett. B} {\bf 2008}, \textit{669},  376--379. 

\bibitem{Wu:2014upa}
Wu, Y.-B.; Zhao, Y.-Y.; Lu, J.-W.; Zhang, X.; Zhang, C.-Y.; Qiao, J.-W. Five-dimensional generalized $f(R)$ gravity with curvature-matter coupling. \textit{Eur. Phys. J. C }{\bf 2014}, \textit{74}, doi:10.1140/\linebreak epjc/s10052-014-2791-9. 

\bibitem{Olmo:2014sra}
Olmo, G.J.; Rubiera-Garcia, D. Brane-world and loop cosmology from a gravity-matter coupling perspective.  {\em ArXiv E-Prints}, {\bf 2014}, 
\href{http://xxx.lanl.gov/abs/1405.7184}{{arXiv:1405.7184}}.

\bibitem{Harko:2010zi}
Harko, T. The matter Lagrangian and the energy-momentum tensor in modified gravity with nonminimal coupling between matter and geometry. \textit{Phys. Rev. D} {\bf 2010}, \textit{81}, doi:10.1103/\linebreak PhysRevD.81.044021. 

 \bibitem{LaLi} 
Landau, L.D.; Lifshitz, E.M. {\it The Classical Theory of Fields}; Butterworth-Heinemann: Oxford, UK, 1998.

\bibitem{Fock} 
Fock, V. {\it The Theory of Space, Time and Gravitation}; Pergamon Press: London, UK, 1959. 
 
 \bibitem{Faraoni:2007sn}
Faraoni, V.  Viability criterion for modified gravity with an extra force. \textit{Phys. Rev. D} {\bf 2007}, \textit{76}, doi:10.1103/PhysRevD.76.127501.  

\bibitem{Puetzfeld:2008xu}
Puetzfeld, D.; Obukhov, Y.N.  Motion of test bodies in theories with nonminimal coupling.  \textit{Phys. Rev. D} {\bf 2008}, \textit{78}, doi:10.1103/PhysRevD.78.121501.


\bibitem{Bertolami:2011rb}
Bertolami, O.; Martins, A. On the dynamics of perfect fluids in non-minimally coupled gravity. \textit{Phys. Rev. D} {\bf 2012}, \textit{85},  doi:10.1103/PhysRevD.85.024012.  

\bibitem{Bertolami:2007vu}
Bertolami, O.; Paramos, J. Do $f(R)$ theories matter? \textit{Phys. Rev. D} {\bf 2008}, 77, doi:10.1103/\linebreak PhysRevD.77.084018. 

\bibitem{Bertolami:2009cd}
Bertolami, O.; Sequeira, M.C. Energy Conditions and Stability in $f(R)$ theories of gravity with non-minimal coupling to matter. \textit{Phys. Rev. D} {\bf 2009}, \textit{79},  doi:10.1103/PhysRevD.79.104010. 


\bibitem{Wang:2010zzr}
Wang, J.; Wu, Y.-B.; Guo, Y.-X.; Yang, W.-Q.; Wang, L. Energy conditions and stability in generalized $f(R)$ gravity with arbitrary coupling between matter and geometry. \textit{Phys. Lett. B} {\bf 2010}, \textit{689}, 133--138.


\bibitem{Wang:2010bh}
Wang, J.; Wu, Y.-B.; Guo, Y.-X.; Qi, F.; Zhao, Y.-Y.; Sun, X.-Y. Conditions and instability in $f(R)$ gravity with non-minimal coupling between matter and geometry. \textit{Eur. Phys. J. C} {\bf 2010}, \textit{69}, {541--546}. 

\bibitem{Wang:2012mws}
Wang, J.; Wu, Y.-B.; Guo, Y.-X.; Yang, W.-Q.; Wang, L. Energy conditions and stability in
generalized $f(R)$ gravity with arbitrary coupling between matter and geometry. \textit{Phys. Lett. B} {\bf 2010}, \textit{689}, {133--138}. 

\bibitem{Sotiriou:2008dh}
 Sotiriou, T.P. The viability of theories with matter coupled to the Ricci scalar. \textit{Phys. Lett. B} {\bf 2008}, \textit{664}, {225--228}.

\bibitem{Sotiriou:2008dh2}
Bertolami, O.; Paramos, J. On the non-trivial gravitational coupling to matter. \textit{Class. Quantum Gravity} {\bf 2008}, \textit{25}, doi:10.1088/0264-9381/25/24/245017.


\bibitem{Tamanini:2013aca}
Tamanini, N.; Koivisto, T.S. Consistency of non-minimally coupled $f(R)$ gravity. \textit{Phys. Rev. D} {\bf 2013}, \textit{88}, doi:10.1103/PhysRevD.88.064052.


\bibitem{Obukhov:2013ona}
Obukhov, Y.N.; Puetzfeld, D. Conservation laws in gravitational theories with general nonminimal coupling. \textit{Phys. Rev. D} {\bf 2013}, \textit{87}, doi:10.1103/PhysRevD.87.081502.

\bibitem{Puetzfeld:2013ir}
Puetzfeld, D.; Obukhov, Y.N. Covariant equations of motion for test bodies in gravitational theories with general nonminimal coupling.  \textit{Phys. Rev. D} {\bf 2013}, \textit{87}, doi:10.1103/\linebreak PhysRevD.87.044045.  

\bibitem{Puetzfeld:2013sca}
Puetzfeld, D.; Obukhov, Y.N. Equations of motion in gravity theories with nonminimal coupling: A loophole to detect torsion macroscopically?  \textit{Phys. Rev. D} {\bf 2013}, \textit{88}, doi:10.1103/\linebreak PhysRevD.88.064025.

\bibitem{Bertolami:2013qaa}
Bertolami, O.; March, R.; P\'{a}ramos, J. Solar System constraints to nonminimally coupled gravity. \textit{Phys. Rev. D} {\bf 2013}, \textit{88}, doi:10.1103/PhysRevD.88.064019.

\bibitem{Castel-Branco:2014exa}
Castel-Branco, N.; P\'{a}ramos, J.; March, R.  Perturbation of the metric around a spherical body from a nonminimal coupling between matter and curvature. {\em ArXiv E-Prints}, {\bf 2014},  \href{http://xxx.lanl.gov/abs/1403.7251}{{arXiv:1403.7251}}.

\bibitem{Garcia:2010xb}
Garcia, N.M.; Lobo, F.S.N. Wormhole geometries supported by a nonminimal curvature-matter coupling. \textit{Phys. Rev. D} {\bf 2010}, \textit{82}, doi:10.1103/PhysRevD.82.104018. 

\bibitem{MontelongoGarcia:2010xd}
Garcia, N.M.; Lobo, F.S.N. Nonminimal curvature-matter coupled wormholes with matter satisfying the null energy condition. \textit{Class. Quantum Gravity} {\bf 2011}, \textit{28}, doi:10.1088/0264-9381/\linebreak28/8/085018.

\bibitem{Bertolami:2012fz}
Bertolami, O.; Ferreira, R.Z. Traversable wormholes and time machines in nonminimally coupled curvature-matter $f(R)$ theories. \textit{Phys. Rev. D} {\bf 2012}, \textit{85}, doi:10.1103/PhysRevD.85.104050. 

\bibitem{Nesseris:2008mq}
Nesseris, S. Matter density perturbations in modified gravity models with arbitrary coupling between matter and geometry. \textit{Phys. Rev. D} {\bf 2009}, \textit{79}, doi:10.1103/PhysRevD.79.044015.

\bibitem{Bertolami:2010cw}
Bertolami, O.; Frazao, P.; Paramos, J. Accelerated expansion from a nonminimal gravitational coupling to matter. \textit{Phys. Rev. D} {\bf 2010}, \textit{81}, doi:10.1103/PhysRevD.81.104046. 

\bibitem{Thakur:2010yv}
Thakur, S.; Sen, A.A.; Seshadri, T.R. Non-minimally coupled $f(R)$ cosmology. \textit{Phys. Lett. B} {\bf 2011}, \textit{696},  {309--314}. 

\bibitem{Bisabr:2012tg}
Bisabr, Y. Modified gravity with a nonminimal gravitational coupling to matter. \textit{Phys. Rev. D} {\bf 2012}, \textit{86}, doi:10.1103/PhysRevD.86.044025.

\bibitem{Bertolami:2010ke}
Bertolami, O.; Frazao, P.; Paramos, J. Reheating via a generalized non-minimal coupling of curvature to matter. \textit{Phys. Rev. D} {\bf 2011}, \textit{83}, doi:10.1103/PhysRevD.83.044010.

 \bibitem{Bertolami:2013uwl}
Bertolami, O.; Paramos, J. Modified Friedmann Equation from Nonminimally Coupled Theories of Gravity. \textit{Phys. Rev. D} {\bf 2014}, \textit{89}, doi:10.1103/PhysRevD.89.044012.

\bibitem{Bertolami:2013kca}
Bertolami, O.; Frazao, P.; Paramos, J. Cosmological perturbations in theories with non-minimal coupling between curvature and matter. \textit{J. Cosmol. Astropart. Phys.} {\bf 2013}, \textit{2013}, doi:10.1088/\linebreak1475-7516/2013/05/029.


\bibitem{Thakur:2013oya}
Thakur, S.; Sen, A.A. Can structure formation distinguish $\Lambda$CDM from nonminimal $f(R)$ gravity? \textit{Phys. Rev. D} {\bf 2013}, \textit{88}, doi:10.1103/PhysRevD.88.044043. 


\bibitem{Bertolami:2009ic}
Bertolami, O.; Paramos, J. Mimicking dark matter through a non-minimal gravitational coupling with matter. \textit{J. Cosmol. Astropart. Phys.} {\bf 2010}, \textit{2010}, doi:10.1088/1475-7516/2010/03/009.

\bibitem{Bertolami:2011ye}
Bertolami, O.; Frazao, P.; Paramos, J. Mimicking dark matter in galaxy clusters through a nonminimal gravitational coupling with matter. \textit{Phys. Rev. D} {\bf 2012}, \textit{86}, doi:10.1103/PhysRevD.86.044034.

\bibitem{Bertolami:2008zh}
Bertolami, O.; Paramos, J.; Harko, T.; Lobo, F.S.N. Non-minimal curvature-matter couplings in modified gravity. {\em ArXiv E-Prints}, {\bf 2008}, \href{http://xxx.lanl.gov/abs/0811.2876}{{arXiv:0811.2876}}.

\bibitem{Bertolami:2013xda}
Bertolami, O.; Paramos, J. Minimal extension of General Relativity: Alternative gravity model
with non-minimal coupling between matter and curvature. \textit{Int. J. Geom. Methods Mod. Phys.} {\bf 2014}, \textit{11}, doi:10.1142/S0219887814600032.


 \bibitem{Harko:2012ve}
Harko, T.; Lobo, F.S.N. Geodesic deviation, Raychaudhuri equation, and tidal forces in
modified gravity with an arbitrary curvature-matter coupling. \textit{Phys. Rev. D} {\bf 2012}, \textit{86}, doi:10.1103/PhysRevD.86.124034.

\bibitem{Harko:2012ve2}
Lobo, F.S.N.; Harko, T.  Extended $f(R,L_m)$ theories of gravity. {\em ArXiv E-Prints}, {\bf 2012},  \href{http://xxx.lanl.gov/abs/1211.0426}{{arXiv:1211.0426}}.

\bibitem{Wang:2012rw}
Wang, J.; Liao, K. Energy conditions in $f(R,L_m)$ gravity. \textit{Class. Quantum Gravity} {\bf 2012}, \textit{29}, doi:10.1088/0264-9381/29/21/215016.

\bibitem{Huang:2013dca}
Huang, R.-N.  The Wheeler-DeWitt equation of $f(R,L_m)$ gravity in minisuperspace.  {\em ArXiv E-Prints}, {\bf 2013},  \href{http://xxx.lanl.gov/abs/1304.5309}{{arXiv:1304.5309}}.

\bibitem{Tian:2014mta}
Tian, D.W.; Booth, I. Lessons from  $f(R,R_c^2,R_m^2,L_m)$ gravity: Smooth Gauss-Bonnet limit, energy-momentum conservation and nonminimal coupling. {\em ArXiv E-Prints}, {\bf 2014},  \href{http://xxx.lanl.gov/abs/1404.7823}{{arXiv:1404.7823}}.

\bibitem{Milgrom1} 
Milgrom, M. A modification of the Newtonian dynamics as a possible alternative to the hidden mass hypothesis. \textit{Astrophys. J.} {\bf 1983}, \textit{270},  365--370. 

\bibitem{Milgrom2} 
Bekenstein, J.D. Relativistic gravitation theory for the modified Newtonian dynamics paradigm. \textit{Phys. Rev. D} {\bf 2004}, \textit{70}, doi.org/10.1103/PhysRevD.70.083509.

\bibitem{Haw} 
Hawking, S.; Ellis, G.F.R. {\it The Large Scale Structure of Space-Time}; Cambridge University Press: Cambridge, UK, 1973.

\bibitem{mas1} 
Mashhoon, B. Tidal Gravitational Radiation. \textit{Astrophys. J.} {\bf 1973}, \textit{185},  83--86.

\bibitem{mas2} 
Mashhoon, B. On tidal phenomena in a strong gravitational field. \textit{Astrophys. J.} {\bf 1975}, \textit{197},  705--716.

\bibitem{mas3}
Mashhoon, B.; Theiss, D.S. Relativistic Tidal Forces and the Possibility of Measuring Them. \textit{Phys. Rev. Lett.} {\bf 1982}, \textit{49}, doi:10.1103/PhysRevLett.49.1542.

\bibitem{Oh} 
Ohanian, H.C. {\it Gravitation and Spacetime}; Norton: New York, NY, USA, 1976.

\bibitem{gasperiniPRD02} 
Gasperini, M.; Piazza, F.; Veneziano, G. Quintessence as a runaway dilaton. \textit{Phys. Rev. D} {\bf 2001}, \textit{65}, doi:10.1103/PhysRevD.65.023508.

\bibitem{damourPRD96} 
Damour, T.; Vilenkin, A. String theory and inflation. \textit{Phys. Rev. D} {\bf 1996}, \textit{53}, doi:10.1103/PhysRevD.53.2981.

\bibitem{armendarizPRD02} 
Armend\'{a}riz-Pic\'{o}n, C. Predictions and observations in theories with varying couplings. \textit{Phys.~Rev.~D} {\bf 2002}, \textit{66}, doi:10.1103/PhysRevD.66.064008.


\bibitem{dilaton2EP} 
Damour, T.; Polyakov, A.M. The string dilation and a least coupling principle. \textit{Nucl. Phys. B} {\bf 1994}, \textit{423}, 532--558.

\bibitem{dilaton2EP2} 
Damour, T.; Polyakov, A.M. String theory and gravity. \textit{Gen. Relativ. Gravit.}
{\bf 1994}, \textit{26},  1171--1176.

\bibitem{dilaton2EP3}
Damour, T.; Piazza, F.; Veneziano, G. Violations of the equivalence principle in a dilaton-runaway scenario. \textit{Phys. Rev. D} {\bf 2002}, \textit{66}, doi:10.1103/PhysRevD.66.046007.


\bibitem{EP} 
Damour, T.; Donoghue, J.F. Equivalence principle violations and couplings of a light dilaton. \textit{Phys. Rev. D} {\bf 2010}, \textit{82}, doi:10.1103/PhysRevD.82.084033.

\bibitem{EP2} 
Damour, T. Theoretical aspects of the equivalence principle. \textit{Class. Quantum Gravity} {\bf 2012}, \textit{29},  doi:10.1088/0264-9381/29/18/184001.

\bibitem{int}
Damour, T.; Gibbons, G.W.; Gundlach, C.C. Dark matter, time-varying $G$, and a dilaton field. \textit{Phys. Rev. Lett.} {\bf 1990}, \textit{64}, doi:10.1103/PhysRevLett.64.123.

\bibitem{int2}
Casas, J.A.; Garc\'{i}a-Bellido, J.; Quiros, M. Scalar-tensor theories of gravity with  Phi-dependent masses. \textit{Class. Quantum Gravity} {\bf 1992}, \textit{9},
doi:10.1088/0264-9381/9/5/018.

\bibitem{int3}
Wetterich, C. An asymptotically vanishing time-dependent cosmological constant. 
\textit{Astron. Astrophys.} {\bf 1995}, \textit{301}, 321--328. 

\bibitem{int4}
Amendola, L. Coupled quintessence. \textit{Phys. Rev. D} {\bf 2000}, \textit{62}, 
doi:10.1103/PhysRevD.62.043511.

\bibitem{NMC} 
Das, S.; Banerjee, N. Brans-Dicke scalar field as a chameleon. \textit{Phys. Rev. D} {\bf 2008}, \textit{78}, doi:10.1103/PhysRevD.78.043512.

\bibitem{min2} 
Minazzoli, O. The $\gamma$ parameter in Brans-Dicke-like (light-)Scalar-Tensor theory with a universal scalar/matter coupling and a new decoupling scenario. {\em ArXiv E-Prints}, {\bf 2012}, \href{http://xxx.lanl.gov/abs/1208.2372}{{arXiv:1208.2372}}.

\bibitem{Cota} 
Aviles, A.; Cervantes-Cota, J.L. Dark degeneracy and interacting cosmic components. \textit{Phys.~Rev.~D} {\bf 2011}, \textit{83}, doi.org/10.1103/PhysRevD.84.083515.

\bibitem{KK} 
Overduin, J.M.; Wesson, P.S. Kaluza-Klein gravity. \textit{Phys. Rep.} \textbf{1997}, {\em 283},  303--378.

Fujii, Y.; Maeda, K.I. {\it The Scalar-Tensor Theory of Gravitation}; Cambridge University Press: Cambridge, UK, 2003.



\bibitem{Min} 
Minazzoli, O.; Harko, T. New derivation of the Lagrangian of a perfect fluid with a barotropic equation of state. \textit{Phys. Rev. D} {\bf 2012}, \textit{86}, 
doi:10.1103/PhysRevD.86.087502.

\bibitem{Harko:2012ar}
Harko, T.; Lobo, F.S.N. Generalized dark gravity. \textit{Int. J. Mod. Phys. D} {\bf 2012}, \textit{21}, doi:10.1142/\linebreak S0218271812420199.


\bibitem{Poplawski:2006ey}
Poplawski, N.J.  A Lagrangian description of interacting dark energy. 
{\em ArXiv E-Prints}, {\bf 2006},  \href{http://xxx.lanl.gov/abs/gr-qc/0608031}{{arXiv:gr-qc/0608031}}.

\bibitem{Houndjo:2011tu}
Houndjo, M.J.S. Reconstruction of $f(R, T)$ gravity describing matter dominated and accelerated phases.\textit{ Int. J. Mod. Phys. D} {\bf 2012}, \textit{21},
doi:10.1142/S0218271812500034.


\bibitem{Momeni:2011am}
Jamil, M.; Momeni, D.; Raza, M.; Myrzakulov, R. Reconstruction of some cosmological models in $f(R,T)$ cosmology. \textit{Eur. Phys. J. C} {\bf 2012}, \textit{72}, doi:10.1140/epjc/s10052-012-1999-9.

\bibitem{Houndjo:2011fb}
Houndjo, M.J.S.; Piattella, O.F. Reconstructing $f(R,T)$ gravity from holographic dark energy. \textit{Int. J. Mod. Phys. D} {\bf 2012}, \textit{21}, doi:10.1142/S0218271812500241.

 \bibitem{Houndjo:2012ij}
Houndjo, M.J.S.; Batista, C.E.M.; Campos, J.P.; Piattella, O.F. Finite-time singularities in $f(R,T)$ gravity and the effect of conformal anomaly. \textit{Can. J. Phys.} {\bf 2013}, \textit{91}, 548--553.

\bibitem{Sharif:2012zzd}
Sharif, M.; Zubair, M. Thermodynamics in $f(R,T)$ Theory of Gravity. \textit{J. Cosmol. Astropart. Phys.} {\bf 2012}, \textit{2012}, doi:10.1088/1475-7516/2012/03/028.

\bibitem{Sharif:2012gz}
Sharif, M.; Rani, S.; Myrzakulov, R. Analysis of $f(R,T)$ gravity models through energy conditions. \textit{Eur. Phys. J. Plus} {\bf 2013}, \textit{128}, 
doi:10.1140/epjp/i2013-13123-0.

 \bibitem{Sharif:2012ce}
Sharif, M.; Zubair, M. Energy Conditions Constraints and Stability of Power Law Solutions in $f(R,T)$ Gravity. \textit{J. Phys. Soc. Jpn.} {\bf 2013}, \textit{82},
doi:10.7566/JPSJ.82.014002.

\bibitem{Alvarenga:2012bt}
Alvarenga, F.G.; Houndjo, M.J.S.; Monwanou, A.V.; Orou, J.B.C. Testing some $f(R,T)$ gravity models from energy conditions. \textit{J. Mod. Phys.} {\bf 2013}, \textit{4},  130--139.


\bibitem{Santos:2013bfa}
Santos, A.F. 
G\"{o}del solution in $f(R,T)$ gravity.
\textit{Mod. Phys. Lett. A} {\bf 2013}, \textit{28},  doi:10.1142/\linebreak S0217732313501411.

\bibitem{Ram:2013kya}
Priyanka, S.R. 
Some Kaluza-Klein cosmological models in $f(R,T)$ gravity theory.
\textit{Astrophys.~Space~Sci.} {\bf 2013}, \textit{347}, 389--397.

\bibitem{Naidu:2013aga}
Naidu, R.L.; Reddy, D.R.K.; Ramprasad, T.; Ramana, K.V. Bianchi type-V bulk viscous string cosmological model in $f(R,T)$ gravity. \textit{Astrophys. Space Sci.} {\bf 2013}, \textit{348},  247--252.

 \bibitem{Sharif:2014cpa}
Sharif, M.; Zubair, M. Study of Bianchi I anisotropic model in $f(R,T)$ gravity. \textit{Astrophys.~Space~Sci.} {\bf 2014}, \textit{349},  457--465. 

\bibitem{Chakraborty:2012kj}
Chakraborty, S. An alternative $f(R,T)$ gravity theory and the dark energy problem. \textit{Gen.~Relativ.~Gravit.} {\bf 2013}, \textit{45}, 2039--2052. 

 \bibitem{Alvarenga:2013syu}
Alvarenga, F.G.; de la Cruz-Dombriz, A.; Houndjo, M.J.S.; Rodrigues, M.E.; S\'{a}ez-G\'{o}mez,~D.
Dynamics of scalar perturbations in $f(R,T)$ gravity. \textit{Phys. Rev. D} {\bf 2013}, \textit{87}, doi.org/10.1103/\linebreak PhysRevD.87.103526.


\bibitem{Sharif:2014jpa}
Sharif, M.; Zubair, M. Reconstruction and stability of $f(R,T)$ gravity with Ricci and modified Ricci dark energy. \textit{Astrophys. Space Sci.} {\bf 2014}, \textit{349}, 529--537.

 \bibitem{dm1} 
Rubin, V.C.; Ford, W.K.; Thonnard, N. 
Rotational properties of 21 SC galaxies with a large range of luminosities and radii, from NGC 4605 ($R = 4$ kpc) to UGC 2885 ($R = 122$ kpc).
\textit{Astrophys.~J.} {\bf 1980}, {\em 238},  471--487.

 \bibitem{dm2} 
Persic, M.; Salucci, P.; Stel, F. 
The universal rotation curve of spiral galaxies---I. The dark matter connection.
\textit{Mon. Not. R. Astron. Soc.} {\bf 1996}, {\em 281},  27--47.

 \bibitem{dm3}  Borriello, A.; Salucci, P.;
Borriello, A.; Salucci, P. The dark matter distribution in disc galaxies. \textit{Mon. Not. R. Astron. Soc.} {\bf 2001}, \textit{323},  285--292.

 \bibitem{BT08} 
Binney, J.; Tremaine, S.
{\it Galactic Dynamics}; Princeton, N.J., Ed.; Princeton University Press: Princeton, NJ, USA, 2008.

\bibitem{OvWe04} 
Overduin, J.M.; Wesson, P.S. Dark matter and background light. \textit{Phys. Rep.} {\bf 2004}, \textit{402}, 267--406.

\bibitem{Boehmer:2007kx}
Boehmer, C.G.; Harko, T.; Lobo, F.S.N. Dark matter as a geometric effect in $f(R)$ gravity. \textit{Astropart. Phys.} {\bf 2008}, \textit{29},  386--392. 

\bibitem{Bohmer:2007fh}
Boehmer, C.G.; Harko, T.; Lobo, F.S.N. Generalized virial theorem in $f(R)$ gravity. \textit{J. Cosmol. Astropart. Phys.} {\bf 2008}, {\em 2008},
doi:10.1088/1475-7516/2008/03/024.


\bibitem{Capozziello:2012qt}
Capozziello, S.; Harko, T.; Koivisto, T.S.; Lobo, F.S.N.; Olmo, G.J. The virial theorem and the dark matter problem in hybrid metric-Palatini gravity. \textit{J. Cosmol. Astropart. Phys.} {\bf 2013}, {\em 2013}, doi:10.1088/1475-7516/2013/07/024.

\bibitem{Capozziello:2013yha}
Capozziello, S.; Harko, T.; Koivisto, T.S.; Lobo, F.S.N.; Olmo, G.J. Galactic rotation curves in hybrid metric-Palatini gravity. \textit{Astropart. Phys.} {\bf 2013}, \textit{50--52}, 65--75. 

\bibitem{Harko:2011nh}
Harko, T.; Koivisto, T.S.; Lobo, F.S.N.; Olmo, G.J. Metric-Palatini gravity unifying local constraints and late-time cosmic acceleration. \textit{Phys. Rev. D} {\bf 2012}, \textit{85}, doi:10.1103/\linebreak PhysRevD.85.084016.

\bibitem{Capozziello:2012ny}
Capozziello, S.; Harko, T.; Koivisto, T.S.; Lobo, F.S.N.; Olmo, G.J. Cosmology of hybrid
metric-Palatini $f(X)$-gravity. \textit{J. Cosmol. Astropart. Phys.} {\bf 2013}, \textit{2013}, doi:10.1088/1475-7516/\linebreak2013/04/011.


\bibitem{Capozziello:2013uya}
Capozziello, S.; Harko, T.; Lobo, F.S.N.; Olmo, G.J. Hybrid modified gravity unifying local tests, galactic dynamics and late-time cosmic acceleration. \textit{Int. J. Mod. Phys. D} {\bf 2013}, \textit{22}, doi:10.1142/\linebreak S0218271813420066.

\bibitem{Capozziello:2013gza}
Capozziello, S.; Harko, T.; Lobo, F.S.N.; Olmo, G.J.; Vignolo, S. The Cauchy problem in hybrid metric-Palatini $f(X)$-gravity. \textit{Int. J. Geom. Methods Mod. Phys.} {\bf 2014}, \textit{11}, doi:10.1142/\linebreak S021988781450042X. 


 \bibitem{Nuc01} 
Nucamendi, U.; Salgado, M.; Sudarsky, D. Alternative approach to the galactic dark matter problem. \textit{Phys. Rev. D} {\bf 2001}, \textit{63}, doi:10.1103/PhysRevD.63.125016.

\bibitem{La03} 
Lake, K. Galactic Potentials. \textit{Phys. Rev. Lett.} {\bf 2004}, \textit{92},
doi:10.1103/PhysRevLett.92.051101.

\bibitem{per} 
Salucci, P.; Persic, M. {\it Dark Halos around Galaxies, in Dark and Visible Matter in Galaxies};
Persic, M., Salucci, P., Eds.; ASP Conference Series;  Astronomical Society of the Pacific: San~Francisco, CA, USA, 1997; Volume 117, p. 1.

\bibitem{pun} 
Pun, C.S.J.; Kovacs, Z.; Harko, T. Thin accretion disks in $f(R)$ modified gravity models. \textit{Phys.~Rev.~D} {\bf 2008}, \textit{78}, doi:10.1103/PhysRevD.78.024043.


\bibitem{MaLo05} 
Mamon, G.A.; Lokas, E.L. Dark matter in elliptical galaxies---II. Estimating the mass within the virial radius. \textit{Mon. Not. R. Astron. Soc.} {\bf 2005}, \textit{363},  705--722. 

\end{thebibliography}

\end{document}